\documentclass[preprint,showpacs,preprintnumbers,amsmath,amssymb]{revtex4-1}

\usepackage{graphicx}
\usepackage{epstopdf}

\usepackage{dcolumn}
\usepackage{bm}
\usepackage{color}

\usepackage{feynmf}
\usepackage{slashed}
\usepackage{enumerate}

\usepackage[scientific-notation=true]{siunitx} 

\usepackage[utf8]{inputenc}

\usepackage[colorlinks=true,urlcolor=blue,anchorcolor=blue,citecolor=blue,filecolor=blue,linkcolor=blue,menucolor=blue,linktocpage=true]{hyperref} 

\usepackage{appendix}
\usepackage{amsmath,amsthm,amssymb,mathrsfs}
\usepackage{bbold}
\usepackage{bbm}

\begin{document}

\title{Electroweak Precision Measurements of a Nearly-Degenerate $Z^\prime$-$Z$ System}

\author{Dayun Qiu}
\thanks{qiudy3@mail2.sysu.edu.cn}
\author{Yi-Lei Tang}
\thanks{tangylei@mail.sysu.edu.cn}
\affiliation{School of Physics, Sun Yat-Sen University, Guangzhou 510275, China}


\begin{abstract}
    In this paper, we discuss the possibility to probe a nearly-degenerate $Z^{\prime}$-$Z$ system by analyzing the $Z$-lineshape at an electron-positron collider. Compared with the usual $Z^{\prime}$ in the literature well separated with the standard model (SM) $Z$ boson in mass, the nearly-degenerate $Z^{\prime}$-$Z$ mixing affects the observed effective ``oblique parameters'' $\tilde{S}$, $\tilde{T}$, $\tilde{U}$, and the effective deviation of ``number of neutrino species'' $\delta \tilde{N}_{\nu}$ in a more complicated way and cannot be simply computed perturbatively up to a particular order. Aiming at solving this problem, we write down a general simplified effective Lagrangian and enumerate some parameter spaces corresponding to some typical models, and suggest a method to extract the constraints by looking into the line-shape of the $Z$-like resonance at an electron-positron collider. 
\end{abstract}



\maketitle

\section{Introduction}

The standard model (SM) of particle physics can be extended with an additional gauge group, thus accommodates various exotic vector bosons. The electromagnetic neutral $Z^{\prime}$ boson, accompanied with an extra $U(1)^{\prime}$ gauge symmetry is the simplest selection\cite{Fayet:1980ad, Fayet:1980rr, Okun:1982xi, Fayet:1990wx}. $Z$-$Z^{\prime}$ mixing\cite{Holdom:1985ag, Galison:1983pa, Foot:1991kb} can arise due to the absence of any unbroken symmetry below the electroweak scale, affecting various phenomenologies for different searching proposals.

Straightforward ``bump'' searches are effective for a $Z^{\prime}$ with a measurable interaction with the SM particles (For a review, see Ref.~\cite{Leike:1998wr}). In the literature, Tevatron\cite{D0:2010kuq, CDF:2011nxq, CDF:2006uji, CDF:2005hun, CDF:2011pih, CDF:2008ieg, CDF:2010zwq} and the ATLAS and CMS collaborations at the LHC\cite{ATLAS:2019erb, CMS:2021ctt, CMS:2019buh, LHCb:2020ysn, ATLAS:2020zms, CMS:2018rmh, ATLAS:2019fgd, CMS:2021xuz, ATLAS:2021suo} published constraints for such kind of $Z^{\prime}$ boson with its mass well-separated with the SM-$Z$ boson. In contrast with this ``direct'' strategy, one of the indirect ways to observe the off-shell $Z^{\prime}$ effects can be found in Ref.~\cite{Komachenko:1989qn}. A more prominent ``oblique'' way is to look into the ``oblique parameters'', e.g., Peskin-Takeuchi parameters $S$, $T$, and $U$\cite{Peskin:1990zt, Peskin:1991sw}, and even $W$, $X$, $Y$, $Z$\cite{Barbieri:2004qk}, etc., to observe the hints imprinted on the electroweak precision measurement parameters from an off-shell $Z^{\prime}$\cite{Holdom:1990xp, Babu:1997st} which might be veiled beneath its faint coupling with, or low decay ratio into the visible SM particles during a straightforward searching process. In the literature, most of the results utilizing the LEP data actually follow this route\cite{Altarelli:1990wt, Richard:2003vc, Peskin:2001rw}. The recent $W$ boson mass data published by the CDF collaboration\cite{CDF:2022hxs} and its deviation from the SM predicted value give rise to the possibility of the existence of an exotic vector boson contributing to the electroweak precision measurement values\cite{Strumia:2022qkt, Asadi:2022xiy, Fan:2022yly, Gu:2022htv, Lu:2022bgw, deBlas:2022hdk, Zhang:2022nnh, Zeng:2022lkk, Thomas:2022gib, Cheng:2022aau, Cai:2022cti, Alguero:2022est, Du:2022fqv, Harigaya:2023uhg}.

Theoretical predictions on both ``direct'' and ``oblique'' ways to find a $Z^{\prime}$ are based upon the perturbative expansions up to a particular order. This is sufficient for the case with a significant mass difference between $Z^{\prime}$ and the SM-$Z$, and the mixing angle is usually extremely small. Yet in the literature, it seems that discussions about a nearly degenerate $Z^{\prime}$-$Z$ system are rare. As we have examined, we only found in Ref.~\cite{L3:1993zvy} some incomplete results based upon a particular model defined in Ref.~\cite{Franzini:1985nt}. In this paper, we will discuss a much more general case.

Inspired by the famous $K^0$-$\overline{K^0}$ system\cite{Donoghue:1992dd, Workman:2022ynf}, we have realized that despite the Hermitian squared mass elements, the non-Hermitian widths might also play important roles in finding the ``mass eigen-states'' of two mixing states. In the case when two vector bosons mix, diagonalizing a mass-squared matrix including the non-Hermitian width contributions is equivalent to a resummation over all possible ``string diagrams'' including the imaginary contributions from all possible one-loop self-energy diagrams. Then these two ``mass eigen-states'' both contribute to the line-shape of a $Z$-boson-like object at an electron-positron collider, which might not be naively considered as a simple composition of two resonances. Compared with a sheer SM-$Z$ line-shape, such an object might appear as a distorted ``resonance'' to affect the electroweak precision measurement results extracted from its appearance.

In this paper, we try to compute the electroweak precision measurement distortions induced by a $\hat{Z}^{\prime}$ field which is nearly degenerate with the SM $\hat{Z}$ field through diagonalizing their mass-squared matrix including their widths. In order to compare our results with the familiar parameters, we define observables $\tilde{S}$, $\tilde{T}$, $\tilde{U}$ and $\delta \tilde{N}_{\nu}$, corresponding to the well-known Peskin-Takeuchi parameters $S$, $T$, $U$, and the deviation of the neutrino species $\delta N_{\nu}$. Unlike $S$, $T$ and $U$, our $\tilde{S}$, $\tilde{T}$, $\tilde{U}$ only reflect the change of the line-shape, and is unable to be attributed to some particular effective operator contributions. Although one can find some incomplete LEP data from the references of the Ref.~\cite{ALEPH:2005ab}, as theorists it is difficult for us to completely restore a LEP simulation environment with higher order contributions considered.  As a compromise, we adopt a set of events simulated in the conditions similar to the LEP environments dubbed ``pseudo-LEP'' data to predict the sensitivity if the real original LEP data are utilized.

Besides the LEP, suggested future leptonic colliders,  e.g., ILC\cite{ILC:2013jhg}, CEPC\cite{CEPCStudyGroup:2018ghi}, FCC-ee\cite{FCC:2018evy}, have been proposed with extremely large integrated luminosities. Usually at least a calibration around the $Z$-scale should be proceeded, at the same time electroweak precision measurement data are updated by the way. For an example, the CEPC takes the potential to produce $\sim 10^{11}$-$10^{12}$ $Z$-bosons, which can be regarded as a super $Z$-factory\cite{Zheng:2018fqv} to significantly improve the sensitivity of the oblique parameters. Simulating such a large ``pseudo-CEPC'' data set is beyond our current computational resources, however its sensitivities can still be estimated by utilizing the pseudo-LEP results.

This paper is organized as follows. In Sec.~\ref{Lagrangian}, we introduce the effective Lagrangian for an exotic vector boson $\hat{Z}^{\prime}$. Other basic concepts are elaborated. Then simulation details and settings are illustrated in Sec.~\ref{Details}. In Sec.~\ref{Numerical}, the numerical results are presented and described in three scenarios. Finally, Sec.~\ref{summary} summarize this paper.

\section{Effective Lagrangian} \label{Lagrangian}

In this paper, we rely on a simplified general effective Lagrangian introduced in Ref.~\cite{Cai:2022cti}, enumerating all the possible kinematic and mass mixing terms that preserve the $U(1)_{\rm{EW}}$ symmetry. Since some of the variables are not straightforwardly connected to the $Z^{\prime}$ sector, we only discuss a subset of them in this paper. We list the corresponding terms as below,
\begin{eqnarray}
\mathcal{L} &\supset& -\frac{1}{4} \hat{Z}^{\prime}_{\mu \nu} \hat{Z}^{\prime \mu \nu} + \frac{1}{2} m_{\hat{Z}^{\prime}}^2 \hat{Z}^{\prime}_{\mu} \hat{Z}^{\prime \mu}  - \frac{ \epsilon_B}{2} \hat{Z}^{\prime}_{\mu \nu} B^{\mu \nu} - \frac{1}{2 \Lambda_{W}^2 } \hat{Z}^{\prime}_{\mu \nu} W^{a \mu \nu} H^{\dagger} \sigma^a H \nonumber \\
&+&  \hat{Z}^{\prime \mu} \left[ i \lambda_{HZ^{\prime}} (D_{\mu} H)^{\dagger} H + \text{H.c.}\right],\label{LKin}
\end{eqnarray}
where $\hat{Z}^{\prime \mu}$ is the exotic neutral vector boson $\hat{Z}^{\prime}$, and $\hat{Z}^{\prime}_{\mu \nu} \equiv \partial_{\mu} \hat{Z}^{\prime}_{\nu}  - \partial_{\nu} \hat{Z}^{\prime}_{\mu}$. $H$ indicates the SM Higgs doublet, $W^{a \mu}$ and $B^{\mu}$ are the $SU(2)_L$ and $U(1)_Y$ gauge fields respectively, and $D_{\mu} = \partial_{\mu} - i \hat{g}^{\prime}_Y B_{\mu}/2 - i \hat{g} \sigma^a W^a_\mu/2$, where $\hat{g}^{\prime}$ and $\hat{g}$ are the ``original'' coupling constants. $\epsilon_B$, $\Lambda_W$, 
$\lambda_{H Z^{\prime}}$ are the corresponding constants. Here the mass term $m_{\hat{Z}^{\prime}}^2 \hat{Z}^{\prime}_{\mu} \hat{Z}^{\prime \mu}$ is put by hand, and might originate from an exotic Higgs carrying a $U(1)^{\prime}$ charge corresponding with the $Z^{\prime}$, or from the Stueckelberg mechanisms. 

After $H$ acquires the vacuum expectation value (VEV),
\begin{eqnarray}
H = \begin{pmatrix}
i \phi^+ \\
\frac{\hat{v}+h+i \phi^0}{\sqrt{2}}
\end{pmatrix}, 
\end{eqnarray}
where $\hat{v} \approx 246 \rm{GeV}$, we therefore acquire the effective kinematic mixing terms
\begin{equation}
    \mathcal{L}_{\mathrm{eff}} \supset -\frac{\epsilon_B}{2}\hat{Z}^{\prime}_{\mu\nu}B^{\mu\nu} - \frac{\epsilon_W}{2}\hat{Z}^{\prime}_{\mu\nu}W^{3\mu\nu}, \label{EpsBEpsW}
\end{equation}
where $\epsilon_W\equiv-\hat{v}^2/(2\Lambda_W^2)$
. Then, the mass terms as well as the kinematic terms can be written in the form of matrices, 
\begin{eqnarray}
\mathcal{L}_{mass}&=& \begin{pmatrix}
\hat{Z}^{\prime}_{\mu}, & B_{\mu}, &W^3_{\mu}
\end{pmatrix} \mathcal{M}_V^2 \begin{pmatrix}
\hat{Z}^{\prime \mu} \\
B^{\mu} \\
W^{3 \mu}
\end{pmatrix}, \\
\mathcal{M}_V^2 &=& \begin{pmatrix}
m_{\hat{Z}^{\prime}}^2 & \hat{g}^{\prime} \delta m^2 & -\hat{g} \delta m^2 \\
\hat{g}^{\prime} \delta m^2  &  \frac{\hat{g}^{\prime 2}}{4} \hat{v}^2 &  -\frac{\hat{g}^{\prime} \hat{g}}{4} \hat{v}^2\\
-\hat{g} \delta m^2 &  -\frac{\hat{g}^{\prime} \hat{g}}{4} \hat{v}^2 & \frac{\hat{g}^2}{4} \hat{v}^2
\end{pmatrix}, \label{MassOrigin}
\end{eqnarray}
and
\begin{eqnarray}
\mathcal{L}_\mathrm{kin}&=&-\frac{1}{4} \begin{pmatrix}
\hat{Z}^{\prime}_{\mu \nu}, & B_{\mu \nu}, & W^3_{\mu \nu}
\end{pmatrix}\mathcal{K}_V \begin{pmatrix}
\hat{Z}^{\prime \mu \nu} \\
B^{\mu \nu} \\
W^{3 \mu \nu}
\end{pmatrix},
\\
\mathcal{K}_V&=& \begin{pmatrix}
1 & \epsilon_B & \epsilon_W \\
\epsilon_B  &  1  &  0 \\
\epsilon_W &  0 & 1 
\end{pmatrix}, \label{KOrigin}
\end{eqnarray}
where $\delta m^2\equiv-\lambda_{HZ'}\hat{v}^2/2$
. Ref.~\cite{Cai:2022cti} aimed at diagonalizing (\ref{MassOrigin}) and (\ref{KOrigin}), and perturbatively expanded the results in the case that $m_{Z^{\prime}}$ is well-separated with the $m_Z$.

The contributions from the self-energy diagrams are usually evaluated through two ways, computed order-by-order perturbatively, or resummed all at once to correct the mass term of every propagator.  Yet near the resonance of each s-channel propagator, the imaginary part of its self-energy diagrams must be resummed prior to any other process and contributes to the Breit-Wigner form of the propagator by adding up an imaginary part in the denominator. Other contributions may be taken into account perturbatively order by order later, and behave like sub-leading corrections in quantities.

\begin{figure}
    \centering
    \includegraphics[width=2.6in]{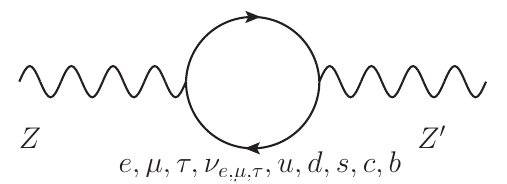}
    \caption{Cross terms between $Z$ and $Z^{\prime}$ particles.}
    \label{CrossTerm}
\end{figure}

Besides the self-energy of each particle, the self-energy among different types of particles, or the ``cross terms'' may also give rise to the imaginary parts, as shown in Fig.~\ref{CrossTerm} as an example. This is not a problem if $Z$ and $Z^{\prime}$ are well separated in mass spectrum, and its contributions are suppressed by a factor $\frac{1}{m_{Z^{\prime}}^2-m_Z^2}$ so they can be taken into account order-by-order perturbatively. However, when $Z$ and $Z^{\prime}$ are nearly degenerate so $m_{Z^{\prime}}^2 \approx m_Z^2$, such a suppression becomes non-viable.

Denote $c_{XY} \equiv \mathrm{Im}\left\lbrace  \Pi_{X \leftrightarrow Y} (p^2 \approx m_{X,Y}^2)\right\rbrace $, where $\Pi_{X \leftrightarrow Y} (p^2)$ is the $g_{\mu\nu}$ coefficient of the self-energy of a particle $X$ transforming into $Y$, one has to correct each element of (\ref{MassOrigin}) with an additional $ic_{XY}$ term before the diagonalizion processes. In fact, as we will illustrate in Appendix \ref{AppResum}, the crossing-terms involving the photon can be negligible, and we utilize the matrix below
\begin{eqnarray} 
    \mathcal{M}_V^{2~\prime} &=& \mathcal{M}_V^2 + i \left( C_{X \leftrightarrow Y} \right)_{3 \times 3} \nonumber \\
    &=& \begin{pmatrix}
    m_{\hat{Z}^{\prime}}^2 + ic_{\hat{Z}^{\prime}\hat{Z}^{\prime}} & \hat{g}^{\prime} (\delta m^2 - \frac{ic_{\hat{Z}^{\prime}\hat{Z}}}{\sqrt{\hat{g}^{\prime 2} + \hat{g}^2}}) & -\hat{g} (\delta m^2 - \frac{ic_{\hat{Z}^{\prime}\hat{Z}}}{\sqrt{\hat{g}^{\prime 2} + \hat{g}^2}}) \\
    \hat{g}^{\prime} (\delta m^2 -\frac{ic_{\hat{Z}^{\prime}\hat{Z}}}{\sqrt{\hat{g}^{\prime 2} + \hat{g}^2}}) &  \frac{\hat{g}^{\prime 2}}{4} (\hat{v}^2 + \frac{4ic_{\hat{Z}\hat{Z}}}{\hat{g}^{\prime 2} + \hat{g}^2})&  -\frac{\hat{g}^{\prime} \hat{g}}{4} (\hat{v}^2 + \frac{4ic_{\hat{Z}\hat{Z}}}{\hat{g}^{\prime 2} + \hat{g}^2})\\
    -\hat{g} (\delta m^2 - \frac{ic_{\hat{Z}^{\prime}\hat{Z}}}{\sqrt{\hat{g}^{\prime 2} + \hat{g}^2}}) &  -\frac{\hat{g}^{\prime} \hat{g}}{4} (\hat{v}^2 + \frac{4ic_{\hat{Z}\hat{Z}}}{\hat{g}^{\prime 2} + \hat{g}^2}) & \frac{\hat{g}^2}{4} (\hat{v}^2 + \frac{4ic_{\hat{Z}\hat{Z}}}{\hat{g}^{\prime 2} + \hat{g}^2})
    \end{pmatrix}. \label{mass-squares}
\end{eqnarray}
Now the mass terms are no longer hermitian, and this can be understood by adding some non-conjugate corrections into the Lagrangian, or the Hamiltonian, just as what happens in a $K^0$-$\overline{K^0}$ system\cite{Donoghue:1992dd, Workman:2022ynf}. We also have to note that the ``hatted'' $\hat{Z}$ denotes the ``SM-$\hat{Z}$'' when all the mixing terms are switched off, that is to say,
\begin{eqnarray}
    \hat{Z}^{\mu} = \frac{1}{\sqrt{\hat{g}^2 + \hat{g}^{\prime 2}}}(-\hat{g}^{\prime} B^{\mu} + \hat{g} W^{3 \mu}).
\end{eqnarray}
In the following of this paper, the hatless ``$Z$'' usually appears within the symbols associated with the aspect of the experimentalists who might be unaware that their observed resonance can accommodate exotic contributions. ``$Z$'' also appears within the definitions from the literature or simulating tools. In this paper, the hatless ``$Z^{\prime}$'' is also associated with a general reference of ``$Z^{\prime}$-model'' or ``$Z^{\prime}$-$Z$ system'', while the hatted ``$\hat{Z^{\prime}}$'' particularly refers to the $\hat{Z^{\prime}}$ field that we introduce in (\ref{LKin}).

Then we can follow Ref.~\cite{Cai:2022cti} to diagonalize the kinetic matrix~\eqref{KOrigin} beforehand:
\begin{eqnarray}
    V_C = V_1 V_2 V_3,\quad
    V_C^\mathrm{T} \mathcal{K}_V V_C = I_{3 \times 3},
\end{eqnarray}
where 
\begin{eqnarray}
    V_1 &=& \begin{pmatrix}
    1 & -\epsilon_B & -\epsilon_W \\
    0 & 1 & 0 \\
    0 & 0 & 1
    \end{pmatrix}, ~V_2 =  \begin{pmatrix}
    1 & 0 & 0 \\
    0 & 1 & \frac{\epsilon_B \epsilon_W}{1-\epsilon_B^2}  \\
    0 & 0 & 1
    \end{pmatrix},
    \nonumber \\
    V_3 &=& \begin{pmatrix}
    1 & 0 & 0 \\
    0 & \frac{1}{\sqrt{1-\epsilon_B^2}} & 0 \\
    0 & 0 & \sqrt{\frac{1-\epsilon_B^2}{1-\epsilon_B^2-\epsilon_W^2}}
    \end{pmatrix}, \label{KinMixingTerms}
\end{eqnarray} 
and then the mass-squared matrix becomes
\begin{eqnarray}
    (V_1 V_2 V_3)^\mathrm{T} \mathcal{M}_V^{2~\prime} V_1 V_2 V_3, \label{DiagonalFirstStep}
\end{eqnarray}
and diagonalizing this matrix gives
\begin{equation}
    V=V_1V_2V_3V_{\mathrm{SM}}V_{\mathrm{f}}, \quad V^{\mathrm{T}}\mathcal{M}_V^{2~\prime}V = \mathrm{diag}(m^2_{1} - \mathrm{i}m_{1}\Gamma_{1},\,m_2^2 - \mathrm{i}m_2\Gamma_2,\,0), \label{diagonalize}
\end{equation}
where $V_{\mathrm{SM}}$ is the familiar EW rotation matrix
\begin{equation}
    V_{\text{SM}} = \begin{pmatrix}
        1 & 0 & 0 \\
        0 & -\frac{\hat{g}^{\prime}}{\sqrt{\hat{g}^{\prime 2} + \hat{g}^2}} & \frac{\hat{g}}{\sqrt{\hat{g}^{\prime 2} + \hat{g}^2}} \\
        0 & \frac{\hat{g}}{\sqrt{\hat{g}^{\prime 2} + \hat{g}^2}} & \frac{\hat{g}^{\prime}}{\sqrt{\hat{g}^{\prime 2} + \hat{g}^2}}
        \end{pmatrix}. \label{VSM}
\end{equation}
$m_1$, $\Gamma_1$, $m_2$, $\Gamma_2$ are the ``masses'' and ``widths'' of two ``mass eigenstates'' of the $Z^{\prime}$-$Z$ system.  These ``mass eigenstates'' are denoted by $Z_{1}$ and $Z_{2}$ in this paper.  Since their masses are nearly degenerate, and their mixing angle might be large, they might altogether form a single SM-$Z$-like object, and which of them is identified to be the $Z$ or $Z^{\prime}$ is unessential.  The appearance of a zero eigenvalue in $V^T \mathcal{M}_V^{2 \prime} V$ might correspond to the photon, and further calculation indicates that such a massless eigenstate coupling with all the fermions exactly mimics a photon, leaving us no problem about its identity. However the vertices involving the $Z_{1,2}$ are extremely complicated with a combination of the continuous products composed with the $V_{1,2,3,\text{SM},\text{f}}$ elements. In this paper, we calculate these couplings numerically without showing their detailed form.


We have to note that although (\ref{mass-squares}) is no longer hermitian, one can still verify that it is symmetric, that is, $\mathcal{M}_V^{2~\prime} = (\mathcal{M}_V^{2~\prime})^{\rm{T}}$. This guarantees the existence of $V_{\mathrm{f}}$ in (\ref{diagonalize}) with the condition $V_{\mathrm{f}}^{\mathrm{T}} V_{\mathrm{f}} = V_{\mathrm{f}} V_{\mathrm{f}}^{\mathrm{T}} = I$. However, the elements of $V_{\mathrm{f}}$ can be complex, which are weird to be understood as the ``mixing terms'' among real vector fields, since the mixed ``eigenstate fields'' are no longer real numbers. This indicates that the usual perceptions of the ``mixing fields'' are non-viable and should be replaced with the concept of the resummmed propagators as described in Appendix \ref{AppResum}.

\section{Details of the event generation and the extraction of the observables} \label{Details}

The standard way to extract the electroweak precision data is to compare the line-shape of the $Z$-resonance with the parameterized functions considering the Breit-Wigner propagators, initial-state radiation (ISR) effects, and the momentum distribution of the beams (See section 55 of ``Reviews, Tables \& Plots'' in Ref.~\cite{Workman:2022ynf} for a review, and for the references therein). The photon mediated s-channel diagrams and all the t- and u-channel contributions with the interference effects should also be considered. After finding out the most fitted parameterized function, $m_Z$, $\Gamma_Z$, $R_{e, \mu, \tau}$, $A_{\rm FB}^{0 e,\mu, \tau}$, and $N_{\nu}$, which are the mass of the $Z$-boson, the width of the $Z$-boson, the ratios of the $Z \rightarrow \rm{hadrons}$ over $Z \rightarrow e/\mu/\tau$ branching ratios, the forward-backward asymmetry parameters, and the effective number of the active neutrinos, respectively, are extracted for further comparison with the SM predicted values.

In this paper, we alternatively adopt what we called the ``SM templates'' to replace the role of the parameterized functions. These are the line-shape data acquired from the event generator based upon the dubbed ``pseudo-SM'' model file, which is a modified variation of the default SM model file provided by FeynRules\cite{Alloul:2013bka}, added with four additional parameters $m_Z$, $s^2_l$, $\Gamma_Z$ and $\xi$ as the input value. Here $s^2_l$ is the effective Weinberg angle affecting the weak coupling constants, independent with the $s^2_W$ associated with the ratio of the $W$ and $Z$ bosons. $\Gamma_Z$ can also be assigned with an arbitrary value, which might not equal to the SM-predicted one. $\xi$ appears in
\begin{eqnarray}
    B_{\mu} &=& -s_W \xi Z_{\mu} + c_W A_{\mu}, \nonumber \\
    W^{3}_{\mu} &=& c_W \xi Z_{\mu} + s_W A_{\mu},
\end{eqnarray}
modifying the definitions of the $B_{\mu}$ and $W^{3}_{\mu}$ in the model file. This parameter aims at rescaling the height of the resonance while keeping the shape of it intact through exerting a universal factor upon the $Z$-$\overline{f}$-$f$ vertices
\begin{eqnarray}
    \mathcal{L}_{\rm F} \supset - \frac{e}{2 s_l c_l} \sum_i \overline{\psi}_i \gamma^{\mu} (\xi g_V^i - \xi g_A^i \gamma^5) \psi_i Z_{\mu}, \label{PseudoSMZCoupling}
\end{eqnarray}
where $e = \sqrt{4 \pi \alpha}$ is the electromagnetic coupling constant,  and $c_l = \sqrt{1-s_l^2}$.  For each $i \in \lbrace \text{all SM fermion species}\rbrace$, $g_V^i = t_{3 L}^i - 2 Q_i s_l^2$ and $g_A^i = t_{3 L}^i$ where $t_{3 L}^i$ is the weak isospin of the fermion $i$, and $Q_i$ is the charge of $\psi_i$. Notice that with the couplings defined in (\ref{PseudoSMZCoupling}), the event generator can automatically compute a width of the $Z$-boson, denoted by $\Gamma^{\rm SM}_Z$ in this paper, which is not necessarily equivalent with the $\Gamma_Z$ which has been enforced into the Breit-Wigner propagator as an input parameter. We are going to utilize both of them in the following of this paper.

Although part of the LEP data can be found in Ref.~\cite{DELPHI:2000wje, OPAL:2000ufp, L3:2000vgx, L3:1993zvy} by the DELPHI, OPAL, and L3 collaboration, a comparison between the theoretical predicted results with these data requires intricate higher-order loop-level contributions as well as the complete Monte-Carlo simulation imitating the detector performances and event selection techniques decades ago. As theorists, we may concern only the sensitivity of a LEP-like collider on such a new physics model as a compromise at current stage, leaving these complexities for future studies if all the LEP raw data from all four detectors can be excavated. Therefore, more familiar Monte-Carlo tools developed in the LHC era are utilized for fast simulation processes in this paper, and comparing our simulated $Z^{\prime}$-model events with the SM templates generated by exactly the same Monte-Carlo tools up to the tree level can reduce both the impacts from the higher-order corrections, and from the difference between the 1990s and current techniques.

To compare the continuous line-shape curves, one should sample a set of discrete $\sqrt{s}$'s, which are defined as the invariant masses, or the total energies of the colliding $e^+ e^-$ system.  LEP published some of the early detailed data including the integrated luminosity and the number of events at each of the sampled $\sqrt{s}$, while later as the integrated luminosity accumulates, in Ref.~\cite{LEP:2000pgt} only the fitted results of the electroweak precision measurements were published. For the integrated luminosities of each sampled $\sqrt{s}$, only the L3 collaboration published them from 1993-1995 in Ref.~\cite{L3:2000vgx}. From the Table 1 of Ref.~\cite{LEP:2000pgt} and the Table 14-17 of Ref.~\cite{L3:2000vgx}, we learned that the $\sqrt{s} \approx 91.2$ GeV occupied most of the integrated luminosity budget. In this paper, we aim at a simpler but clearer analogy rather than a perfect and precise restoration of the LEP experimental environments, so we adopt $\sqrt{s}=[88.2, 89.2, 90.2, 91.2, 92.2, 93.2, 94.2] \rm{GeV}$ as our samples as a reference, and equalize their integrated luminosities as 300 $\mathrm{pb}^{-1}$ for each $\sqrt{s}$ when generating the $Z^{\prime}$-model events.  Since in Ref.~\cite{LEP:2000pgt}, the total luminosity is 60 $\mathrm{pb}^{-1}$, we therefore multiply a $\sqrt{5}$ when evaluating the statistical uncertainties of the $\sqrt{s}=91.2$ GeV events. According to Ref.~\cite{L3:2000vgx}, other $\sqrt{s}$'s are still assigned with considerable integrated luminosities, so we multiply a $\sqrt{30}$ at the $\sqrt{s}=[88.2, 89.2, 90.2, 92.2, 93.2, 94.2]$ GeV uncertainty data for the equivalence with the effective 10 $\mathrm{pb}^{-1}$ integrated luminosities. We call our data with such settings the ``pseudo-LEP'' results.  Since from the knowledge of statistics, the sensitivity of a measurement is usually proportional to the square root of the luminosity, so our assignments of the integrated luminosities indicate a virtual collider with its performances a little bit better than the authentic LEP. The results acquired from the pseudo-LEP are therefore still expected to reflect the authentic LEP sensitivities.

For the pseudo-SM templates, we generate $10^7$ events for each $\sqrt{s}=[88.2$, $89.2$, $90.2$, $91.2$, $92.2$, $93.2$, $94.2] \rm{GeV}$ sample with various input values of $m_Z$, $s_l^2$, $\Gamma_Z$, and $\xi$. Then we use the polynomial
\begin{equation}
    \begin{aligned}[b]
        & \sigma_t^{\text{PSM}}(m_Z, \Gamma_Z, s_l^2, \xi, \sqrt{s})  = a_{0t}(\sqrt{s}) + (m_Z - m_{Z0})^2 a_{1t}(\sqrt{s}) + (s_l^2 - s_{l0}^2)^2 a_{2t}(\sqrt{s}) + \\
        & (\Gamma_Z - \Gamma_{Z0})^2 a_{3t}(\sqrt{s}) + (\xi - \xi_0)^2 a_{4t}(\sqrt{s})  + (m_Z - m_{Z0}) (s_l^2 - s_{l0}^2) a_{5t}(\sqrt{s}) +  \\
        &(m_Z - m_{Z0}) (\Gamma_Z - \Gamma_{Z0}) a_{6t}(\sqrt{s}) + (m_Z - m_{Z0}) (\xi - \xi_0) a_{7t}(\sqrt{s}) + (s_l^2 - s_{l0}^2) (\Gamma_Z - \Gamma_{Z0}) a_{8t}(\sqrt{s}) \\
        &+ (s_l^2 - s_{l0}^2) (\xi - \xi_0) a_{9t}(\sqrt{s}) + (\Gamma_Z - \Gamma_{Z0}) (\xi - \xi_0) a_{10t}(\sqrt{s})  + (m_Z - m_{Z0}) a_{11t}(\sqrt{s}) + \\
        &(s_l^2 - s_{l0}^2) a_{12t}(\sqrt{s})  + (\Gamma_Z - \Gamma_{Z0}) a_{13t}(\sqrt{s})  +  (\xi - \xi_0) a_{14t}(\sqrt{s}),
    \end{aligned} \label{FittedLineShape}
\end{equation}
to fit the pseudo-SM template events by quadratic fitting for each $\sqrt{s}=[88.2$, $89.2$, $90.2$, $91.2$, $92.2$, $93.2$, $94.2] \rm{GeV}$ in each final product channel $t \in \lbrace (e^+e^-)_{\text{F}/\text{B}}$,$(\mu^+\mu^-)_{\text{F}/\text{B}}$,$\rm{hadrons}\rbrace$. Here the subscript of the ``$(e^+e^-)_{\text{F}}$'' or ``$(e^+e^-)_{\text{B}}$'' denotes the ``forward'' or ``backward'' directions that the positive charged product outcomes parallels or anti-parallels to the incoming positron beam respectively. The $a_{(0-14)t}$ are the factors to be determined by the fitting processes.  $m_{Z0}$, $s_{l0}^2$, $\Gamma_{Z0}$, $\xi_0$ are the ``central values'' of the corresponding input parameters. Here we assign $m_{Z0}$ with the current measured value of the $Z$-boson mass. The $s_{l0}$ is assigned by solving $s_{l0}^2 c_{l0}^2 = s_W^2 c_W^2 =\frac{\pi \alpha}{\sqrt{2} m_{Z0}^2 G_F}$, and $\Gamma_{Z0}$ is computed from the pure SM theory with the previously assigned $m_{Z0}$ and $s_{l0}$ values, both neglecting the beyond-tree-level contributions as we have mentioned before. Here $G_F$ is the Fermi constant, $s_{W}$ is the Weinberg angle defined in the SM model file accommodated in the FeynRules. The $s_{l0} = s_{W}$ assignment eliminates all the $Z^{\prime}$ contributions at the reference central point. We should note that the precise absolute value of these three parameters are not important on the stage of this paper, since what we concern is the tiny ``correction'' upon them from the new physics contributions, rather than their original ``background values''. In the future, if the algorithm proposed in this paper is practically adopted, corrections including the loop-level contributions should be further taken into account. Finally, $\xi_0$ is appointed to be the value 1. The charge asymmetry of the quarks also affect the charge unbalance of the final hadrons, however analyzing such an asymmetry is beyond our ability of simulation, so we evade taking this into account.

In principle the $e^+ e^- \rightarrow \tau^+ \tau^-$ channels should also be considered. $\tau$ might decay into muons, electrons, or hadrons to fake the corresponding channels. The leptonic decay channels can be simply distinguished by the additional missing energy/momentum criterion during the event selections, and the hadronic decay channels might be problematic. At a lepton collider, the hadronic $\tau$-decay products can be well discriminated from the hadronic-jet events\cite{Grunewald:1995ny}.  As a comparison, the famous ``tau-tagging'' techniques seem to have started drawing attention since the ``LHC-era'' because of the miscellaneous QCD-backgrounds which are difficult to be eliminated, while during the ``LEP-era'', probably due to the extremely low QCD-background so that the hadronic decaying $\tau$ can be well-separated from the QCD jets, the ``$\tau$-contamination'' is therefore only briefly discussed in the published papers. For an example, in Ref.~\cite{DELPHI:1994bya}, the $\tau$-contamination was found to occupy typically $0.45\pm0.05\%$ times the QCD-jet cross section, indicating so tiny effects on our discussions. Further more, during the ``LHC-era'',  the efficiency of the ``$\tau$-tagging'' techniques at a hadronic collider seems to have been significantly improved within these years\cite{Bagliesi:2007qx, CMS:2018jrd}, and it is reasonable to expect a similar improvement not only at the future lepton collider programs,  but also for the previous LEP data if the raw data can be excavated to be reanalyzed.  Considering the complexity of the $\tau$-tagging algorithms for a fast simulation in this paper, and the potential of the technology improvement in the future, we just assume that all the $\tau$ products can be well-discriminated and negligible in this paper.

In this work, we apply WHIZARD~\cite{Kilian:2007gr,Moretti:2001zz,Christensen:2010wz} as our event generator. LHAPDF6~\cite{Buckley:2014ana,Andersen:2014efa}, PYTHIA6~\cite{Bierlich:2022pfr}, FastJet~\cite{Cacciari:2005hq,Cacciari:2011ma} and DELPHES~\cite{deFavereau:2013fsa,Selvaggi:2014mya,Mertens:2015kba} are connected for the detector-level data. 
In DELPHES, we utilize the CEPC card regardless our purpose for a LEP-like prediction, and not so many differences are expected.
The beams structure is chosen to be the Gaussian distribution. The WHIZARD parameter $\sigma$ of both the ``gaussian\_spread1'' and ``gaussian\_spread2'' are set to be 250/90000\cite{OPAL:2004xxz}, which indicates the relative Gaussian-width of the energy of each beam. ISR is also switched on.  The event selection rules are set as 
$P_T > 7$~GeV and $|\eta|<2.4$ for each isolated (anti-)muons.  Here, the ``isolated'' muons or electrons mean that these objects do not appear as a part of a hadronic jet.  Clustered jets which is not identified as the isolated electrons/positrons/(anti)-muons are also regarded as ``single objects'' to fit the criterion above. Here ``E'', ``$P_T$'', and ``$\eta$'' are the energy, the transverse momentum, the pseudo-rapidity of the particle/jet in the argument respectively.  We should also note that the isolated electron or muon identification criteria are set as the DELPHES default standards, that the total $P_T$ for all the exotic objects appeared within a circle with the radius $R = \sqrt{\Delta \eta^2 + \Delta \phi^2} = 0.5$  around the electron or muon candidate should be less than 0.12 or 0.25 times the $P_T$ of the candidate electron or muon respectively. Here $\Delta \eta$ and $\Delta \phi$ are the difference of the pseudo-rapidity and the azimuth-angle.  Additionally $M^2_{f \overline{f}} \geq 1$~$\mathrm{GeV}^2$ where ``$M^2_{f \overline{f}}$'' means the invariant mass squared of the $e^{\pm}/\mu^{\pm}/$jet pairs. For the isolated electron/positron pairs, the $|\eta|<2.4$ criterion becomes too loose to accommodate too many nearly co-linear events originating from the $t$- or $u$-channel internal photon propagators. Therefore we add a $|\cos(\theta)|\leq 0.8$ cut on each isolated electron/positron to eliminate such events, where ``$\theta$'' is the absolute polar angle in the lab frame. 


We then compare the line-shape cross sections in the $Z^{\prime}$-model for each parameter point with the pseudo-SM template cross sections fitted in (\ref{FittedLineShape}) to find the best fitted $(m_Z^*, s_l^{*2}, \Gamma_Z^*, \xi^*)$ by minimizing the $\chi_F^2$ defined as
\begin{equation}
    \chi_F^2(m_Z, \Gamma_Z, s_l^2, \xi) = \sum_{t, \sqrt{s}}\frac{(\sigma_t^{Z^{\prime}}(\sqrt{s}) - \sigma_t^{\mathrm{PSM}}(m_Z, \Gamma_Z, s_l^2, \xi, \sqrt{s}))^2}{(\Delta_t^{Z^{\prime}}(\sqrt{s}))^2 + (\Delta_i^{\mathrm{PSM}}(m_Z, \Gamma_Z, s_l^2, \xi, \sqrt{s}))^2}, \label{chi-square}
\end{equation}
where $\sqrt{s}=[88.2$, $89.2$, $90.2$, $91.2$, $92.2$, $93.2$, $94.2]$ GeV, $\sigma_t^{\mathrm{PSM}}(m_Z, \Gamma_Z, s_l^2, \xi, \sqrt{s})$ is defined in (\ref{FittedLineShape}). $\sigma_t^{Z^{\prime}}(\sqrt{s})$ are line-shape sample cross sections for the $Z^{\prime}$-model with a particular set of parameters. $\Delta_i^{Z^{\prime}}$ and $\Delta_i^{\mathrm{PSM}}(m_Z, \Gamma_Z, s_l^2, \xi, \sqrt{s})$ are the statistical uncertainties of the pseudo-SM template cross sections and the $Z^{\prime}$-model cross sections respectively. The statistical uncertainty $\Delta_{\rm X}$ for each cross section $\sigma_{\rm X}$ at channel $\rm X$ is evaluated by
\begin{eqnarray}
    \Delta_{\rm X}^{Z^{\prime}/{\rm PSM}} = \frac{f_{\rm X}^{Z^{\prime}/{\rm PSM}}}{\sqrt{n_{\rm X}^{Z^{\prime}/{\rm PSM}}}}\sigma_{\rm X},
\end{eqnarray}
where $n_{\rm X}^{Z^{\prime}/{\rm PSM}}$ is the number of events passing through all selection criterion. As we have mentioned, $f^{Z^\prime}_{\rm X}=\sqrt{5}$ for all $\sqrt{s}=91.2$ GeV channels, and $f^{Z^{\prime}}_X=\sqrt{30}$ for $\sqrt{s}=[88.2, 89.2, 90.2, 92.2, 93.2, 94.2]$ GeV channels. For all $f^{\rm PSM}_{\rm X}$, they are assigned with 1. The total $\chi_F^2$ includes seven points of center-of-mass energy, three channels, and the leptonic channels are separated into forward-back parts, so the total degree of freedom is counted to be $N_{\text{df}}=$35.

Then the best-fitted $(m_Z^*, s_l^{*2}, \Gamma_Z^*, \xi^*)$ will be converted into effective oblique parameters $\tilde{S}$, $\tilde{T}$, $\tilde{U}$, and $\delta \tilde{N}_{\nu}$, which are the effective Peskin-Takeuchi oblique parameters and the effective deviation of the species of neutrinos respectively. 
Ref.~\cite{Peskin:1990zt, Peskin:1991sw} derived the $S$, $T$, $U$ expressions depending on $m_W, s_l, \Gamma_Z$.  Here similar to Ref.~\cite{Cai:2022cti}, we have followed the steps in Ref.~\cite{Burgess:1993vc} to derive and reverse these formulas as the definitions of the effective $\tilde{S}$, $\tilde{T}$, $\tilde{U}$ so that the constraints of $S$, $T$, $U$ in the literature can be straightforwardly cast here. One subtle thing is that in the usual electroweak precision measurement discussions, $m_Z$ is the input parameter, so $\delta m_W$ is given by
\begin{equation}
    \delta m_W = -\frac{\alpha m_W^{\text{ ``SM''}}}{4(c_{\mathrm{W}}^2-s_{\mathrm{W}}^2)}(S-2c^2_{\mathrm{W}}T-\frac{c_{\mathrm{W}}^2-s_{\mathrm{W}}^2}{2s^2_{\mathrm{W}}}U), \label{deltamW}
\end{equation}
where $\delta m_W = m_W - m_W^{\text{``SM''}}$. Here, the actual meaning of the quoted superscript the ``SM'' in $m_W^{\text{``SM''}}$ means that it is predicted from the SM formula $m_W^{\text{``SM''}}=m_Z^{\text{MD}} c_{\mathrm{W}}$, where the ``MD'' superscript in $m_Z^{\text{MD}}$ indicates the abbreviation for ``measured'', which means that its value is straightforwardly extracted from the experimental data of the $Z$-like resonance peak. In this paper, the ``$m_Z^{\text{MD}}$'' is actually the best-fitted $m_Z^*$. The $m_W$, however, is the ``measured'' $W$-boson mass and can be computed by $m_W=m_Z^{\text{input}} c_{\mathrm{W}}$, where the $m_Z^{\text{input}} = \frac{\sqrt{\hat{g}^2 + \hat{g}^{\prime 2}}}{2} \hat{v}$ indicating the ``SM-$Z$'' boson's mass if all the $Z^{\prime}$ effects are switched off in the model. Therefore, one can realize that the correction on $m_W$ is equivalent to a correction on $m_Z^{\text{input}}$ towards the opposite direction, so $\delta m_W = m_W - m_W^{\text{``SM''}} = m_Z^{\text{input}} c_{\mathrm{W}} - m_Z^* c_{\mathrm{W}} \simeq -(m_Z^* - m_Z^{\text{input}}) \frac{m_W^{\text{``SM''}}}{m_Z^{\text{input}}}$. Together with the formulas in  Ref.~\cite{Peskin:1990zt, Peskin:1991sw,Ciuchini:2013pca}, we define $\tilde{S}$, $\tilde{T}$, $\tilde{U}$, as well as the deviation of the neutrino species number $\delta \tilde{N}_{\nu}$ by solving the following equations.
\begin{eqnarray}
    -\delta m_W \frac{m_Z^{\text{input}}}{m_W^{\text{``SM''}}} &=& m_Z^* - m_Z^{\text{input}} = \frac{\alpha m_Z^{\text{input}}}{4(c_{\mathrm{W}}^2-s_{\mathrm{W}}^2)}(\tilde{S}-2c^2_{\mathrm{W}}\tilde{T}-\frac{c_{\mathrm{W}}^2-s_{\mathrm{W}}^2}{2s^2_{\mathrm{W}}}\tilde{U}), \nonumber \\
    s_l^{*2} - (s_l^2)^{\mathrm{SM}} &=& \frac{\alpha}{4(c_{\mathrm{W}}^2-s_{\mathrm{W}}^2)}(\tilde{S}-4s_{\mathrm{W}}^2c_{\mathrm{W}}^2\tilde{T}), \nonumber \\
    \Gamma_Z^* - (\Gamma_Z)^{\mathrm{SM}} &=& \frac{\alpha^2m_Z^{\text{input}}}{72s^2_{\mathrm{W}}c^2_{\mathrm{W}}(c_{\mathrm{W}}^2-s_{\mathrm{W}}^2)}(-10(3-8s^2_{\mathrm{W}})\tilde{S}+(63-126s^2_{\mathrm{W}}-40s^4_{\mathrm{W}})\tilde{T}), \nonumber \\
    \delta \tilde{N}_{\nu} &=& \frac{\Gamma_Z^* - (\Gamma_Z)^{\mathrm{pSM}}}{(\Gamma_{\nu\nu})^{\mathrm{pSM}}}, \label{STU_expression}
\end{eqnarray}
where the first equation is actually equivalent to the (\ref{deltamW}) by substituting the $\delta m_W$ with $-(m_Z^* - m_Z^{\text{input}}) \frac{m_W^{\text{``SM''}}}{m_Z^{\text{input}}}$. $(s_l^2)^{\mathrm{SM}}$ and $(\Gamma_Z)^{\mathrm{SM}}$ are also the calculated values of $s_l^2$ and $\Gamma_Z$ regarding the $m_Z^*$ as the input value in this case. $(\Gamma_Z)^{\mathrm{pSM}}$ and $(\Gamma_{\nu\nu})^{\mathrm{pSM}}$ are calculated according to the pseudo-SM model files with the best fitted $m_Z^*$, $s_l^{*2}$ and $\xi^*$ as the input parameter. Solving the three equations in the front of \eqref{STU_expression}, we obtain
\begin{equation}
    \begin{aligned}[b]
        \tilde{S} &= 367.29\delta s_l^2 + 37.92\delta \Gamma_Z, \\
        \tilde{T} &= 132.35\delta s_l^2 + 52.94\delta \Gamma_Z, \\
        \tilde{U} &= -2.62\delta m_Z + 144.31\delta s_l^2 - 37.92\delta \Gamma_Z,
    \end{aligned}
\end{equation}
where $\delta m_Z = m_Z^* - m_Z^{\mathrm{input}}$, $\delta s_l^2=s_l^{*2} - (s_l^2)^{\mathrm{SM}}$ and $\delta \Gamma_Z=\Gamma_Z^* - (\Gamma_Z)^{\mathrm{SM}}$. The fourth equation in (\ref{STU_expression}) is also applied to compute $\delta \tilde{N}_{\nu}$, in which the best fitted $\xi^*$ is implicitly accommodated through calculating the $(\Gamma_Z)^{\text{pSM}}$ and the $(\Gamma_{\nu \nu})^{\text{pSM}}$ with the corrected vertices in (\ref{PseudoSMZCoupling}).

\begin{table}[!t]
    \setlength{\tabcolsep}{.5em}
    \begin{tabular}{|c|c|ccc|}
    \hline
      &  Result & \multicolumn{3}{c|}{Correlation} \\
    \hline
    $S$ & $0.06 \pm 0.1$ & $1.00$ & & \\
    $T$ & $0.11 \pm 0.12$ & $0.90$ & $1.00$ & \\
    $U$ & $-0.02 \pm 0.09$ & $-0.57$ & $-0.82$ & $1.00$ \\
    \hline
    \end{tabular}
    \caption{Global fit results of the oblique parameters $S$, $T$, and $U$ adopted from Ref.~\cite{Lu:2022bgw}.} \label{STU_Fitted}
\end{table}

The $\chi^2_{\rm STU}$ comparing the $\tilde{S}$, $\tilde{T}$ and $\tilde{U}$ with the global fitted data is defined by
\begin{eqnarray}
    \chi^2_{\rm STU} = \begin{pmatrix}
    \tilde{S}-S_0, & \tilde{T}-T_0, &\tilde{U}-U_0 
    \end{pmatrix} C^{-1} \begin{pmatrix}
    \tilde{S}-S_0 \\
    \tilde{T}-T_0 \\
    \tilde{U}-U_0
    \end{pmatrix},
\end{eqnarray}
where $S_0$, $T_0$ and $U_0$ are the best global-fitted values, $C^{-1}$ represents the inverse of the covariance matrix of $S_0$, $T_0$ and $U_0$. In this paper, the $S_0$, $T_0$, $U_0$ and $C^{-1}$ are adopted from Tab.~\ref{STU_Fitted}. 

In this paper, based upon some typical models (See an enumeration of the $Z^{\prime}$-models in Ref.~\cite{Langacker:2008yv}, and the references therein), we are going to show our calculated $\tilde{S}$, $\tilde{T}$, $\tilde{U}$, $\delta \tilde{N}_{\nu}$, as well as the estimated $e^+ e^-$ collider sensitivity in three scenarios. They are
\begin{itemize}
    \item Scenario I: $\hat{Z}^{\prime}$ only couples to invisible particles.

    This scenario is inspired from the dark matter model associated with a $Z^{\prime}$, in which $Z^{\prime}$ plays a crucial role connecting the visible world with the dark sector. $Z^{\prime}$ might couple with the dark matter, and the dark matter particles decayed from it are invisible at a collider. Besides the dark matter, $Z^{\prime}$ might also couple with the sterile neutrino, which might also be invisible if they are long-lived enough to decay outside the detector.

    This scenario is accomplished by straightforwardly assigning a $c_{\hat{Z}^{\prime} \hat{Z}^{\prime}}$ value in (\ref{mass-squares}) for convenience, rather than introducing some invisible fields for the $\hat{Z}^{\prime}$ to decay into them.

    \item Scenario II: $\hat{Z}^{\prime}$ couples with the SM fermions universally among all three generations.

    Charged under the $U(1)^{\prime}$ gauge symmetry universally among all three generations, the coupling constants are stringently constrained by the off-shell $\hat{Z}^{\prime}$ mediated processes, leading to a particularly small $c_{\hat{Z}^{\prime} \hat{Z}^{\prime}}$ to give a narrow but sharp valley-like structure imposed on the resonance. The initial momentum distribution in the beams and the ISR effect smear this structure to give a relatively ``smooth'' curve. 
    Therefore, the electroweak observables are then extract by comparing the $Z^{\prime}$ induced pseudo-LEP data with the pseudo-SM template results.

    Here we define the coupling constants as
    \begin{eqnarray}
    g_{u_R}\hat{Z}^{\prime}_{\mu}\bar{u}_{R}\gamma^{\mu}u_{R} + g_{u_L}\hat{Z}^{\prime}_{\mu}\bar{u}_{L}\gamma^{\mu}u_{L} &=& \hat{Z}^{\prime}_{\mu}\bar{u}\gamma^{\mu}(\frac{g_{u_L}+g_{u_R}}{2}-\frac{g_{u_L}-g_{u_R}}{2}\gamma^5)u, \nonumber \\
    g_{d_R}\hat{Z}^{\prime}_{\mu}\bar{d}_{R}\gamma^{\mu}d_{R} + g_{d_L}\hat{Z}^{\prime}_{\mu}\bar{d}_{L}\gamma^{\mu}d_{L} &=& \hat{Z}^{\prime}_{\mu}\bar{d}\gamma^{\mu}(\frac{g_{d_L}+g_{d_R}}{2}-\frac{g_{d_L}-g_{d_R}}{2}\gamma^5)u, \nonumber \\
    g_{l_R}\hat{Z}^{\prime}_{\mu}\bar{l}_{R}\gamma^{\mu}l_{R} + g_{l_L}\hat{Z}^{\prime}_{\mu}\bar{l}_{L}\gamma^{\mu}l_{L} &=& \hat{Z}^{\prime}_{\mu}\bar{l}\gamma^{\mu}(\frac{g_{l_L}+g_{l_R}}{2}-\frac{g_{l_L}-g_{l_R}}{2}\gamma^5)u, \nonumber \\
    g_{\nu_L}\hat{Z}^{\prime}_{\mu}\bar{\nu}_{L}\gamma^{\mu}\nu_{L} &=& \hat{Z}^{\prime}_{\mu}\bar{\nu}(\frac{g_{\nu_L}}{2}-\frac{g_{\nu_L}}{2}\gamma^5)\nu. \label{coupling_Zp}
    \end{eqnarray} 

    \item Scenario III: $\hat{Z}^{\prime}$ couples with the SM fermions depending on their generations.

    For some particular models (Ref.~\cite{Salvioni:2009jp} enumerated such models, and references can be found therein), a particular generation of particles might be charged under the $U(1)^{\prime}$ group, or two generations of particles takes a particular combination of the $U(1)^{\prime}$ charge (For an example, the $U(1)_{L_{\mu}-L_{\tau}}$ models\cite{He:1990pn, He:1991qd, Ma:2001md, Foot:1994vd, Baek:2001kca}). This not only affects the shape of the $Z$-resonance, but also breaks the universality of the $l^+ l^-$ branching ratios, where $l=e, \mu, \tau$. In this paper, we only discuss the $e$-$\mu$ asymmetry, and utilize the
    \begin{equation}
      \frac{R_e}{R_{\mu}} \label{R_Ratio} 
    \end{equation} 
    where
    \begin{equation}
      R_l \equiv \frac{\Gamma_{\mathrm{had}}}{\Gamma_{l^{+}l^{-}}}
    \end{equation}
    to observe such an asymmetry.
\end{itemize}

\section{Numerical Results} \label{Numerical}

The effective Lagrangian \eqref{LKin} takes such parameters, $m_{\hat{Z}^{\prime}}$, $\epsilon_B$, $\Lambda_W$, 
and $\lambda_{H Z^{\prime}}${. For the convenience of} a relatively intuitive presentation, equivalently,  $\epsilon_B$, $\epsilon_W$, 
$m_{\hat{Z}^{\prime}}$, and $\delta m^2$ appeared in \eqref{mass-squares} and \eqref{KOrigin} and can be treated as the free parameters for further discussions. Among them, $\epsilon_B$, $\epsilon_W$ and $m_{\hat{Z}^{\prime}}$ are the most important. 
$\delta m^2$ can also give rise to non-perturbative mixings.  However, if one switch off the $\delta m^2$ and diagonalize the kinematic terms by \eqref{KinMixingTerms}, off-diagonal terms arise in \eqref{DiagonalFirstStep}. Expand \eqref{DiagonalFirstStep} up to the lowest non-trivial order, it becomes
\begin{eqnarray}
    (V_1 V_2 V_3)^T \mathcal{M}_V^{2 ~\prime} V_1 V_2 V_3 \simeq \left( 
    \begin{array}{ccc}
    m_{\hat{Z}^{\prime}}^2 & -m_{\hat{Z}^{\prime}}^2 \epsilon_B & -m_{\hat{Z}^{\prime}}^2 \epsilon_W \\
    -m_{\hat{Z}^{\prime}}^2 \epsilon_B & \frac{\hat{g}^{\prime}}{4} \hat{v}^2 & -\frac{\hat{g}^{\prime} \hat{g}}{4} {\hat{v}^2} \\
    -m_{\hat{Z}^{\prime}}^2 \epsilon_W & -\frac{\hat{g}^{\prime} \hat{g}}{4} {\hat{v}^2} & \frac{\hat{g}^2}{4} \hat{v}^2
    \end{array}\right). \label{EquivalentCross}
\end{eqnarray}
Therefore, a particular choice of $\frac{\epsilon_B}{\epsilon_W} = -\frac{\hat{g}^{\prime}}{\hat{g}}$ generates the similar off-diagonal terms in \eqref{MassOrigin} which are proportional to $\delta m^2$. Therefore, the effects from $\delta m^2$ can be alternatively estimated by choosing an appropriate combination of $\epsilon_B$ and $\epsilon_W$ approximately, so in the rest of this paper, we neglect $\delta m^2$.

To further understand this approximate equivalence prompted by (\ref{EquivalentCross}), notice that changing into the momentum representation, the $-\frac{\epsilon_B}{2} \hat{Z}^{\prime}_{\mu \nu} B^{\mu \nu} - \frac{\epsilon_W}{2} \hat{Z}^{\prime}_{\mu \nu} W^{3 \mu \nu}$ in (\ref{EpsBEpsW}) can be reduced by substituting the $\partial_{\mu}$ with $-i p_{\mu}$, where $p_{\mu}$ is the momentum of the propagating vector boson. Since we concentrate in the $s$-channel vector propagators around $\sqrt{p^2} \approx m_{\hat{Z}} \approx m_{\hat{Z}^{\prime}}$, the equivalent mass crossing terms $-\epsilon_B m_{\hat{Z}^{\prime}}^2 \hat{Z}^{\prime}_{\mu} B^{\mu} - \epsilon_W m_{\hat{Z}^{\prime}}^2 \hat{Z}^{\prime}_{\mu} W^{3 \mu}$ arise in this case. The remained $\propto \epsilon_B (p \cdot \hat{Z}^{\prime}) (p \cdot B)  + \epsilon_W (p \cdot \hat{Z}^{\prime}) (p \cdot W^3)$ terms will finally be connected with the fermionic final states, and the $p$ will be inevitably dotted into a fermionic external line to summon the Goldstone coupling constants, which are small enough to be neglected in this paper, just as the similar mechanism to neglect the $p^{\mu} p^{\nu}$ terms in (\ref{PiMuNu}) to be discussed in Appendix \ref{AppResum}.  Therefore, up to the linear order, the contributions from $\delta m^2$ is equivalent to a particular combination of $\epsilon_B$ and $\epsilon_W$ if all of them are sufficiently small.

In the following of this section, we plot the $\tilde{S}$, $\tilde{T}$, $\tilde{U}$ and $\delta \tilde{N}_{\nu}$ for Scenario I-III on various plains. 
For Scenario III when universality among generations is broken, $\frac{R_e}{R_{\mu}}-1$ is also plotted.  
Besides these familiar observables which might not be able to summarize the complete distortion of the $Z$-like resonance, the $\chi_F^2$ defined in \eqref{chi-square} can also reflect the deviation of the lineshape from a standard single-particle resonance, which can also be regarded as a hint of the new physics. In this paper, we also take best-fitted $\chi_F^2$ into account, and the details will be addressed later.

The Monte-Carlo algorithm that we utilize inevitably introduces statistical fluctuations in our results, thus smoothness is lost in the plotted figures. We use a second-order polynomial of $\epsilon_B$, $\epsilon_W$ and the coupling constants defined in \eqref{coupling_Zp} to fit each of the fluctuated $\tilde{S}$, $\tilde{T}$, $\tilde{U}$, $\delta \tilde{N}_{\nu}$ and $\frac{R_e}{R_{\mu}}-1$ by least square method to smooth the results. Since all new physics corrections are destined to vanish if one switches off $\epsilon_{B}$, $\epsilon_{W}$ and all additional coupling constants between the $\hat{Z}^{\prime}$ and the SM-fermions, we utilize this condition to fix the constant terms of the polynomials. The similar results from several independent runs with different random seeds verify the reliability of this fitting algorithm, so in this paper, we show the fitted results in the figures.

In principle, we should compare our results with the global fitted results of all $\tilde{S}$, $\tilde{T}$, $\tilde{U}$, $\delta \tilde{N}_{\nu}$, and $\frac{R_e}{R_{\mu}}-1$ parameters. Unfortunately, in the literature $\tilde{S}$, $\tilde{T}$, $\tilde{U}$ are fitted with the assumption of universality and $\delta \tilde{N}_{\nu}=0$. $\delta \tilde{N}_{\nu}$ is also fitted with the assumption that the visible $Z$-propagator is not distorted, that is to say, $\tilde{S}=0$, $\tilde{T}=0$, $\tilde{U}=0$. A complete global fitting including all these parameters is far beyond our target, and remember that we are only able to show readers the ``sensitivity'' of a LEP-like electron-positron collider without performing a real fitting process due to the lack of the published data, so in this paper, we still show the corresponding ``STU-$(1,2)$-$\sigma$'', ``$\delta \tilde{N}_{\nu}$-$(1,2)$-$\sigma$'', ``$R_e/R_{\mu}$-$(1,2)$-$\sigma$'' contours in each of the figures. The prefixes ``STU-'', ``$\delta \tilde{N}_{\nu}$-'', ``$R_e/R_{\mu}$-'' indicate that the $1$-$\sigma$ and $2$-$\sigma$ fitted results originate from the global fitted oblique parameter in Tab.~\ref{STU_Fitted}, indirect $N_{\nu}$ results from Ref.~\cite{Workman:2022ynf}, $R_e$ and $R_{\mu}$ data with their uncertainties adopted from Ref.~\cite{Workman:2022ynf}. However we should note that these contours only characterize the sensitivity of the collider on this model, and should not be regarded as real constraints.

The best-fitted $\chi_F^2$'s are also computed. Practically, the LEP collaborations published the best-fitted $\chi^2/N_{\rm df}$'s in Ref.~\cite{LEP:2000pgt}, where $N_{\rm df}$ means the number of degrees of freedom. For a real experiment, the $\chi^2$ is usually dominated by statistical fluctuations so the expectation value of the $\chi^2/N_{\rm df}$ becomes 1, which is compatible with the results published in Ref.~\cite{LEP:2000pgt}. According to the central limit theorem, the distribution of $\chi^2/N_{\rm df}$ should approach a normal distribution with the standard deviation $\sigma=\sqrt{2/N_{\rm df}}$ when $N_{\rm df} \gg 1$. Therefore, if the $\chi^2/N_{\rm df}$ acquired from the experimental data deflect from the value of 1 too ``far away'', e.g.,  $|\chi^2/N_{\rm df}-1| \gtrsim {\color{green}2} \sigma$, one can reasonably doubt the validity of the assumptions that the experimental collaborations utilize to fit the data. In Ref.~\cite{LEP:2000pgt} specifically, if such a case happens, the validity of a SM $Z$-resonance is doubted and the hint of some new physics arises. Our $\chi_F^2$ defined in \eqref{chi-square} also includes the contributions from the Monte-Carlo statistical fluctuations, however as we have mentioned, we have generated our events in a much larger integrated luminosity to suppress these fluctuations, so in this paper, the statistical fluctuations in $\chi_F^2/N_{\rm df}$ are almost deducted, and it can reflect the systematic deviation of the distorted $Z^{\prime}$-$Z$ resonance from the pure $Z$ lineshape. Therefore,  considering that the $\chi_F^2$ is non-negative, we plot the best-fitted $\chi_F^2/N_{\rm df} = 1.69 \sqrt{2/N_{\rm df}}$ contour tagged by ``$\chi_F^2$-$2 \sigma$'' to show the sensitivity of a LEP-like collider to find the difference between the $Z^{\prime}$-$Z$ resonance and a SM-$Z$ lineshape.  Here the ``$2 \sigma$'' means that the single-sided $p$-value of a $\chi^2$ distribution (or the approximate normal distribution) is equivalent to the double-sided $p$-value of a normal distribution corresponding to the $2 \sigma$ sensitivity when $N_{\rm{df}} \gg 1$. As we have mentioned before, all through this paper we adopt $N_{\rm{df}}=35$, except we will mention that in our Scenario III, for the particular $\frac{R_{e}}{R_{\mu}}$ fitting processes we deleted some channels, therefore $N_{\rm{df}}=21$.

Besides, people are more interested about the sensitivity of some proposed future colliders. According to Ref.~\cite{CEPCStudyGroup:2018ghi}, the CEPC is expected to significantly improve the uncertainties of the electroweak precision measurements. With the expected sensitivities published in Ref.~\cite{CEPCStudyGroup:2018ghi} around the $Z$-pole, we also plot the ``CEPC-ST-$(1,2)$-$\sigma$'', ``CEPC-$\delta \tilde{N}_{\nu}$-$(1,2)$-$\sigma$'', ``CEPC-$R_e/R_{\mu}$-$(1,2)$-$\sigma$'' contours within one plot for each of the $\epsilon_B$ and $\epsilon_W$ plain. Note that Ref.~\cite{CEPCStudyGroup:2018ghi} only shows the expected $S$-$T$ results in its Figure 11.18, so we are forced to give up the $U$. Although in principle $U$-sensitivity can be extracted from the Tab.~11.16 of Ref.~\cite{CEPCStudyGroup:2018ghi}, however the complete covariance matrix is missing for a complete fitting. Since we only target at the collider's potential sensitivity, neglecting the $U$ is expected not to affect the final results significantly. Another subtle thing is that Ref.~\cite{CEPCStudyGroup:2018ghi} only gives $R_{\mu}$, which is insufficient to estimate the $R_{e}/R_{\mu}$ uncertainties. In fact, at a lepton collider, the $e^+ e^- \rightarrow e^+ e^-$ channel data are slightly less precise than the corresponding $e^+ e^- \rightarrow \mu^+ \mu^-$ data because of the less accuracy of the electron/positron trajectory measurements. 
Fortunately, Ref.~\cite{CEPCPhysicsStudyGroup:2022uwl, Belloni:2022due} give both the expected $R_e$ and $R_{\mu}$ sensitivities at the CEPC. In this paper, we adopt the values there to estimate the expected CEPC $R_{e}/R_{\mu}$ uncertainties as a reference in this paper.

Now we show our results for the three scenarios respectively.

\subsection{Scenario I: $\hat{Z}^{\prime}$ couples with invisible matters} \label{ZprimetoDM}

In this scenario, the width $\Gamma_{\hat{Z}^{\prime}}$ that $\hat{Z}^{\prime}$ only decays to invisible matter is regarded as an input parameter, so $c_{\hat{Z}^{\prime}\hat{Z}^{\prime}}$ and $c_{\hat{Z}\hat{Z}}$ are given by
\begin{equation}
    c_{\hat{Z}^{\prime}\hat{Z}^{\prime}} = -m_{\hat{Z}^{\prime}}\Gamma_{\hat{Z}^{\prime}}, \quad c_{\hat{Z}\hat{Z}} = -m_{\hat{Z}}\Gamma_{\hat{Z}},
\end{equation}
where $m_{\hat{Z}}=\frac{\hat{g}^2 + \hat{g}^{\prime 2}}{4}\hat{v}^2$ and the $\Gamma_{\hat{Z}}$ which is the width of the $\hat{Z}$-boson can be computed by the event generator if all mixing parameters switch-off.
Since the $\hat{Z}^{\prime}$ has no coupling to SM fermions, there is no self energy diagram of $\hat{Z}^{\prime}$-$\hat{Z}$. Therefore we have 
\begin{equation}
    c_{\hat{Z}^{\prime}\hat{Z}} = 0.
\end{equation}

If the $\Gamma_{\hat{Z}^{\prime}}$ is close to $\Gamma_{\hat{Z}}$, and the $m_{\hat{Z}^{\prime}}$ is also close to $m_{\hat{Z}}$, maximum mixings between $Z^{\prime}$-$Z$ might arise, however the overlapped and interfered peaks still look like a single $Z$-pole. 

\begin{figure}[!t]
    \includegraphics[width=0.32\textwidth]{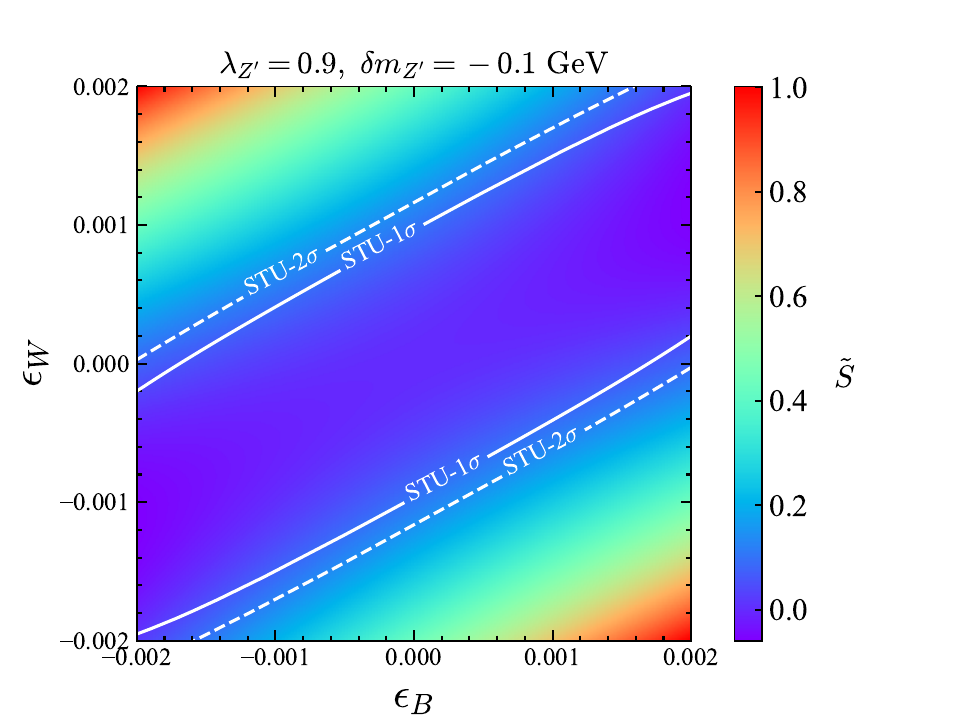}
    \includegraphics[width=0.32\textwidth]{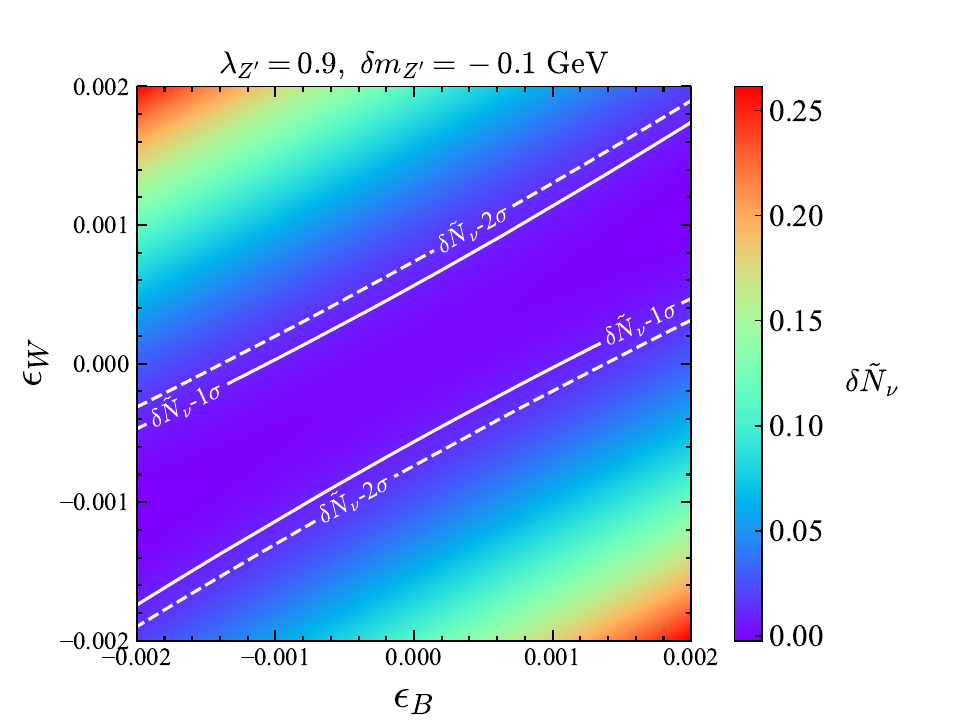}
    \includegraphics[width=0.24\textwidth]{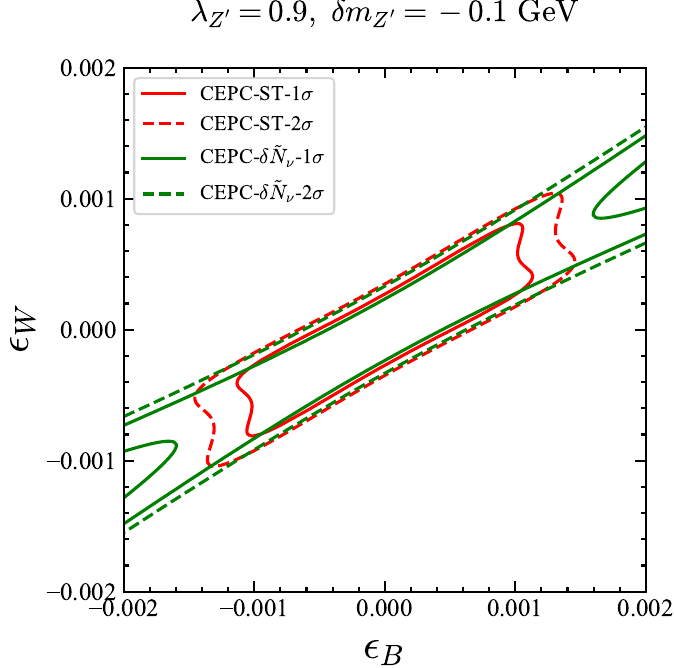}
    \caption{Plots for $\delta m_{Z^{\prime}} = -0.1$~GeV, $\lambda_{Z^{\prime}} = 0.9$ and $\delta m^2 = 0$ on the $\epsilon_B$-$\epsilon_W$ plain. The right panel shows the estimated CEPC sensitivity. The remaining two panels display the heat maps of $\tilde{S}$ and $\delta \tilde{N}_{\nu}$. 1-$\sigma$ and 2-$\sigma$ contours are displayed on the $\epsilon_B$-$\epsilon_W$ plain. The $\tilde{T}$ and $\tilde{U}$ can be obtained by multiplying $\tilde{S}$ by the coefficients $1.8$ and $-1.6$, respectively.  Current $N_{\nu}$ data is extracted from Ref.~\cite{Workman:2022ynf}.} \label{dmp-0.1_alpha0.9} 
\end{figure}

\begin{figure}[!t]
    \includegraphics[width=0.32\textwidth]{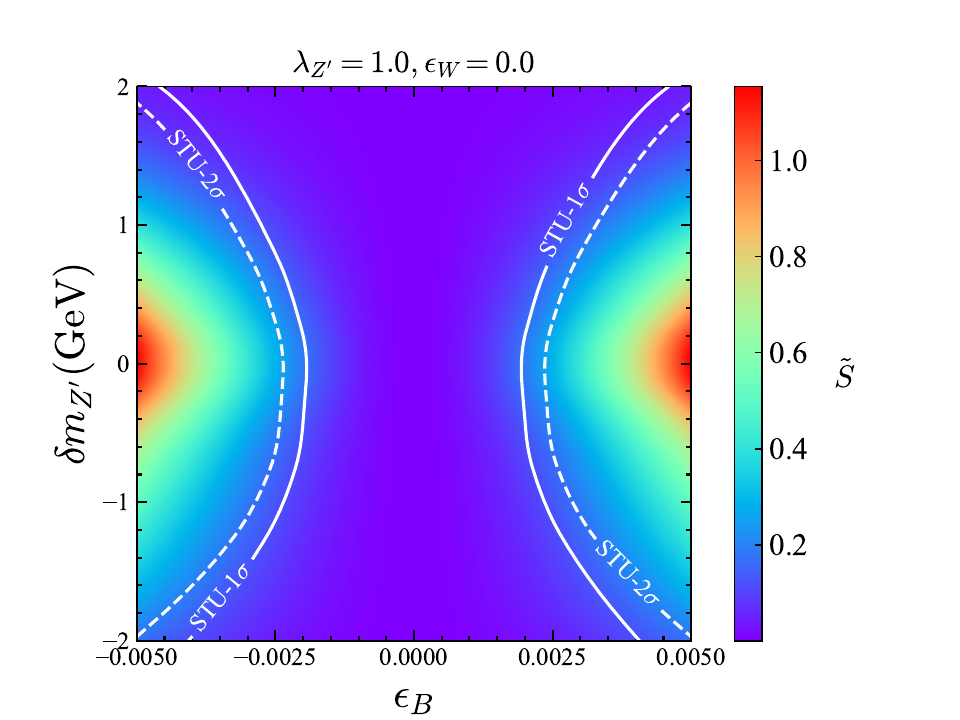}
    \includegraphics[width=0.32\textwidth]{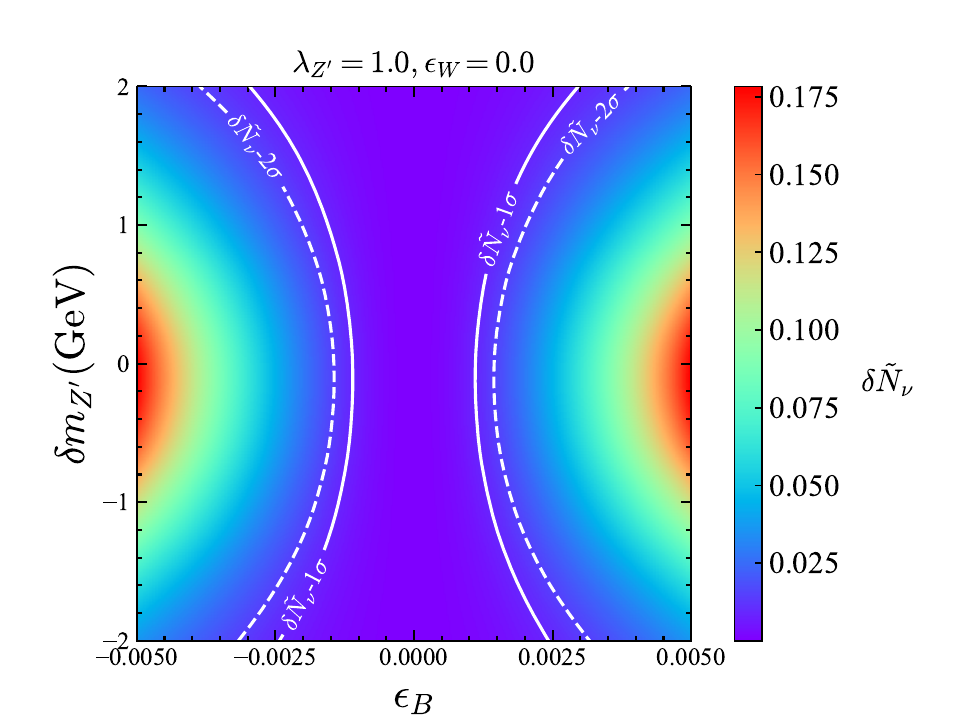}
    \includegraphics[width=0.24\textwidth]{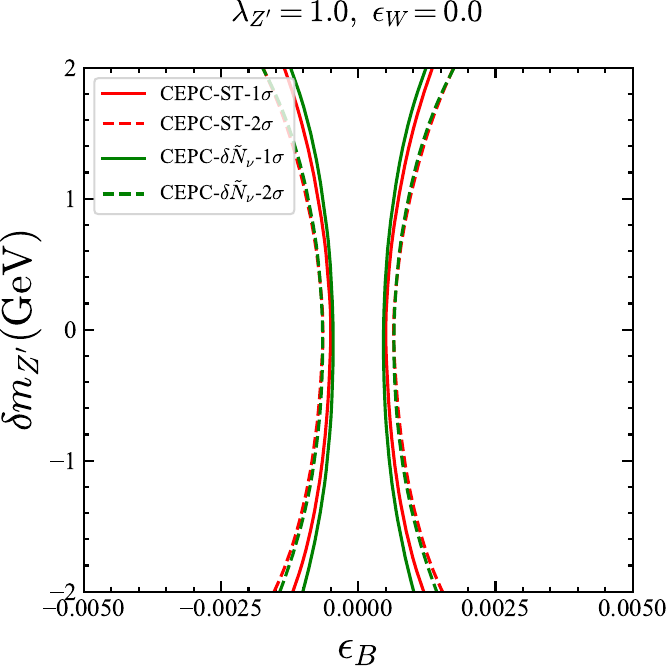}
    \caption{Plots for $\epsilon_W = 0$, $\lambda_{Z^{\prime}} = 1.0$ and $\delta m^2 = 0$ on the $\epsilon_B$-$\delta m_{Z^{\prime}}$ plain. The right panel shows the estimated CEPC sensitivity. The remaining two panels display the heat maps of $\tilde{S}$ and $\delta \tilde{N}_{\nu}$, and plot the 1-$\sigma$ and 2-$\sigma$ contours on the $\epsilon_B$-$\epsilon_W$ plain. The $\tilde{T}$ and $\tilde{U}$ can be obtained by multiplying $\tilde{S}$ by the coefficients $1.5$ and $-1.2$, respectively. Current $N_{\nu}$ data is extracted from Ref.~\cite{Workman:2022ynf}.} \label{dmp-0.1_alpha0.9-delZp} 
\end{figure}

\begin{figure}[!t]
    \includegraphics[width=0.32\textwidth]{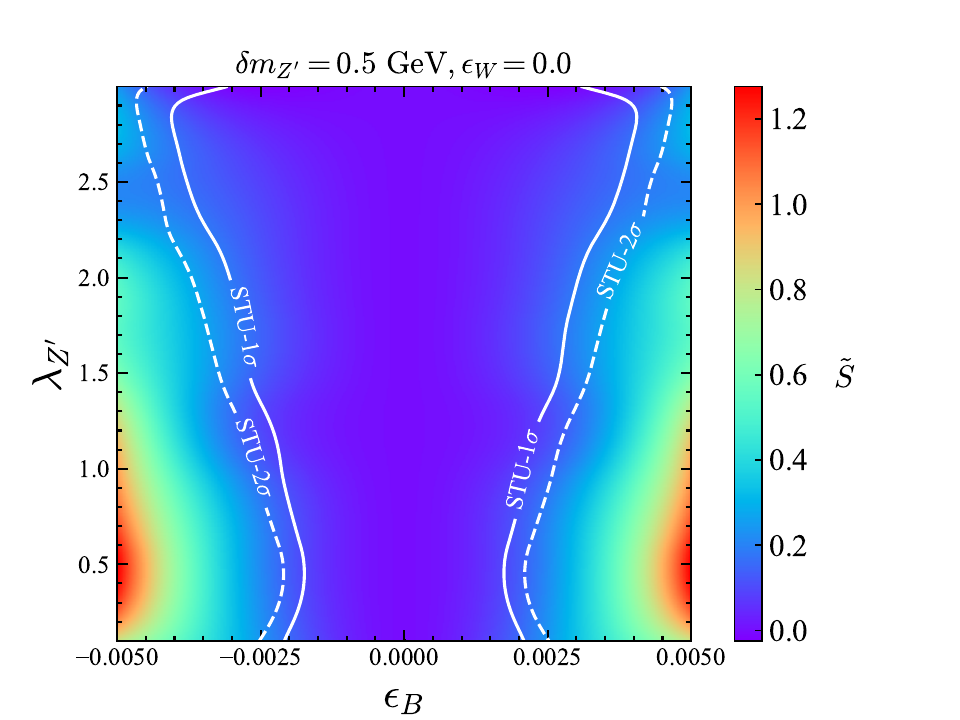}
    \includegraphics[width=0.32\textwidth]{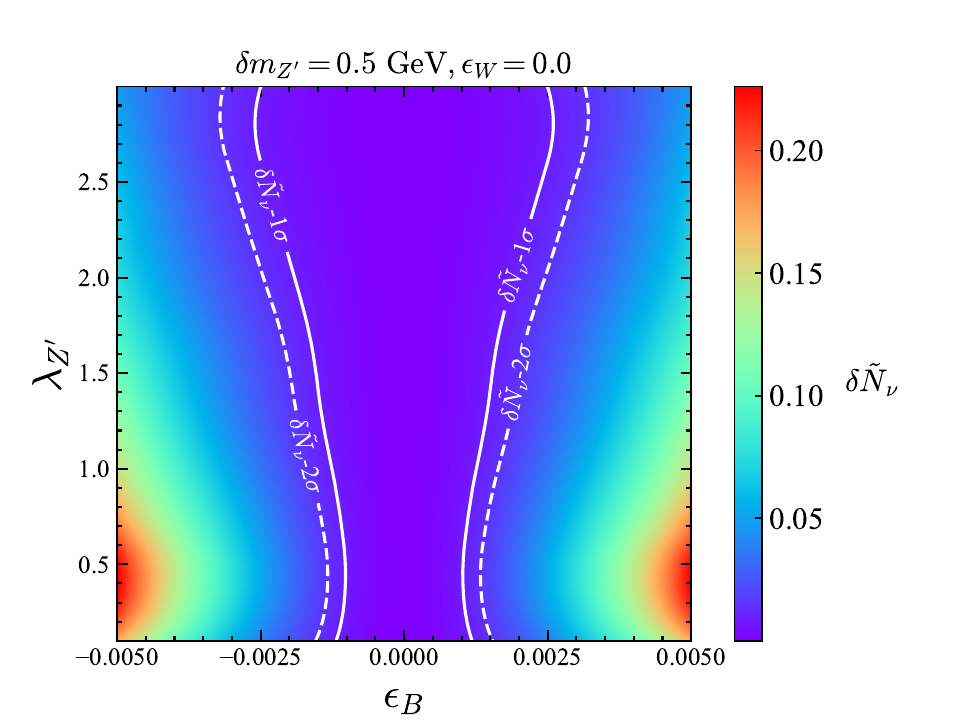}
    \includegraphics[width=0.24\textwidth]{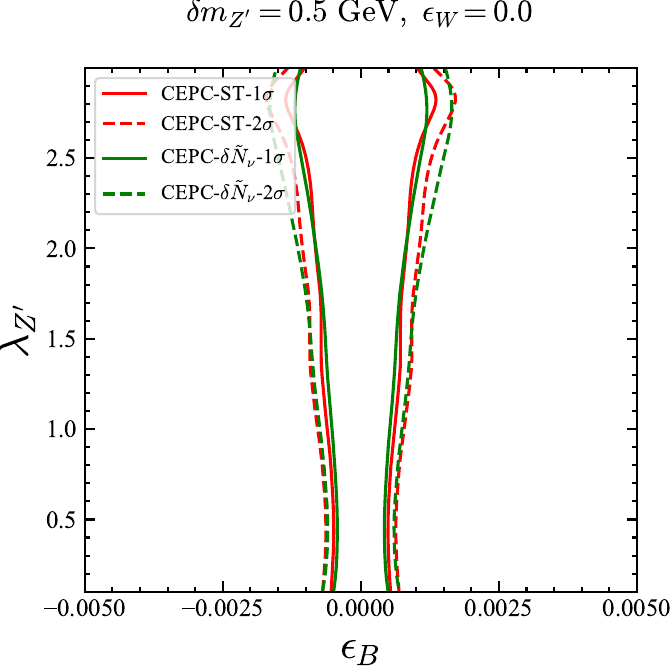}
    \caption{Plots for $\delta m_{Z^{\prime}} = 0.5$~GeV, $\epsilon_W=0$ and $\delta m^2 = 0$ on the $\epsilon_B$-$\lambda_{Z^{\prime}}$ plain. The right panel shows the estimated CEPC sensitivity. The remaining two panels display the heat maps of $\tilde{S}$ and $\delta \tilde{N}_{\nu}$, and plot the 1-$\sigma$ and 2-$\sigma$ contours on the $\epsilon_B$-$\epsilon_W$ plain. The $\tilde{T}$ and $\tilde{U}$ can be obtained by multiplying $\tilde{S}$ by the coefficients $1.5$ and $-1.1$, respectively. Current $N_{\nu}$ data is extracted from Ref.~\cite{Workman:2022ynf}.} \label{dmp-0.1_alpha0.9-lamZp} 
\end{figure}

For convenience we define
\begin{equation}
    \lambda_{Z^{\prime}} = \frac{\Gamma_{\hat{Z}^{\prime}}}{\Gamma_{\hat{Z}}},\quad \delta m_{Z^{\prime}}=m_{\hat{Z}^{\prime}}-m_{\hat{Z}},
\end{equation}
which is the ratio of the widths of the two ``interaction-eigenstates'' and the mass difference between them, respectively. The $\delta m_{Z^{
\prime}}$ parameter is also used in the following two scenarios. 
We have tried several combinations of parameters like $\delta m_{Z^{\prime}} = 0, -0.1$ GeV, and $\lambda_{Z^{\prime}} = 0.5, 0.9$ on the $\epsilon_B$-$\epsilon_W$ plain, and found that the results are quite similar if $\hat{Z}^{\prime}$ and $\hat{Z}$ are nearly-degenerate. Finally, we choose to plot the results when $\lambda_{Z^{\prime}}=0.9$ and $\delta m_{Z^{\prime}}=-0.1$ GeV in Fig.~\ref{dmp-0.1_alpha0.9} as a paradigm on the $\epsilon_B$-$\epsilon_W$ plain.  
In Fig.~\ref{dmp-0.1_alpha0.9-delZp}, \ref{dmp-0.1_alpha0.9-lamZp}, we also plot the similar results on the $\epsilon_B$-$\delta m_{Z^{\prime}}$ and $\epsilon_B$-$\lambda_{Z^{\prime}}$ plains. Other conditions are addressed in the corresponding captions. In these figures we neglect the $\tilde{T}$ and $\tilde{U}$ results for brevity, since they can be roughly estimated by multiplying the universal factors upon the $\tilde{S}$ results within the parameter range of the plots. These factors are shown in the figure captions. In the following of this paper, we will also omit some pictures if such a universal factor exists. We also have to note that in Fig.~\ref{dmp-0.1_alpha0.9-delZp}, a simple polynomial fitting with the $\epsilon_B$, $\delta m_{Z^{\prime}}$ parameters fails along the $\delta m_{Z^{\prime}}$ direction, so we only use polynomials to fit the results along the $\epsilon_B$ directions for each fixed $\delta m_{Z^{\prime}}$, then interpolate the coefficients of the polynomials along the $\delta m_{Z^{\prime}}$ to finally smooth the results. Notice that the ``$\chi_F^2$-$2 \sigma$'' line is also missing, which means that in this scenario, generally the shape of the $Z$-$Z^{\prime}$ system looks very similar to a SM-like $Z$ resonance, making the familiar $\tilde{S}$, $\tilde{T}$, $\tilde{U}$, $\delta \tilde{N}_{\nu}$ observables much more sensitive than the $\chi_F^2$ parameters.


\subsection{Scenario II: $\hat{Z}^{\prime}$ couples with the SM fermions universally among all three generations} \label{ZprimetoSM}

In this scenario, $\hat{Z}^{\prime}$ couples directly with the SM quarks and leptons. General definitions of the coupling constants have been listed in \eqref{coupling_Zp}. To compute the $ \text{Im} \Pi_{\hat{Z}^{\prime} \hat{Z}}$, the following fermions' couplings with the $\hat{Z}_{\mu}$ fields are required,
\begin{eqnarray}
    \mathcal{L} \supset \hat{Z}_{\mu}\bar{\psi}^{f}\gamma^{\mu}(V_{\hat{Z}}^f - A_{\hat{Z}}^f\gamma^5)\psi^f, \quad f = u, d, l, \nu,
\end{eqnarray}
where according to the SM,
\begin{eqnarray}
 &V_{\hat{Z}}^u = \frac{1}{4c_{\mathrm{W}}} - \frac{2s^2_{\mathrm{W}}}{3c_{\mathrm{W}}}, \quad A_{\hat{Z}}^u = \frac{1}{4c_{\mathrm{W}}}, \quad V_{\hat{Z}}^d = -\frac{1}{4c_{\mathrm{W}}} + \frac{s^2_{\mathrm{W}}}{3c_{\mathrm{W}}}, \quad A_{\hat{Z}}^d = -\frac{1}{4c_{\mathrm{W}}}, \nonumber \\
    &V_{\hat{Z}}^l = -\frac{1}{4c_{\mathrm{W}}} + \frac{s^2_{\mathrm{W}}}{c_{\mathrm{W}}}, \quad A_{\hat{Z}}^l = -\frac{1}{4c_{\mathrm{W}}}, \quad V_{\hat{Z}}^{\nu} = \frac{1}{4c_{\mathrm{W}}}, \quad A_{\hat{Z}}^{\nu} = \frac{1}{4c_{\mathrm{W}}}.
\end{eqnarray}
Denote $V_{\hat{Z}^{\prime}}^f \equiv \frac{g_{f_L}+g_{f_R}}{2}$ and $A_{\hat{Z}^{\prime}}^f \equiv \frac{g_{f_L}-g_{f_R}}{2}$,
under the approximation that the decay products are massless, $c_{\hat{Z} \hat{Z}}$, $c_{\hat{Z}^{\prime} \hat{Z}}$ and $c_{\hat{Z}^{\prime} \hat{Z}^{\prime}}$ are calculated to be
\begin{eqnarray}
    c_{\hat{Z} \hat{Z}} &=& -m_{\hat{Z}}\Gamma_{\hat{Z}}, \nonumber \\
    c_{\hat{Z}^{\prime} \hat{Z}^{\prime}} &=& -\frac{m_{\hat{Z}}^2}{12\pi}\sum_{f\neq t}(V^f_{\hat{Z}^{\prime}}V^f_{\hat{Z}^{\prime}} + A^f_{\hat{Z}^{\prime}}A^f_{\hat{Z}^{\prime}}) , \nonumber \\
    c_{\hat{Z}^{\prime} \hat{Z}} &=& -\frac{m_{\hat{Z}}^2}{12\pi}\sum_{f\neq t}(V^f_{\hat{Z}^{\prime}}V^f_{\hat{Z}} + A^f_{\hat{Z}^{\prime}}A^f_{\hat{Z}}). 
\end{eqnarray}

\begin{figure}
    \centering
    \includegraphics[width=0.48\textwidth]{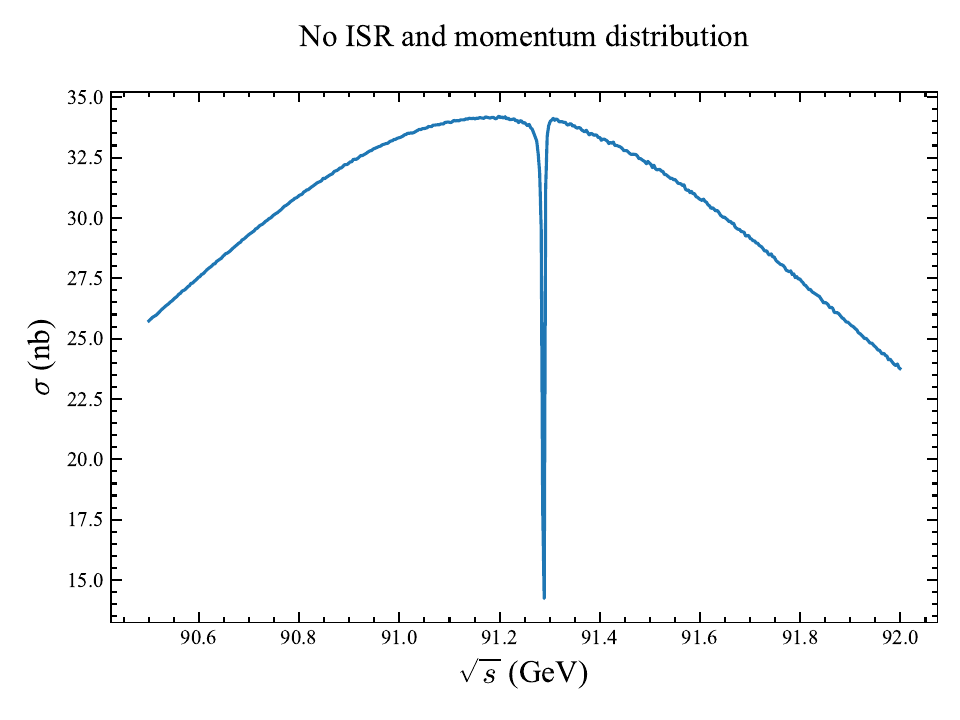}
    \includegraphics[width=0.48\textwidth]{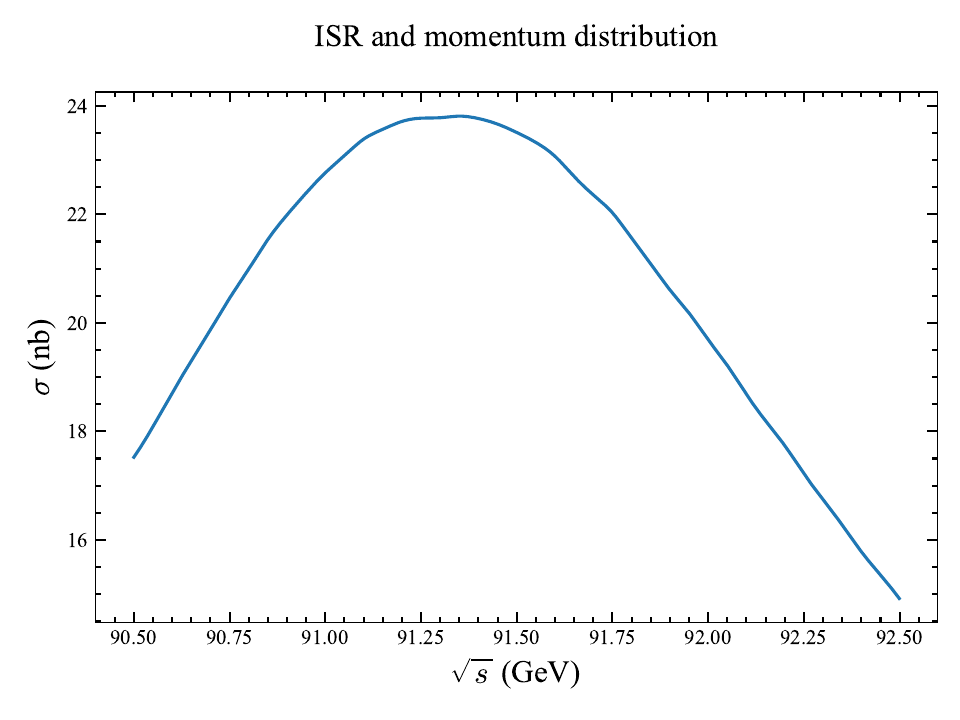}
    \caption{$Z$ line-shape predicted by an example model that $\hat{Z}^{\prime}$ couples with the SM fermions. The left panel shows the results when the electron/positron is ``bare'', while the right panel considers the ISR and beam momentum distribution effects.}
    \label{Smearing}
\end{figure}

In this scenario, the direct coupling constants between $\hat{Z}^{\prime}$ and the SM particles should be stringently constrained. Therefore the $c_{\hat{Z}^{\prime} \hat{Z}^{\prime}}$ and $c_{\hat{Z}^{\prime} \hat{Z}}$ become much smaller than $c_{\hat{Z} \hat{Z}}$. This actually suppresses the mixing angle and in this case, the $\hat{Z}^{\prime}$-like object cleaves a deep but narrow valley within the resonance induced by the $Z$-like object, which will later be smeared by the ISR and beam momentum distribution effects, just as we have plotted in Fig.~\ref{Smearing} as an example.

It is actually impossible for a practical lepton collider to scan every $\sqrt{s}$ in a sufficient resolution to depict such inconspicuous structure, so what we can do is still adopt the scattered $\sqrt{s}=[88.2$, $89.2$, $90.2$, $91.2$, $92.2$, $93.2$, $94.2] \rm{GeV}$ samples to extract the $\tilde{S}$, $\tilde{T}$, $\tilde{U}$, $\delta \tilde{N}_{\nu}$ parameters.

\begin{figure}[!t]
    \includegraphics[width=0.32\textwidth]{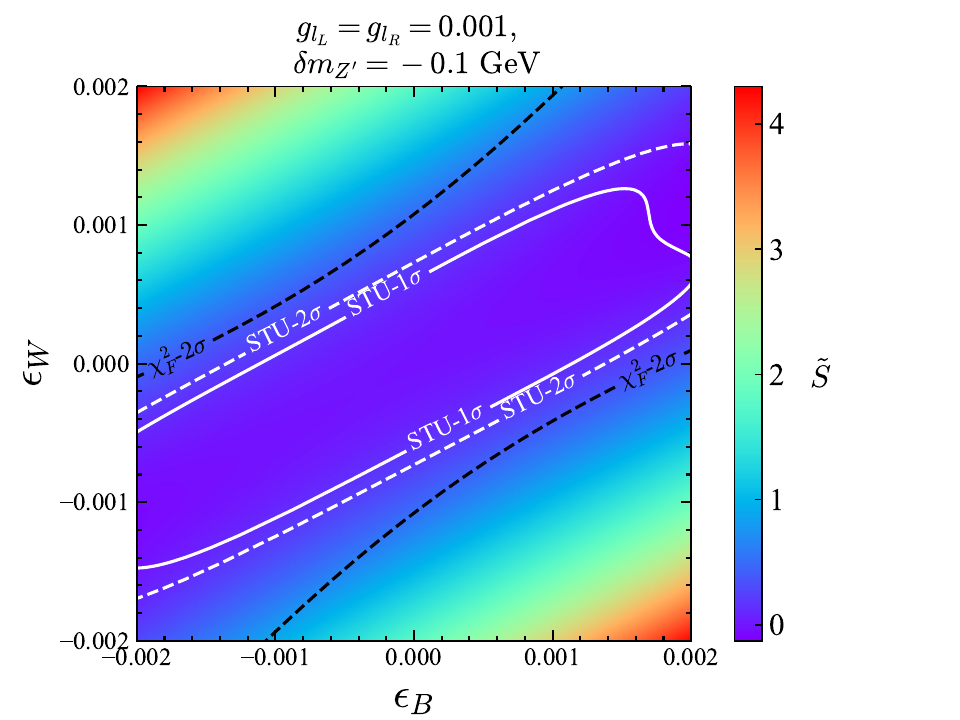}
    \includegraphics[width=0.32\textwidth]{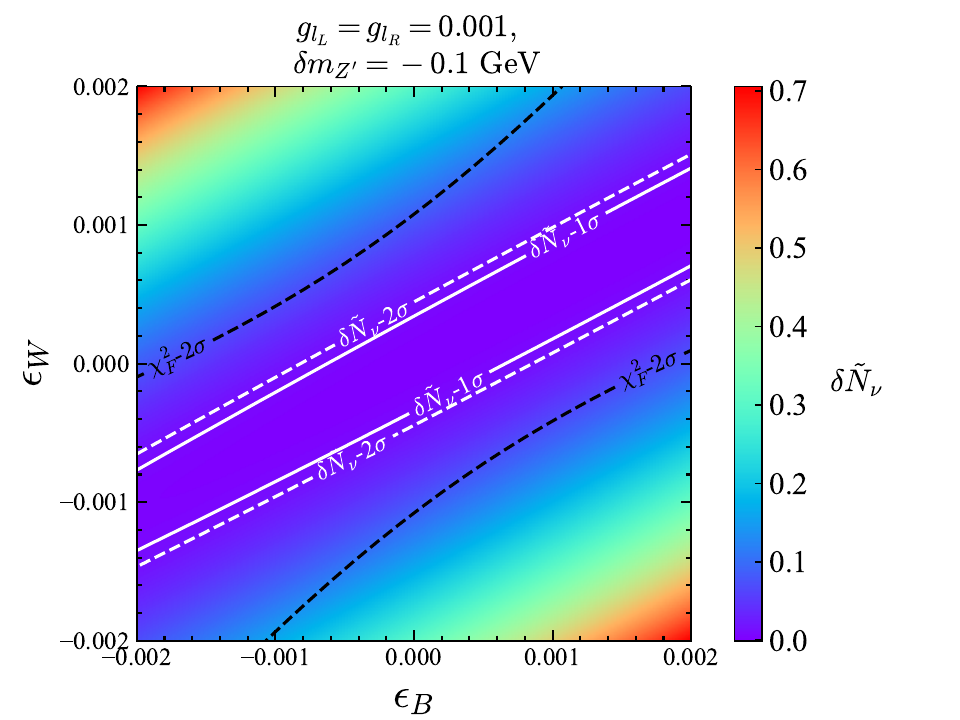}
    \includegraphics[width=0.24\textwidth]{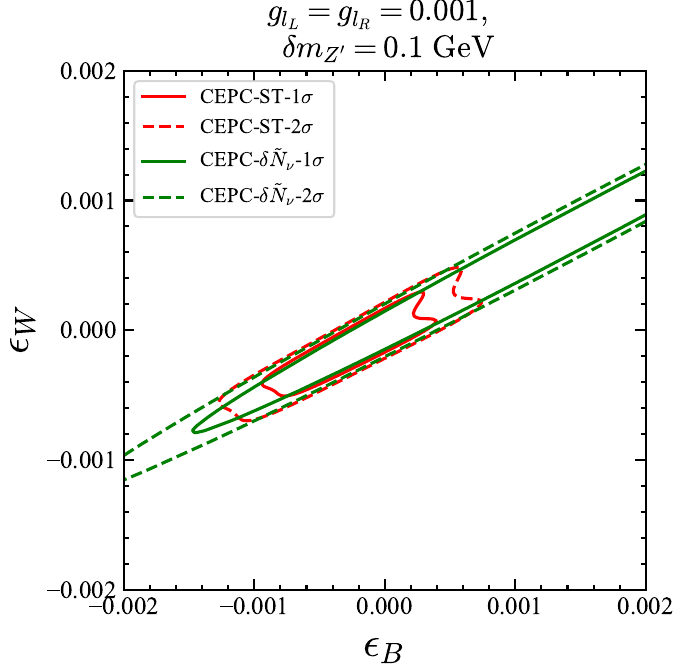}
    \caption{Plots for $\delta m_{Z^{\prime}} = 0.1$~GeV, $g_{l_L} = g_{l_R} = 0.001$ and $\delta m^2 = g_{u_L} = g_{u_R} = g_{d_L} = g_{d_R} = g_{\nu_L} = 0$ on the $\epsilon_B$-$\epsilon_W$ plain. The $\tilde{T}$ and $\tilde{U}$ can be obtained by multiplying $\tilde{S}$ by the coefficients $1.4$ and $-1.1$, respectively. The other symbols are the same as in Fig.~\ref{dmp-0.1_alpha0.9}.} \label{LV0.1}
\end{figure}

\begin{figure}[!t]
    \includegraphics[width=0.32\textwidth]{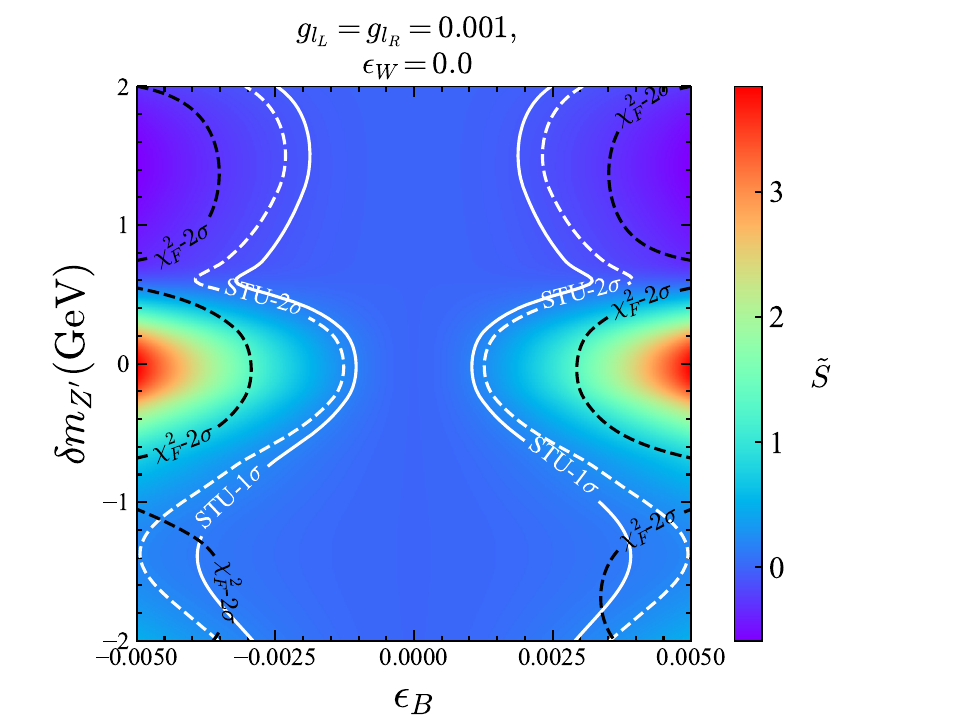}
    \includegraphics[width=0.32\textwidth]{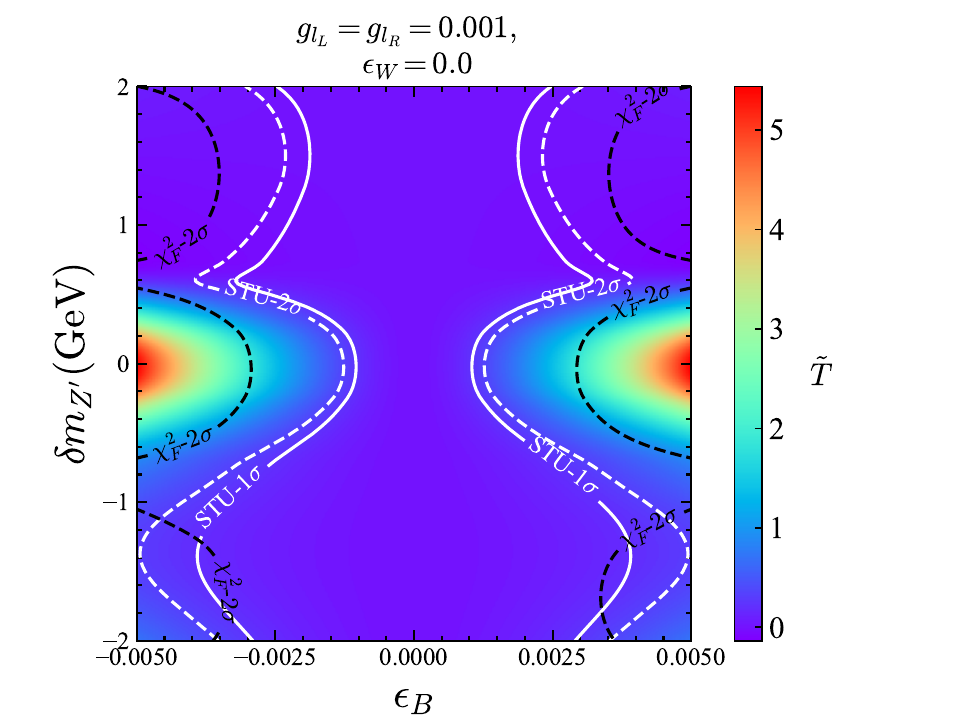}
    \includegraphics[width=0.32\textwidth]{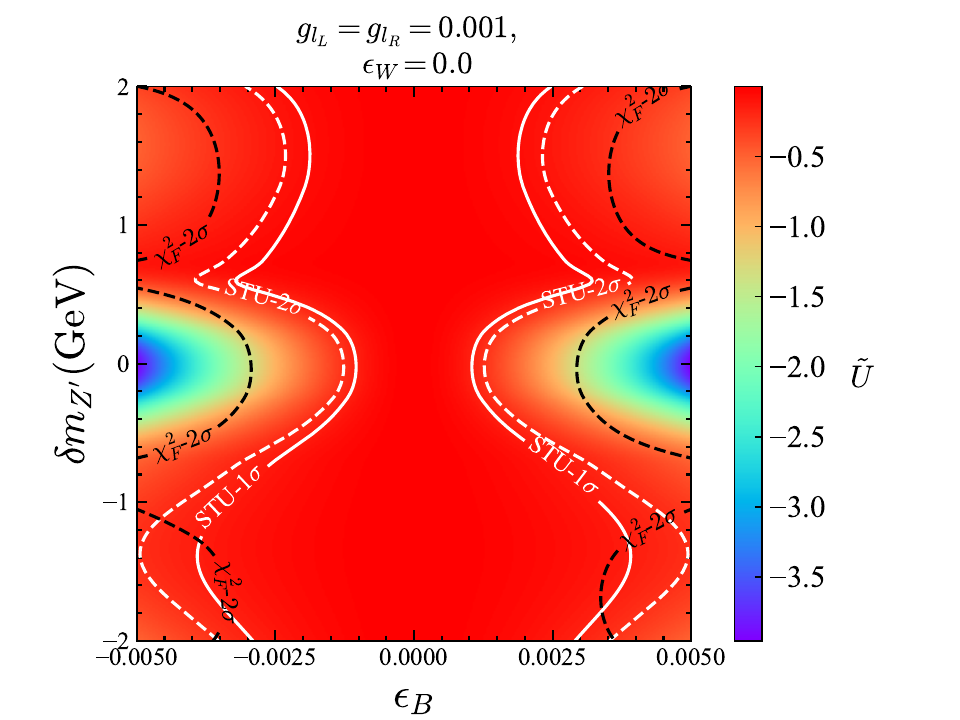}
    \includegraphics[width=0.32\textwidth]{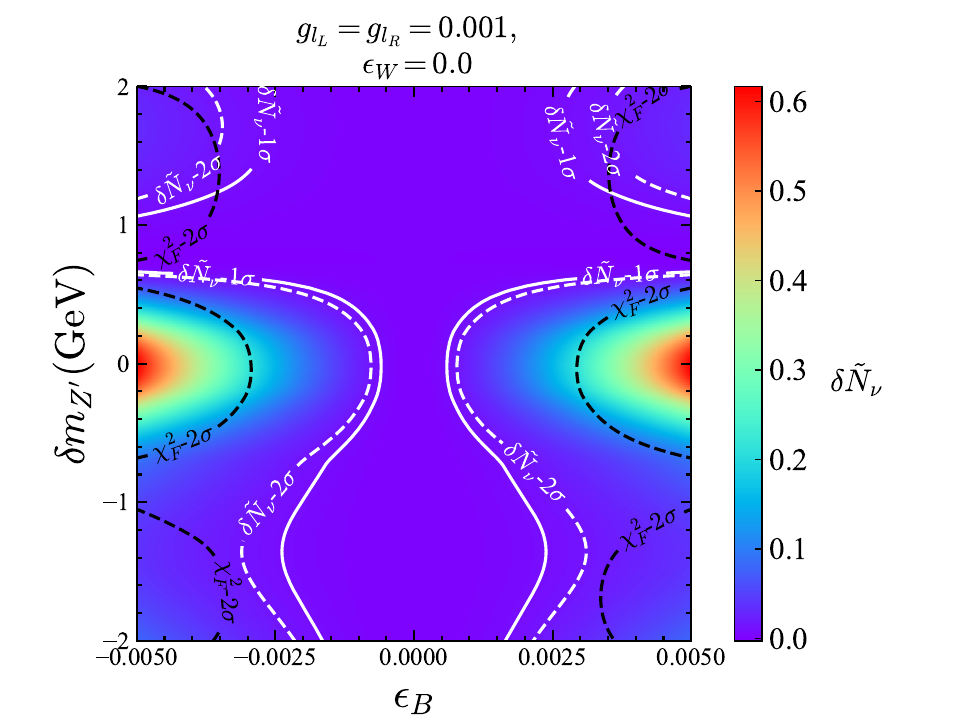}
    \includegraphics[width=0.24\textwidth]{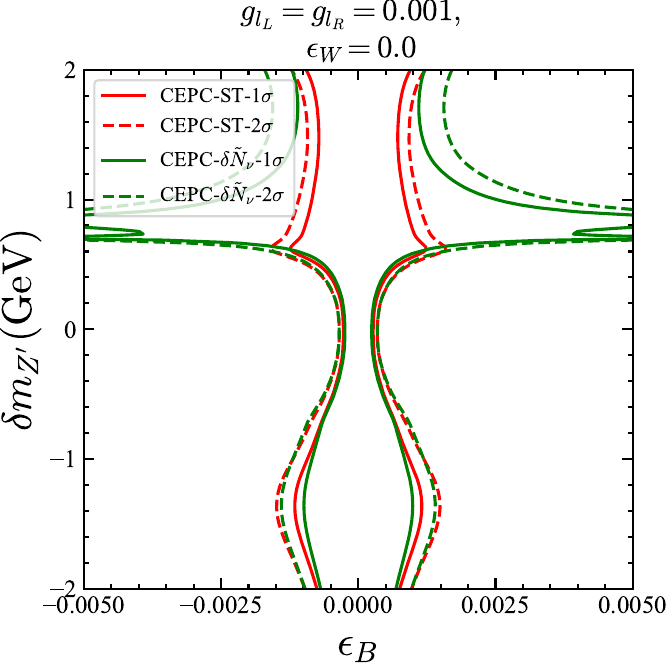}
    \caption{Plots for $g_{l_L} = g_{l_R} = 0.001$ and $\epsilon_W = \delta m^2 = g_{u_L} = g_{u_R} = g_{d_L} = g_{d_R} = g_{\nu_L} = 0$ on the $\epsilon_B$-$\delta m_{Z^{\prime}}$ plain. The other symbols are the same as in Fig.~\ref{dmp-0.1_alpha0.9-delZp}.} \label{LV0.1-delZp}
\end{figure}

\begin{figure}[!t]
    \includegraphics[width=0.32\textwidth]{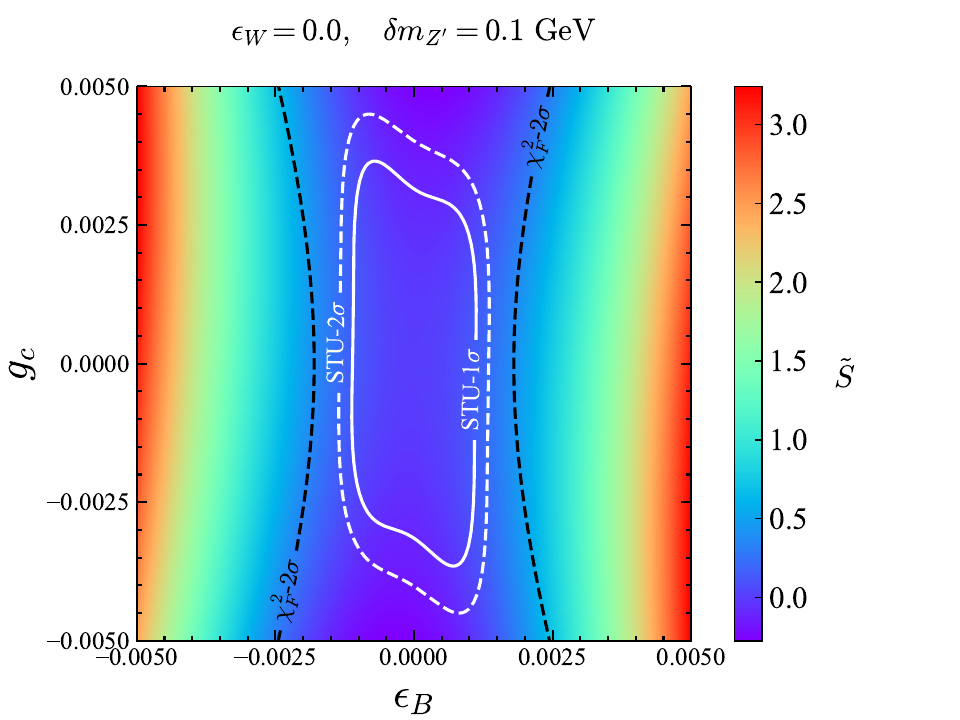}
    \includegraphics[width=0.32\textwidth]{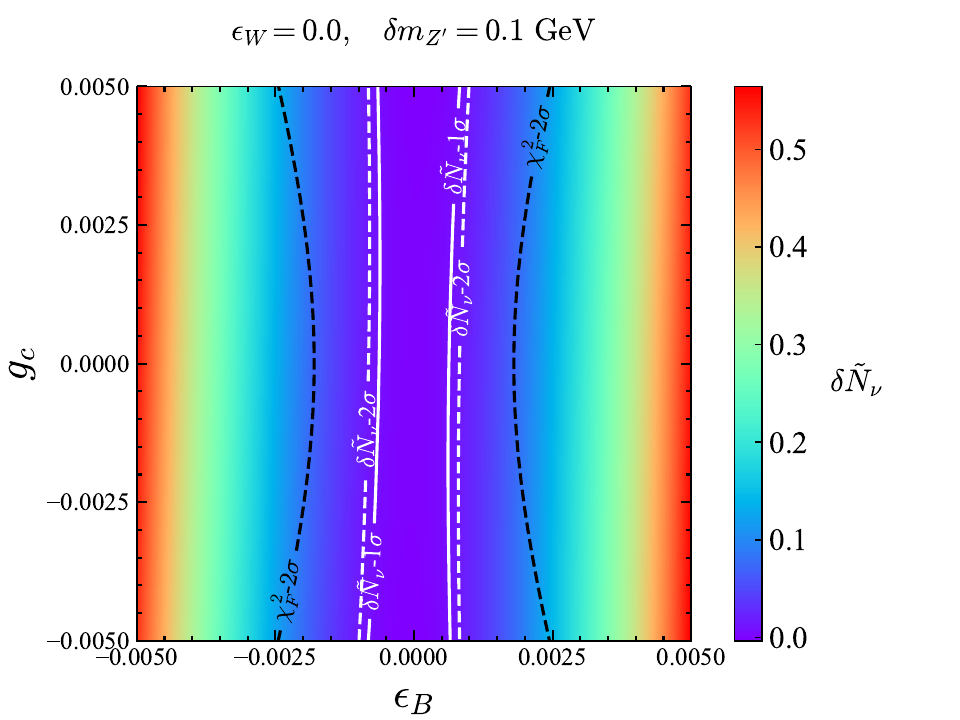}
    \includegraphics[width=0.24\textwidth]{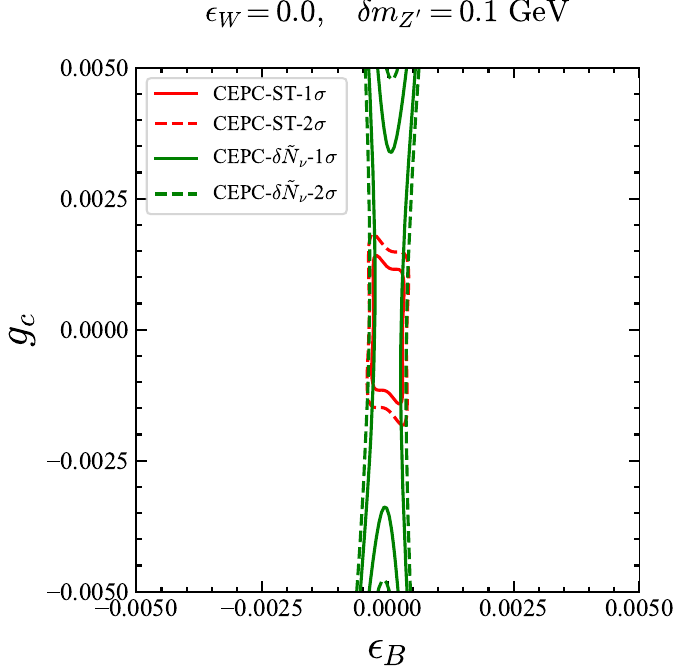}
    \caption{Plots for $\delta m_{Z^{\prime}} = 0.1$~GeV, $g_{l_L} = g_{l_R} = g_c$ and $\epsilon_W = \delta m^2 = g_{u_L} = g_{u_R} = g_{d_L} = g_{d_R} = g_{\nu_L} = 0$ on the $\epsilon_B$-$g_c$ plain. The $\tilde{T}$ and $\tilde{U}$ can be obtained by multiplying $\tilde{S}$ by the coefficients $1.6$ and $-1.2$, respectively. The other symbols are the same as in Fig.~\ref{dmp-0.1_alpha0.9-lamZp}.} \label{LV0.1-cp}
\end{figure}

\begin{figure}[!t]
    \includegraphics[width=0.32\textwidth]{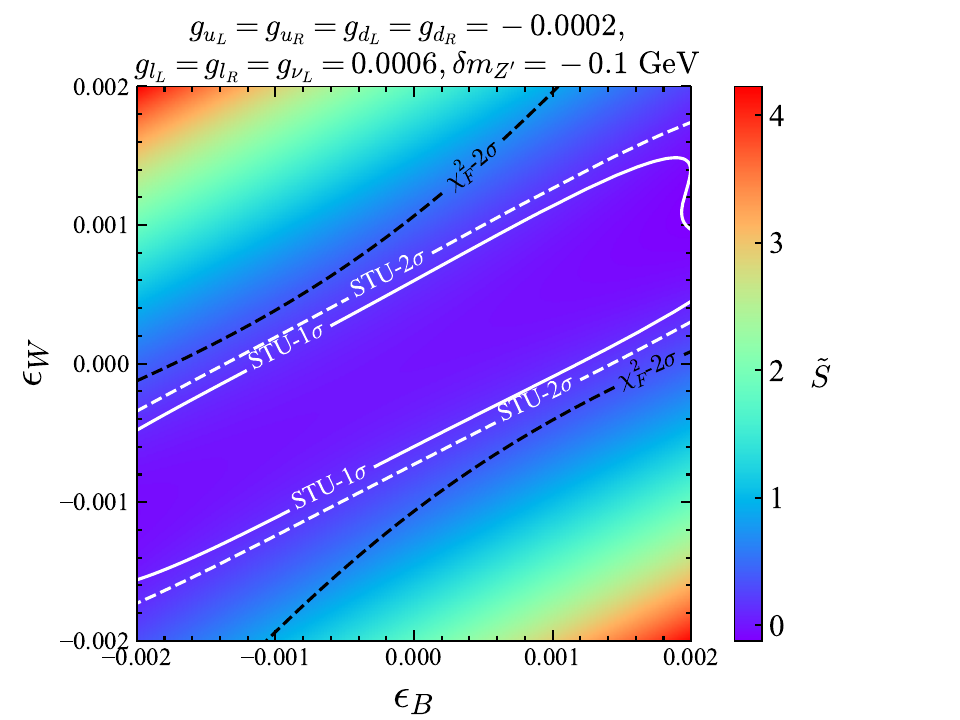}
    \includegraphics[width=0.32\textwidth]{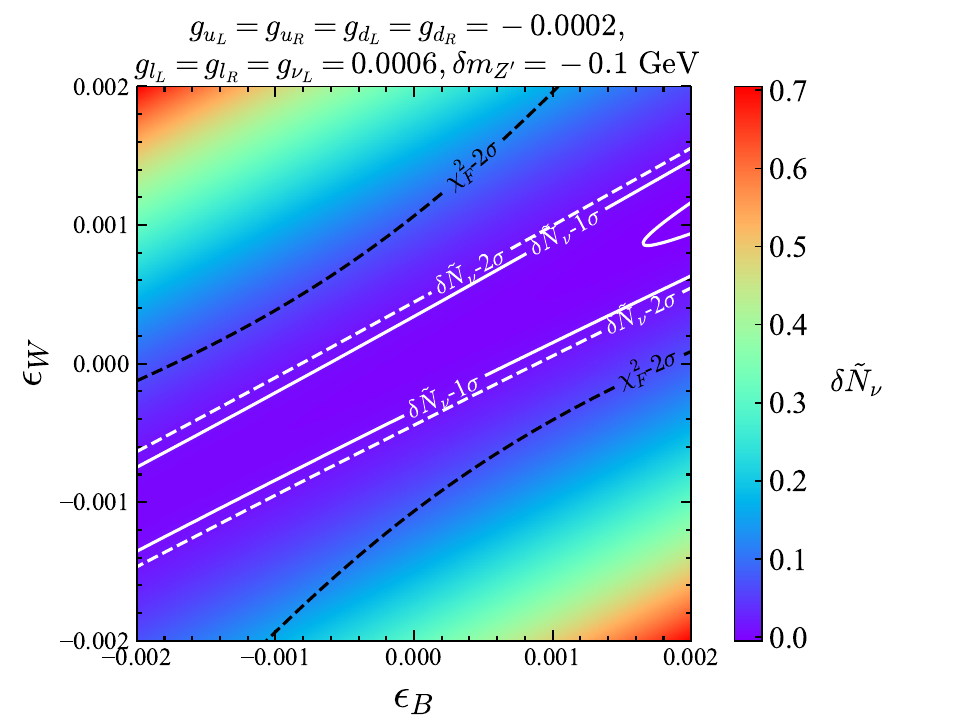}
    \includegraphics[width=0.24\textwidth]{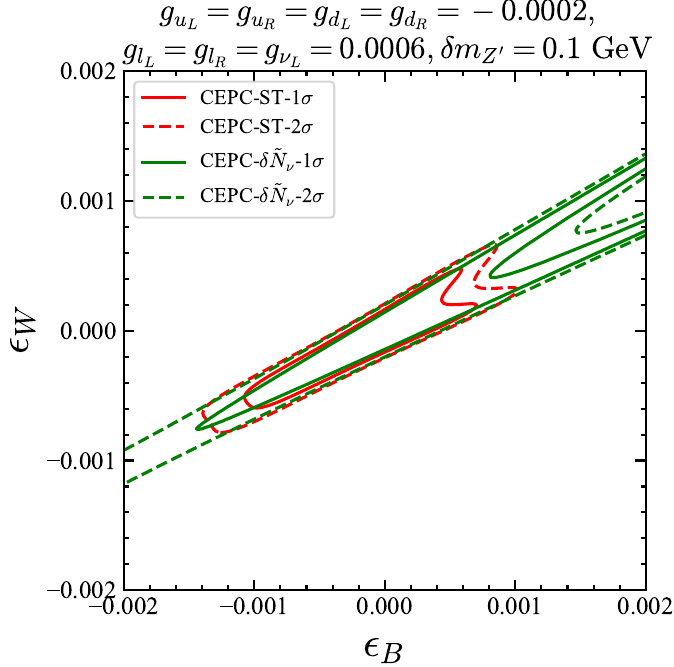}
    \caption{Plots for $\delta m_{Z^{\prime}}=0.1$~GeV, $g_{u_L} = g_{u_R} = g_{d_L} = g_{d_R} = -0.0002$, $g_{l_L} = g_{l_R} = g_{\nu_L} = 0.0006$ and $ \delta m^2 = 0$ on the $\epsilon_B$-$\epsilon_W$ plain. The $\tilde{T}$ and $\tilde{U}$ can be obtained by multiplying $\tilde{S}$ by the coefficients $1.4$ and $-1.1$, respectively. The other symbols are the same as in Fig.~\ref{dmp-0.1_alpha0.9}.} \label{B-L0.1}
\end{figure}

\begin{figure}[!t]
    \includegraphics[width=0.32\textwidth]{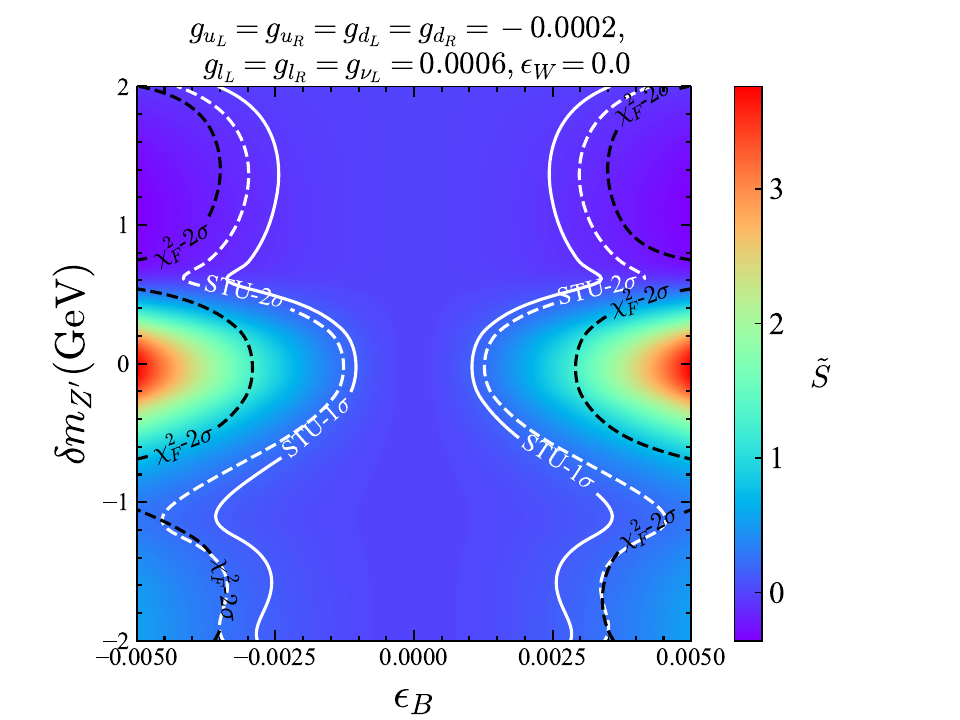}
    \includegraphics[width=0.32\textwidth]{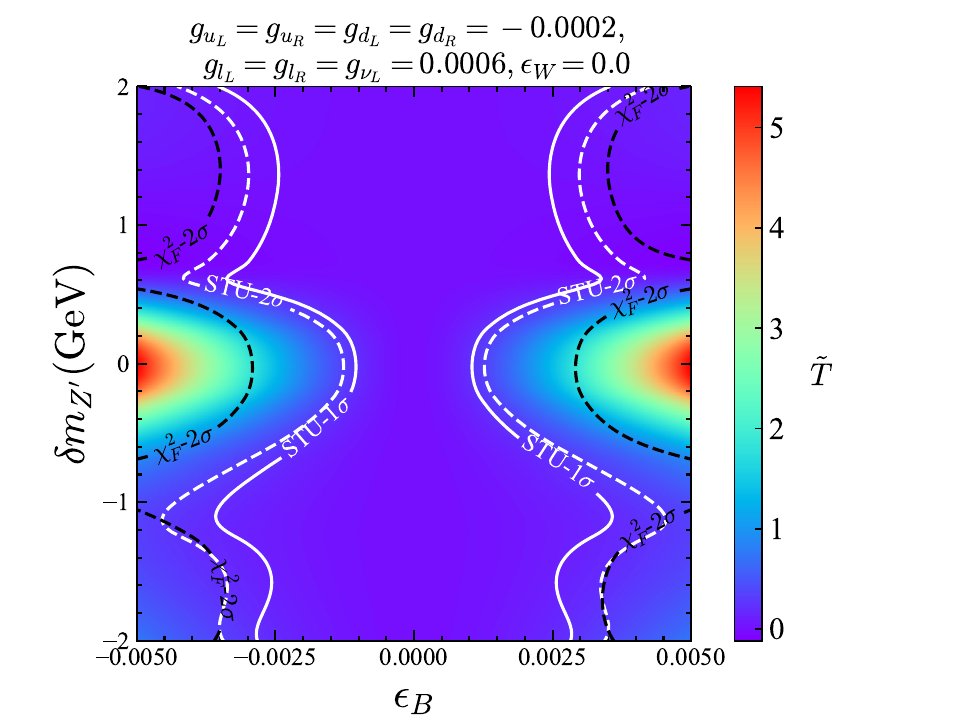}
    \includegraphics[width=0.32\textwidth]{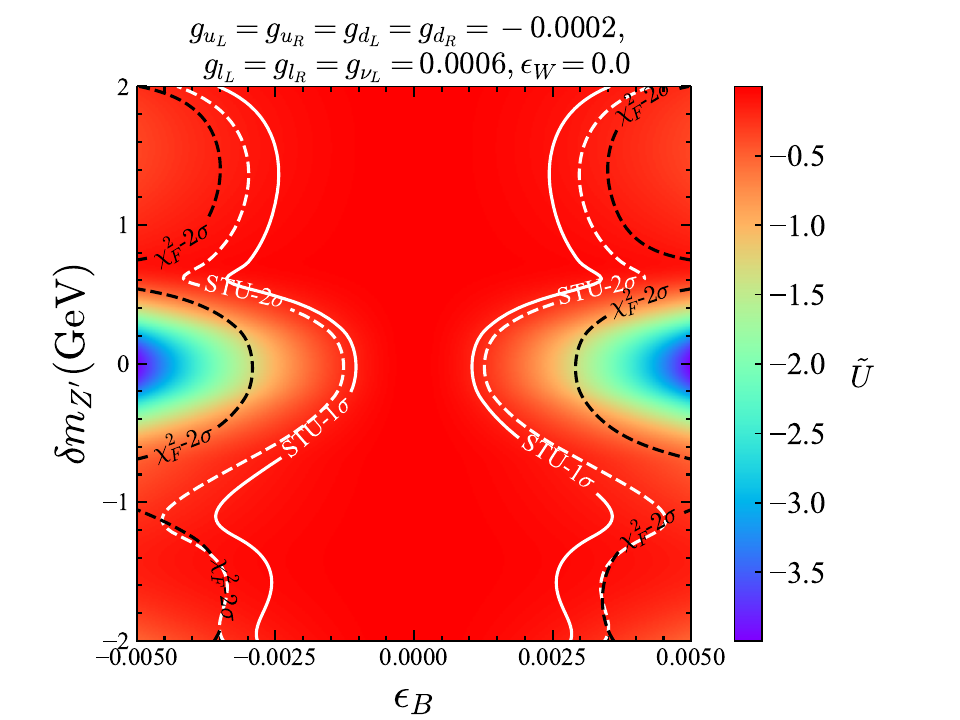}
    \includegraphics[width=0.32\textwidth]{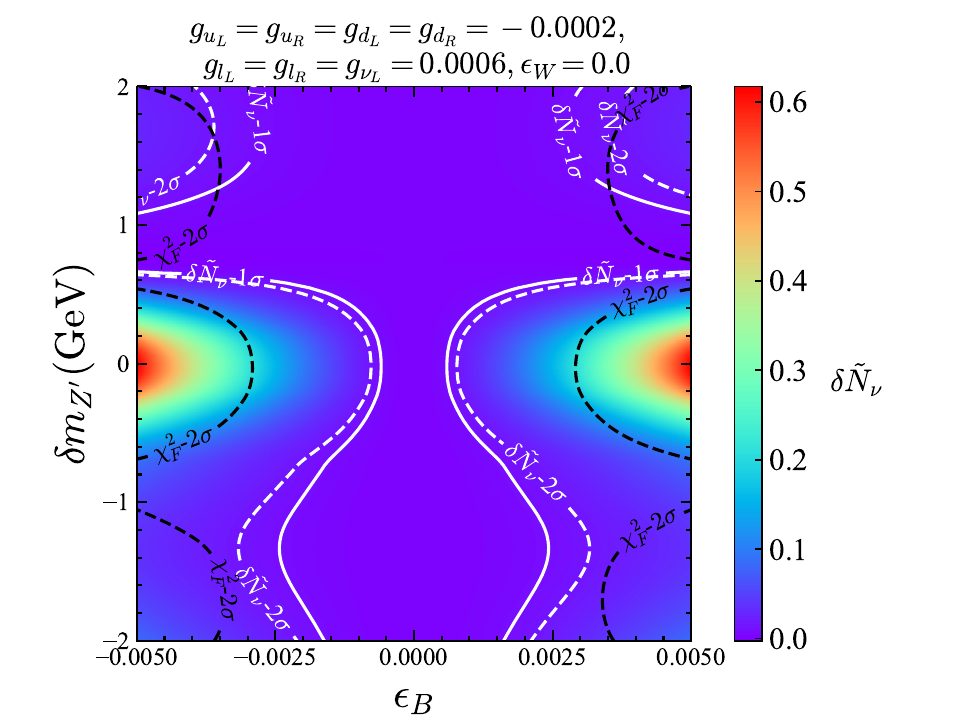}
    \includegraphics[width=0.24\textwidth]{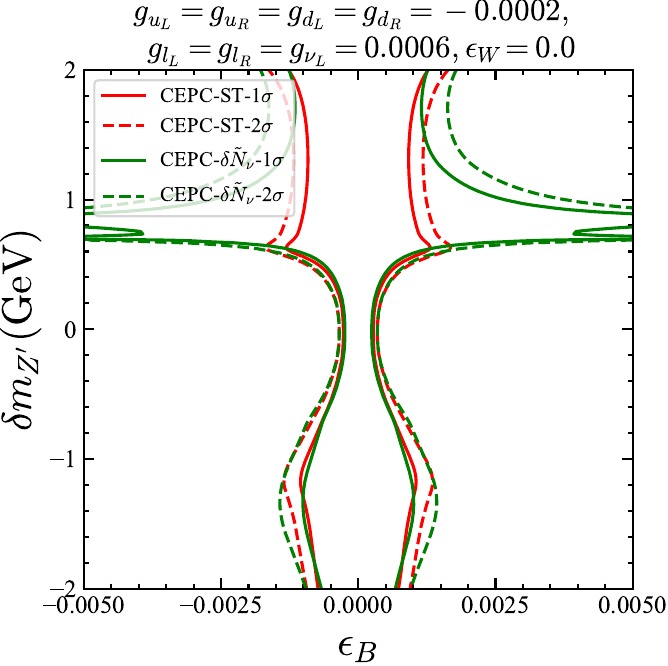}
    \caption{Plots for $g_{u_L} = g_{u_R} = g_{d_L} = g_{d_R} = -0.0002$, $g_{l_L} = g_{l_R} = g_{\nu_L} = 0.0006$ and $\epsilon_W = \delta m^2 = 0$ on the $\epsilon_B$-$\delta m_{Z^{\prime}}$ plain. The other symbols are the same as in Fig.~\ref{dmp-0.1_alpha0.9-delZp}. }\label{B-L0.1-delZp}
\end{figure}

\begin{figure}[!t]
    \includegraphics[width=0.32\textwidth]{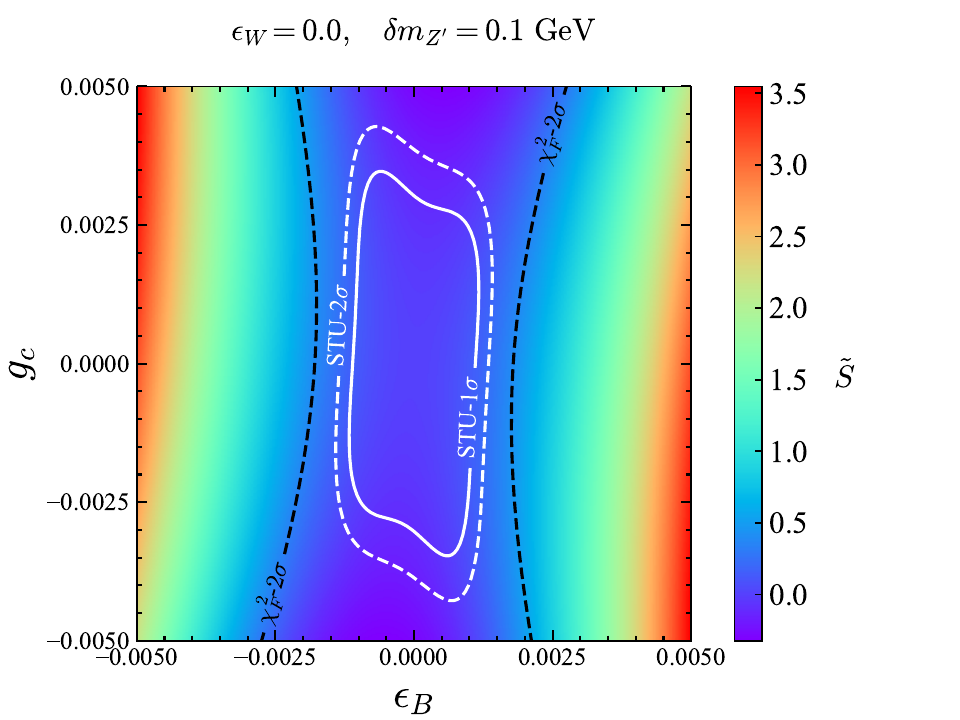}
    \includegraphics[width=0.32\textwidth]{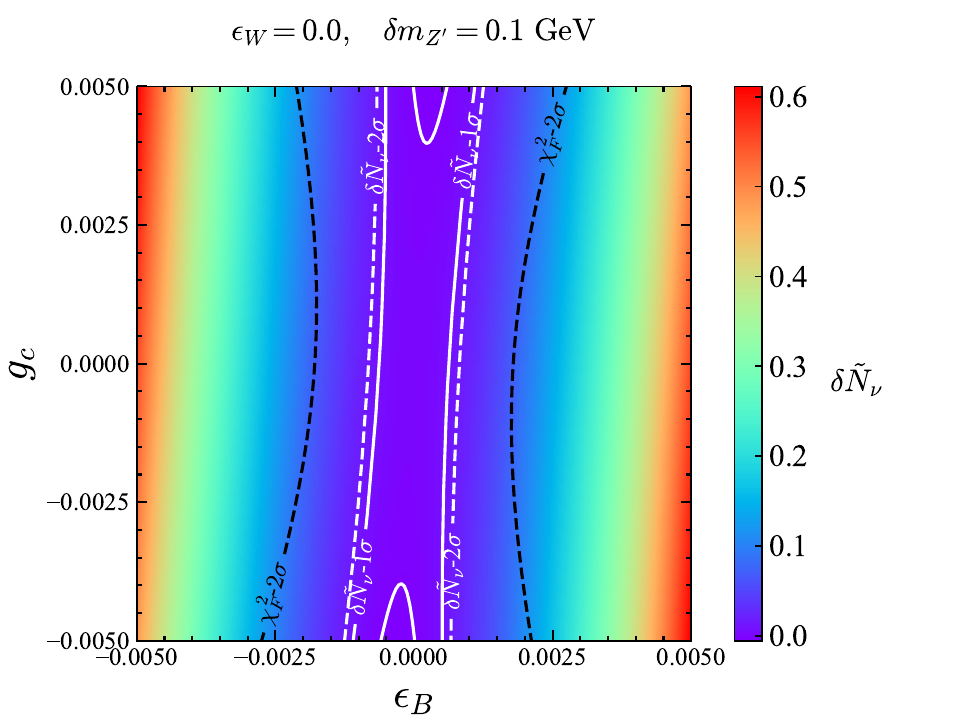}
    \includegraphics[width=0.24\textwidth]{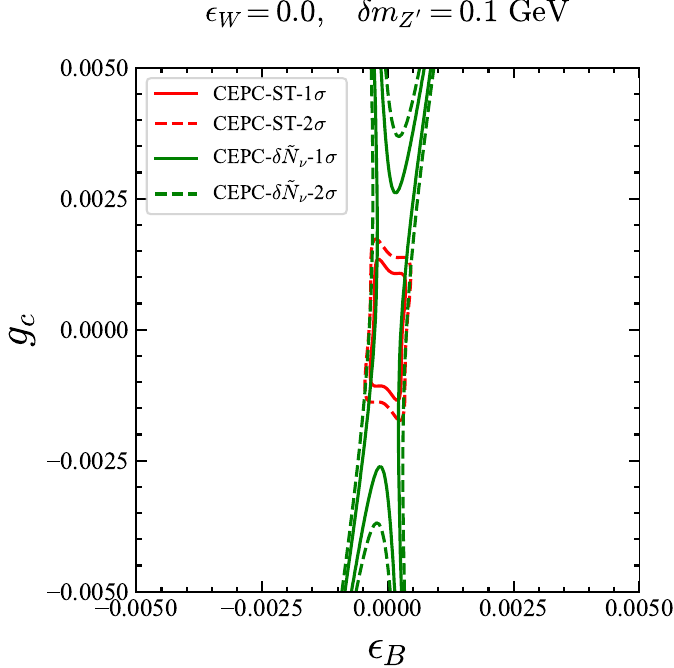}
    \caption{Plots for $\delta m_{Z^{\prime}} = 0.1$~GeV, $-3 g_{u_L} = -3 g_{u_R} = -3 g_{d_L} = -3 g_{d_R} = g_{l_L} = g_{l_R} = g_{\nu_L} = g_c$ and $\delta_W = \delta m^2 = 0$ on the $\epsilon_B$-$g_c$ plain. The $\tilde{T}$ and $\tilde{U}$ can be obtained by multiplying $\tilde{S}$ by the coefficients $1.5$ and $-1.2$, respectively. The other symbols are the same as in Fig.~\ref{dmp-0.1_alpha0.9-lamZp}.} \label{B-L0.1-gc}
\end{figure}

In this paper, we only consider two cases, one is that the $\hat{Z}^{\prime}$ only couples universally with all the leptons, and the other is that the $\hat{Z}^{\prime}$ couple with leptons and quarks with the ratio of the coupling constants to be $3:(-1)$, inspired by the $U(1)_{B-L}$ model. For the first case, since we find out that the chirality of the interaction terms defined in \eqref{coupling_Zp} does not affect the results to a significant extent, we only adopt the vector-type interaction pattern $g_{l_L}=g_{l_R}$, and show plain examples in Fig.~\ref{LV0.1}, \ref{LV0.1-delZp} and \ref{LV0.1-cp}. 
Inspired by the $U(1)_{B-L}$ models, we plot the Fig.~\ref{B-L0.1}, \ref{B-L0.1-delZp} and \ref{B-L0.1-gc} as the example plains when $g_{l_L}=g_{l_R}=-3g_{u_L} = -3g_{u_R} = -3g_{d_L} = -3g_{d_R}$. One can also learn about the LEP sensitivity as well as the estimated CEPC sensitivity in these figures. 



\subsection{Scenario III: $\hat{Z}^{\prime}$ couples with the SM fermions depending on the generations}

 The $Z^{\prime}$ couplings to fermions might be generation-dependent, in which case coefficients in \eqref{coupling_Zp} need to be changed into the generation-dependent version. 
 Constrained by our current calculation ability and inspired by some eminent models, we only consider the case that $\hat{Z}^{\prime}$ only couples with $\mu$ and/or $\tau$. The interaction terms are parametrized by
\begin{equation}
   \mathcal{L}_{\mathrm{eff}} \supset \hat{Z}^{\prime}_{\mu}\bar{\mu}\gamma^{\mu}(\frac{g_{\mu_L}+g_{\mu_R}}{2}-\frac{g_{\mu_L}-g_{\mu_R}}{2}\gamma^5)\mu + \hat{Z}^{\prime}_{\mu}\bar{\tau}\gamma^{\mu}(\frac{g_{\tau_L}+g_{\tau_R}}{2}-\frac{g_{\tau_L}-g_{\tau_R}}{2}\gamma^5)\tau. \label{Zp_mu_tau}
\end{equation}
Although the $e^{+}e^{-}\rightarrow\mu^{+}\mu^{-}$ might be affected by (\ref{Zp_mu_tau}), breaking the universality among the $e^{\pm}$ and $\mu^{\pm}$ channels, however in the literature, all the $S$, $T$, $U$ parameters are extracted with the assumption of the leptonic universality. To keep in pace with these fitting processes as possible as we can, we still follow our previous steps to fit the $\tilde{S}$, $\tilde{T}$, $\tilde{U}$, $\delta \tilde{N}_{\nu}$, pretending to be blind to the non-universality. 
On the contrary, in order to compute the (\ref{R_Ratio}), we should follow another route by deleting all the $\mu^+ \mu^-$ terms in (\ref{chi-square}) to find the best-fitted point in our pseudo-SM template again.  As we have mentioned before, in this case the $N_{\rm df}$ becomes 21.
With the newly independent fitted results, (\ref{R_Ratio}) can be expressed by
\begin{equation}
   \frac{R_e}{R_{\mu}} = \frac{\sigma^{\mathrm{NP}}_{\mu^{+}\mu^{-}}(\sqrt{s}=m_Z)}{\sigma^{\mathrm{NP}}_{se^{+}e^{-}}(\sqrt{s}=m_Z)},
\end{equation} 
where subscript ``$s$'' in $\sigma^{\mathrm{NP}}_{se^{+}e^{-}}(m_Z)$ indicates the $s$-channel cross section the $e^+ e^- \rightarrow e^+ e^-$ channel. Certainly, $\sigma^{\mathrm{NP}}_{se^{+}e^{-}}(m_Z)$ is impossible to be acquired straightforwardly due to the contamination from the $t$-, $u$-channels. 
Fortunately, in the pseudo-SM template, the universality is preserved, and the $\sigma^{\mathrm{PSM}}_{\mu^{+}\mu^{-}}$ does not include the $t$-, $u$-channel contributions. Therefore,  the $\sigma^{\mathrm{PSM}}_{\mu^{+}\mu^{-}}(\sqrt{s}=91.2~\mathrm{GeV})$ adopted from the pseudo-SM template can be utilized in place of $\sigma^{\mathrm{NP}}_{se^{+}e^{-}}(m_Z)$, and we also take $\sqrt{s}=91.2 \rm{GeV}$ to replace the $m_Z$, so
\begin{equation}
    \frac{R_e}{R_{\mu}} = \frac{\sigma^{\mathrm{NP}}_{\mu^{+}\mu^{-}}(\sqrt{s}=91.2~\mathrm{GeV})}{\sigma^{\mathrm{PSM}}_{\mu^{+}\mu^{-}}(\sqrt{s}=91.2~\mathrm{GeV})}.
\end{equation}

We will extract the current $\frac{R_e}{R_{\mu}}$ data from Ref.~\cite{Workman:2022ynf}, and plot the $1$-$\sigma$ and $2$-$\sigma$ contour in the $\epsilon_{B}$-$\epsilon_{W}$ and $\epsilon_{B}$-$g_c$ plain. As we have mentioned, the CEPC sensitivity is extracted from Ref.~\cite{CEPCPhysicsStudyGroup:2022uwl, Belloni:2022due}.

\begin{figure}[!t]
\begin{tabular}{ccc}
    \includegraphics[width=0.32\textwidth]{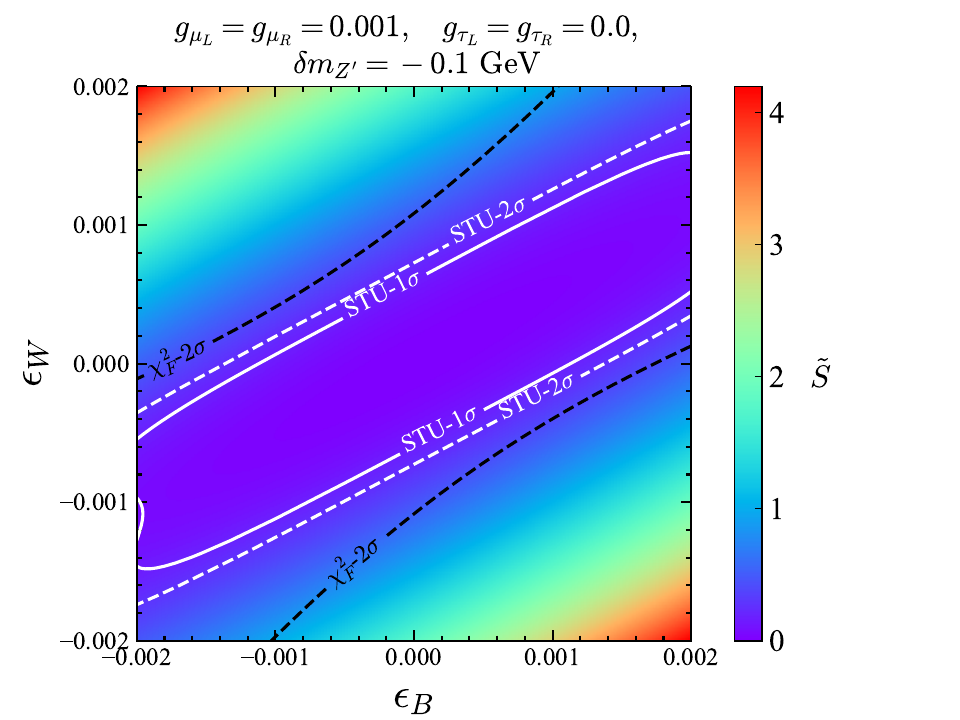} 
    \includegraphics[width=0.32\textwidth]{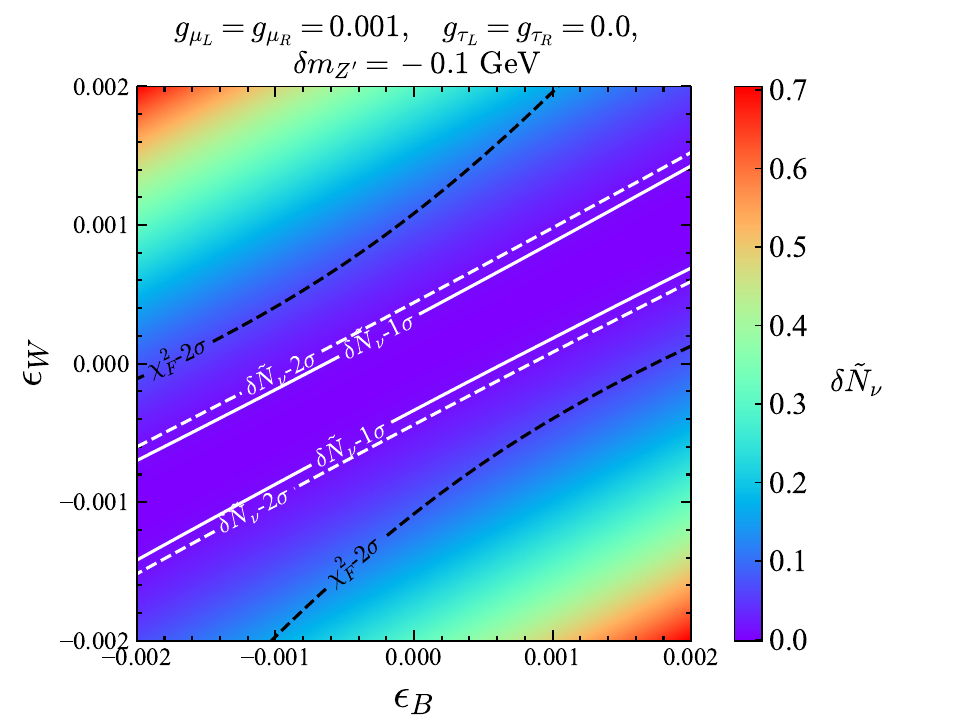} 
    \includegraphics[width=0.32\textwidth]{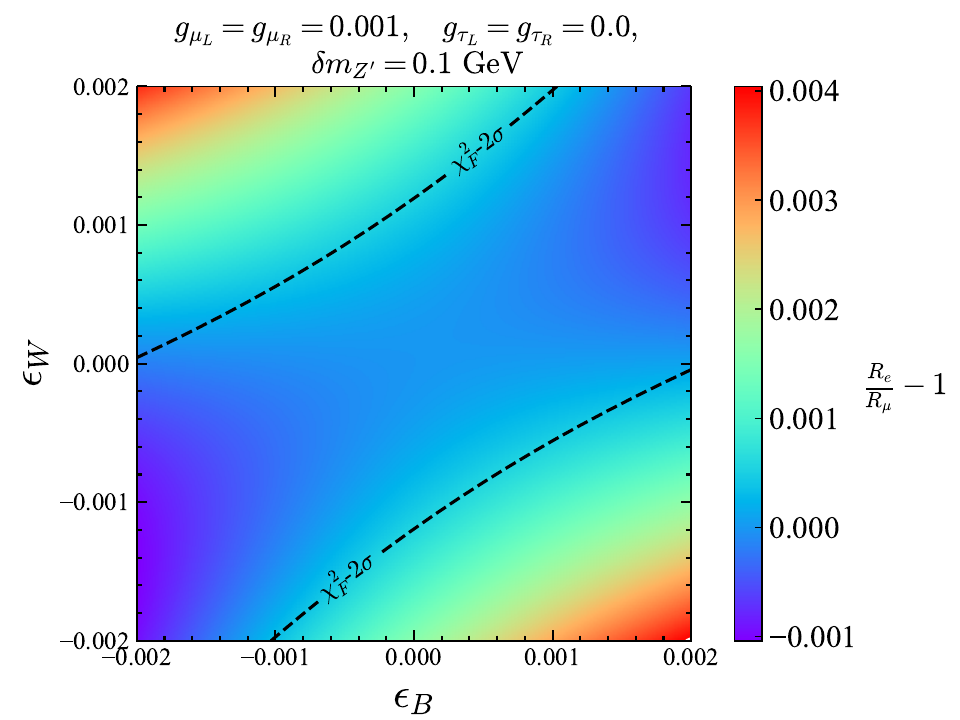} \\
    \includegraphics[width=0.24\textwidth]{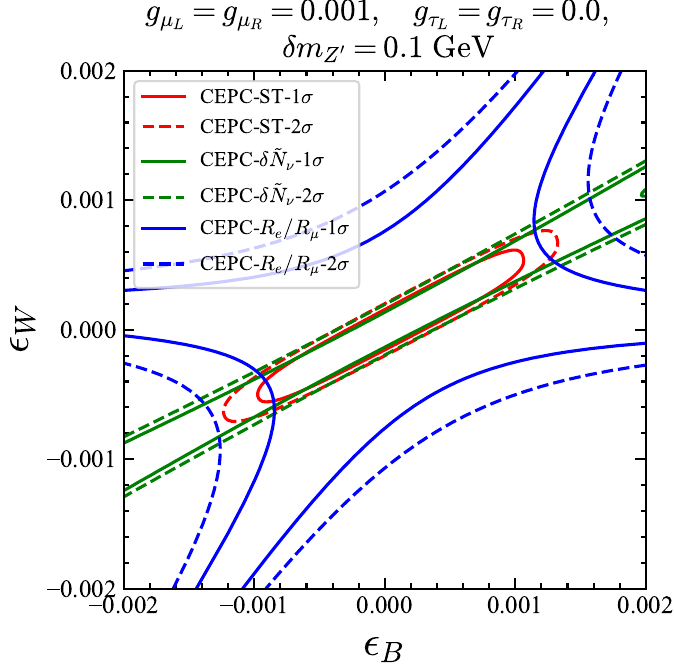} 
\end{tabular}
    \caption{Plots for $\delta m_{Z^{\prime}} = 0.1$~GeV, $g_{\mu_L} = g_{\mu_R} = 0.001$ and $g_{\tau_L} = g_{\tau_R} = \delta m^2 = 0$ on the $\epsilon_B$-$\epsilon_W$ plain. The $\tilde{T}$ and $\tilde{U}$ can be obtained by multiplying $\tilde{S}$ by the coefficients $1.4$ and $-1.1$, respectively. Compared with the Fig.~\ref{dmp-0.1_alpha0.9}, $\frac{R_{e}}{R_{\mu}}-1$ is additionally plotted.} \label{decaymu0.1}
\end{figure}

\begin{figure}[!t]
\begin{tabular}{lll}
    \includegraphics[width=0.32\textwidth]{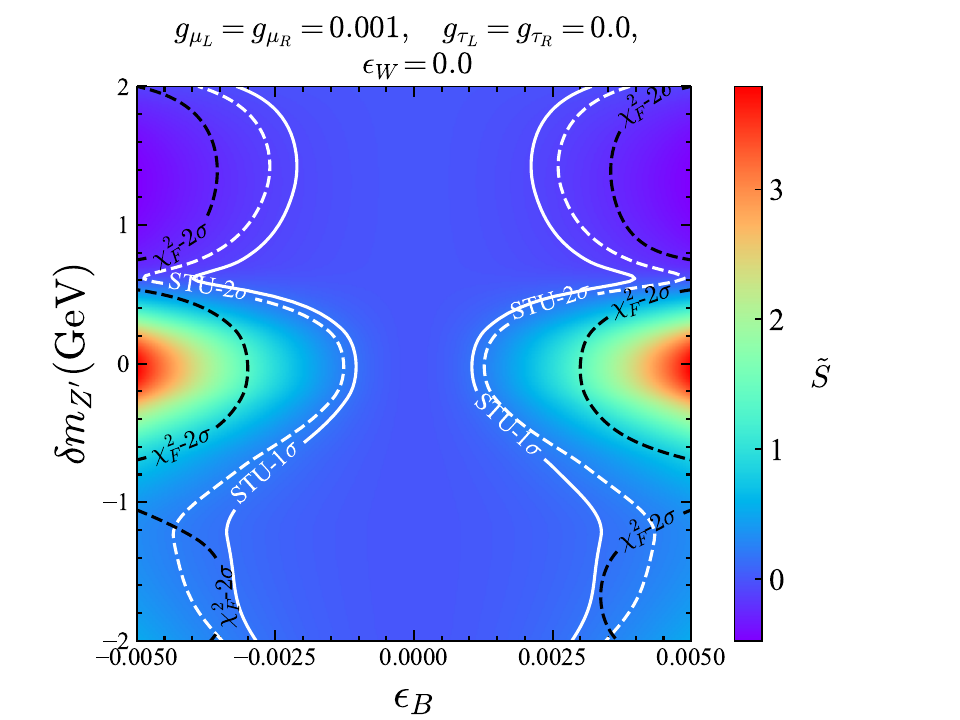} &
    \includegraphics[width=0.32\textwidth]{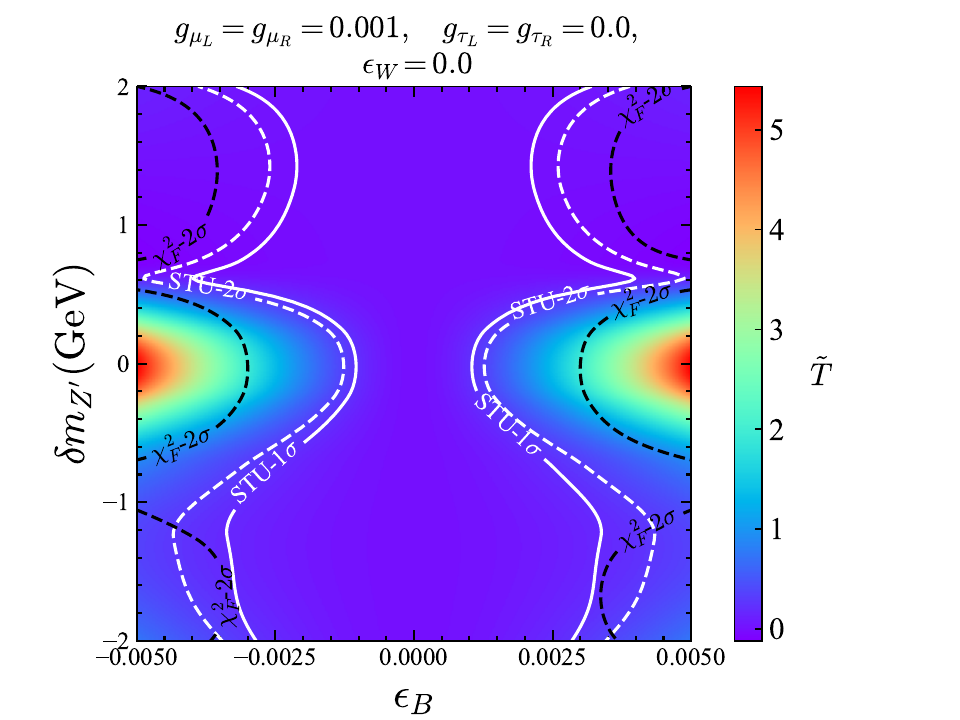} &
    \includegraphics[width=0.32\textwidth]{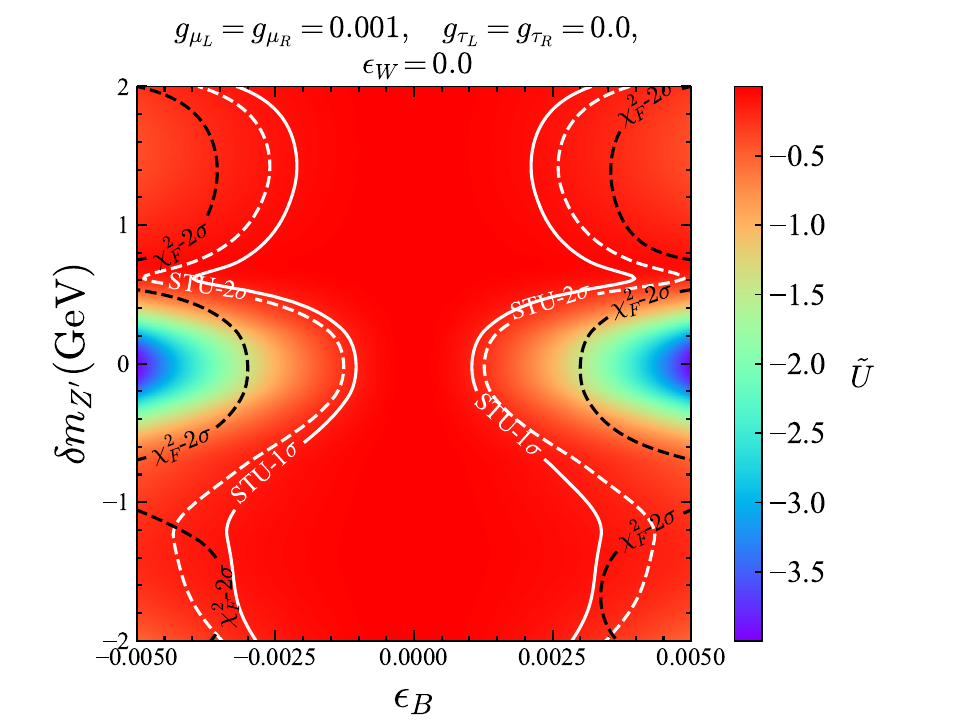} \\
    \includegraphics[width=0.32\textwidth]{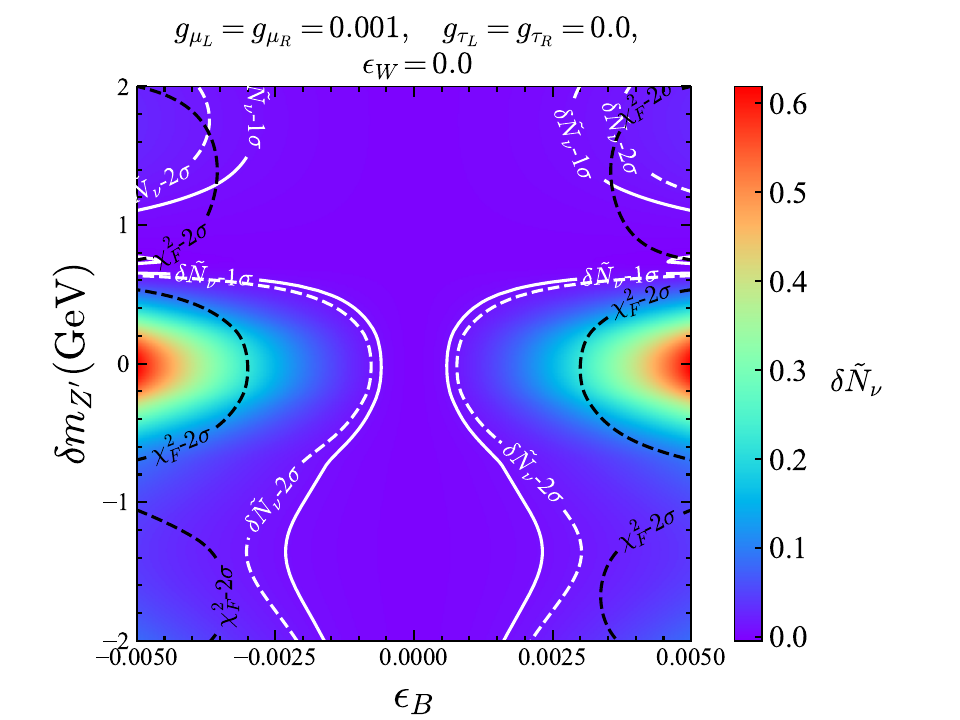} &
    \includegraphics[width=0.32\textwidth]{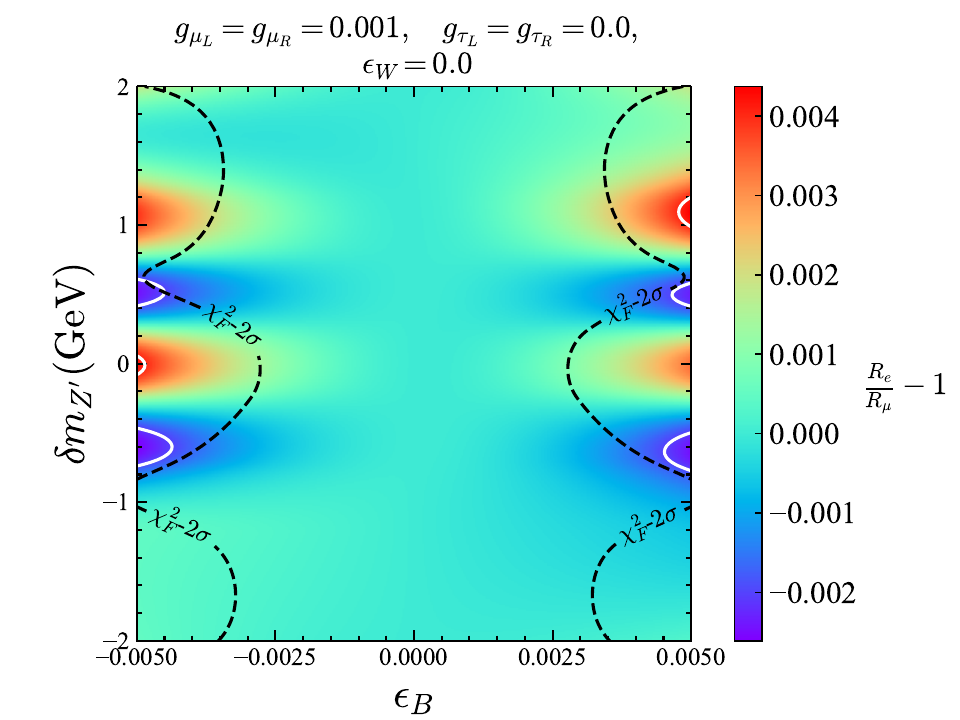} &
   \ \  \includegraphics[width=0.24\textwidth]{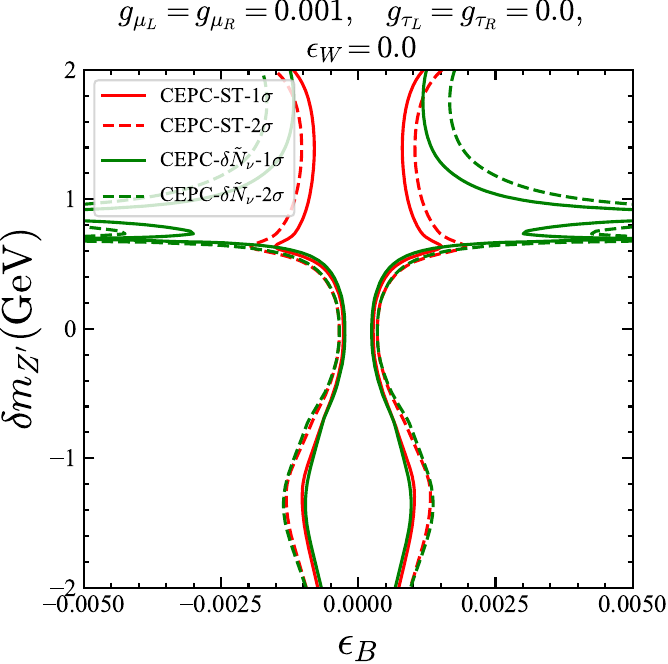}
\end{tabular}
    \caption{Plots for $g_{\mu_L} = g_{\mu_R} = 0.001$ and $g_{\tau_L} = g_{\tau_R} = \epsilon_W = \delta m^2 = 0$ on the $\epsilon_B$-$\delta m_{Z^{\prime}}$ plain. Compared with the Fig.~\ref{dmp-0.1_alpha0.9-delZp}, $\frac{R_{e}}{R_{\mu}}-1$ is additionally plotted.} \label{decaymu0.1-delZp}
\end{figure}

\begin{figure}[!t]
\begin{tabular}{ccc}
    \includegraphics[width=0.32\textwidth]{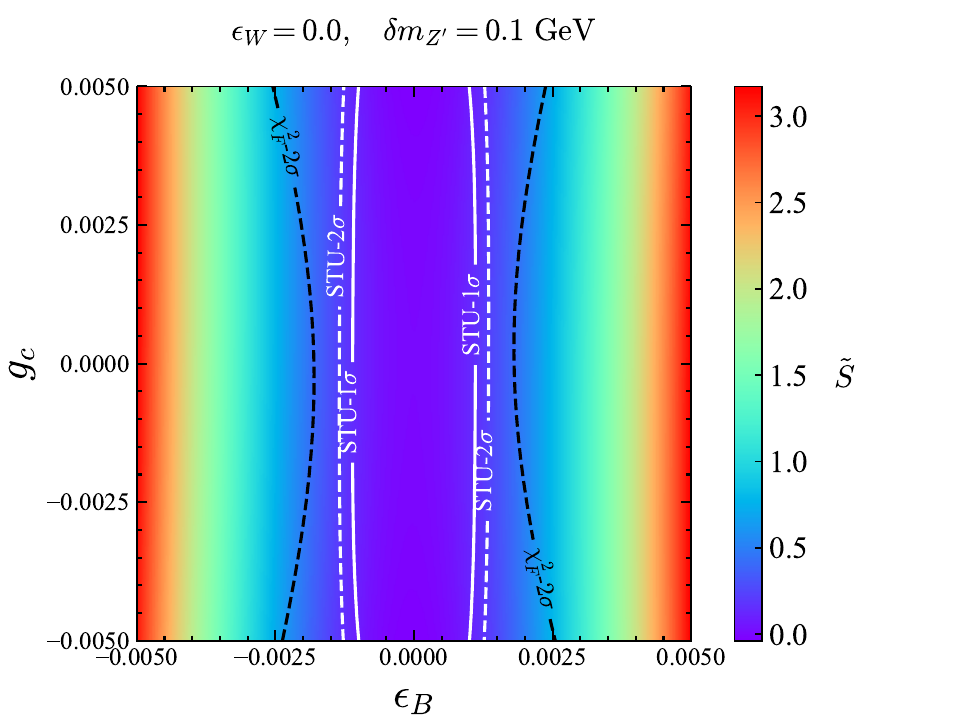} 
    \includegraphics[width=0.32\textwidth]{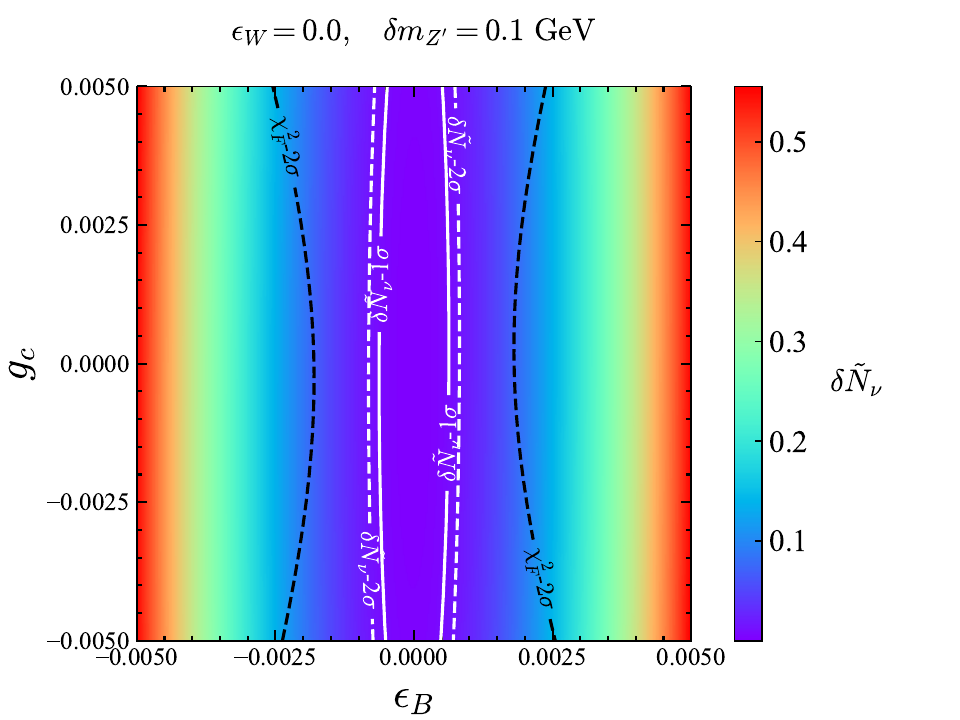} 
    \includegraphics[width=0.32\textwidth]{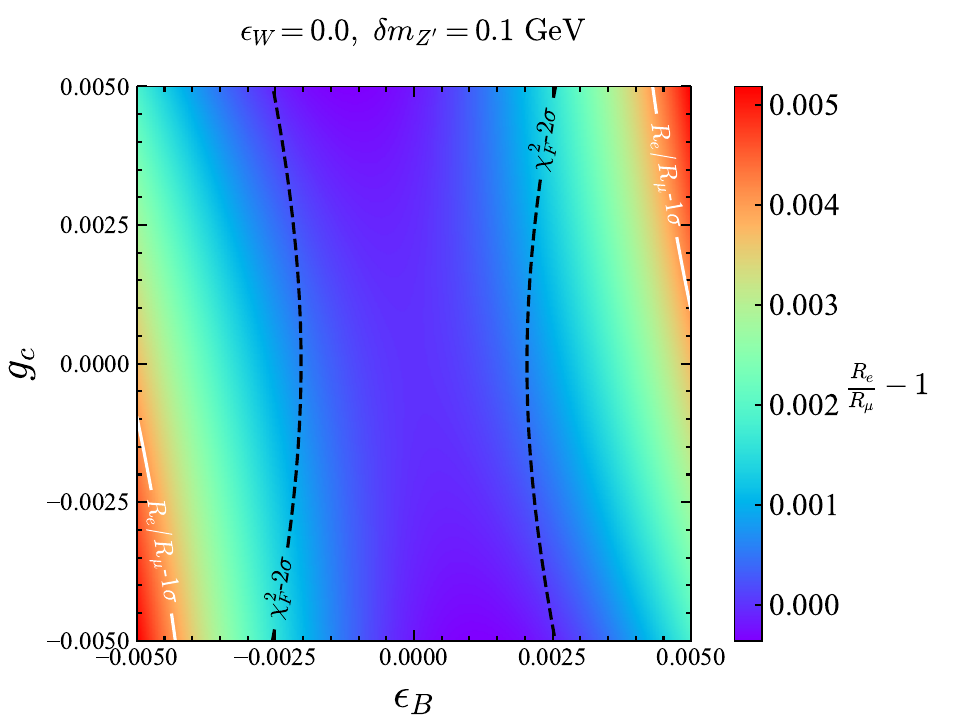} \\
    \includegraphics[width=0.24\textwidth]{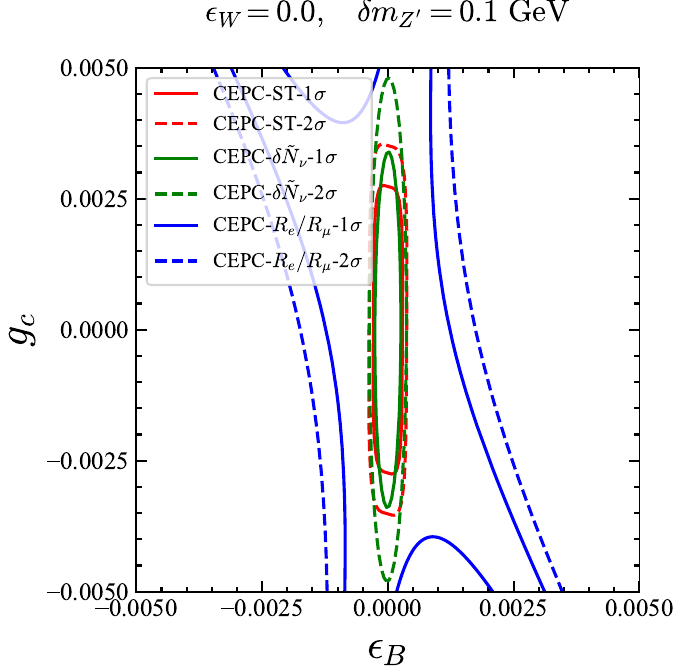}
\end{tabular}
    \caption{Plots for $g_{\mu_L} = g_{\mu_R} = g_c$ and $g_{\tau_L} = g_{\tau_R} = \epsilon_W = \delta m^2 = 0$ on the $\epsilon_B$-$g_c$ plain. The $\tilde{T}$ and $\tilde{U}$ can be obtained by multiplying $\tilde{S}$ by the coefficients $1.5$ and $-1.2$, respectively.  Compared with the Fig.~\ref{dmp-0.1_alpha0.9-lamZp}, $\frac{R_{e}}{R_{\mu}}-1$ is additionally plotted.} \label{decaymu0.1-gc}
\end{figure}

\begin{figure}[!t]
    \begin{tabular}{ccc}
    \includegraphics[width=0.32\textwidth]{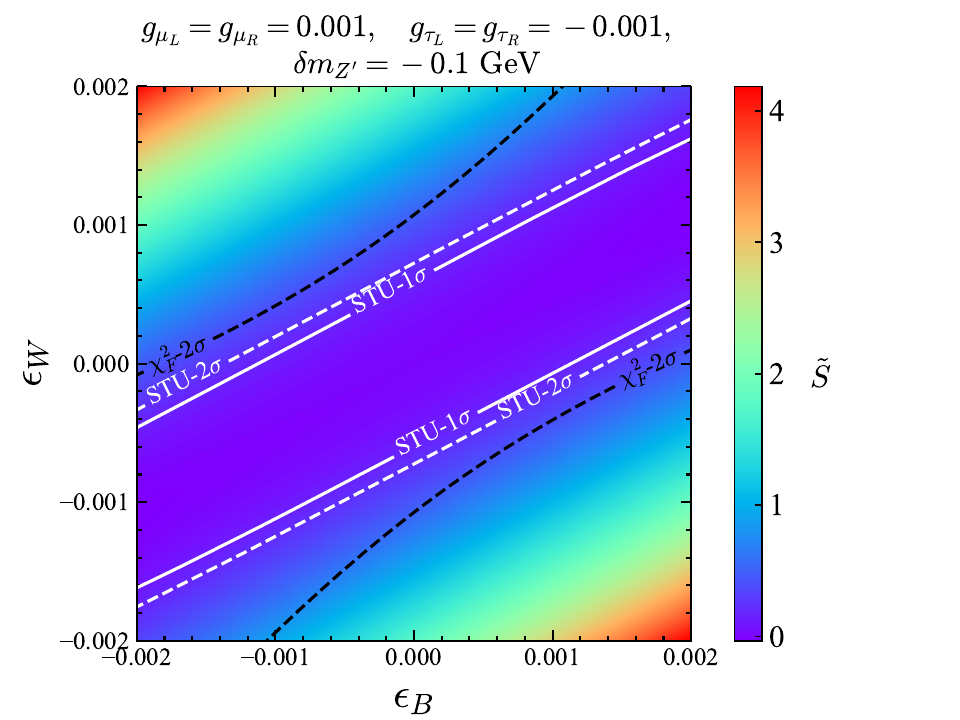} 
    \includegraphics[width=0.32\textwidth]{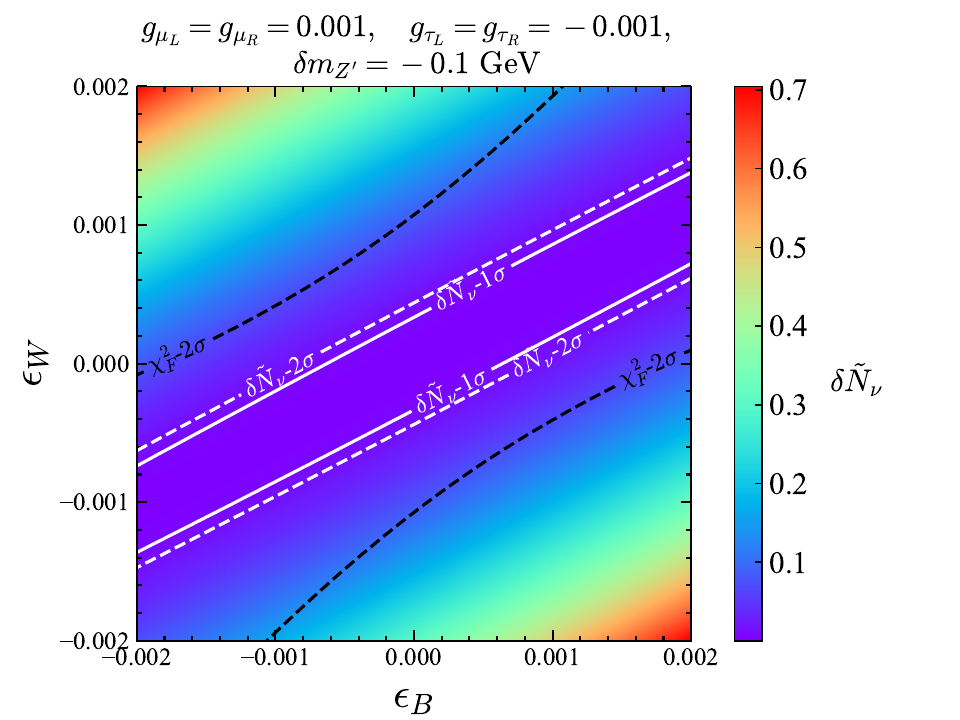} 
    \includegraphics[width=0.32\textwidth]{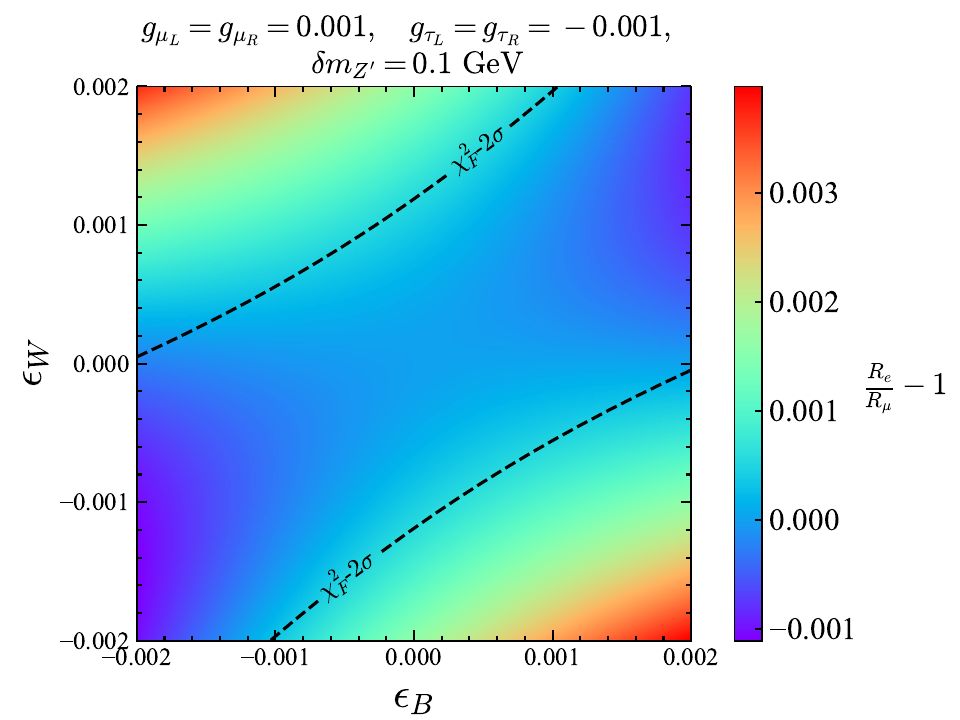} \\
    \includegraphics[width=0.24\textwidth]{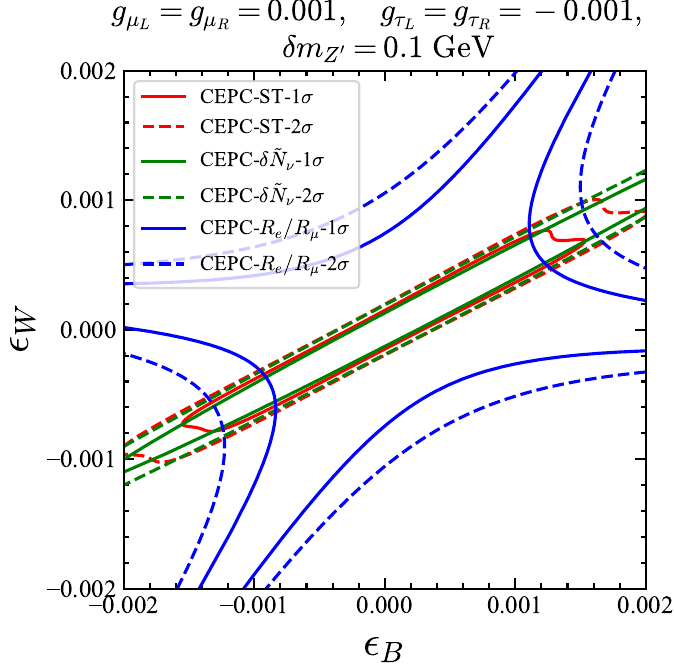} 
    \end{tabular}
    \caption{Plots for $\delta m_{Z^{\prime}} = 0.1$~GeV, $g_{\mu_L} = g_{\mu_R} = 0.001$, $g_{\tau_L} = g_{\tau_R} = -0.001$ and $\delta m^2  = 0$ on the $\epsilon_B$-$\epsilon_W$ plain. The $\tilde{T}$ and $\tilde{U}$ can be obtained by multiplying $\tilde{S}$ by the coefficients $1.4$ and $-1.1$, respectively. Compared with the Fig.~\ref{dmp-0.1_alpha0.9}, $\frac{R_{e}}{R_{\mu}}-1$ is additionally plotted.} \label{decaymuta0.1}
\end{figure}

\begin{figure}[!t]
    \begin{tabular}{lll}
    \includegraphics[width=0.32\textwidth]{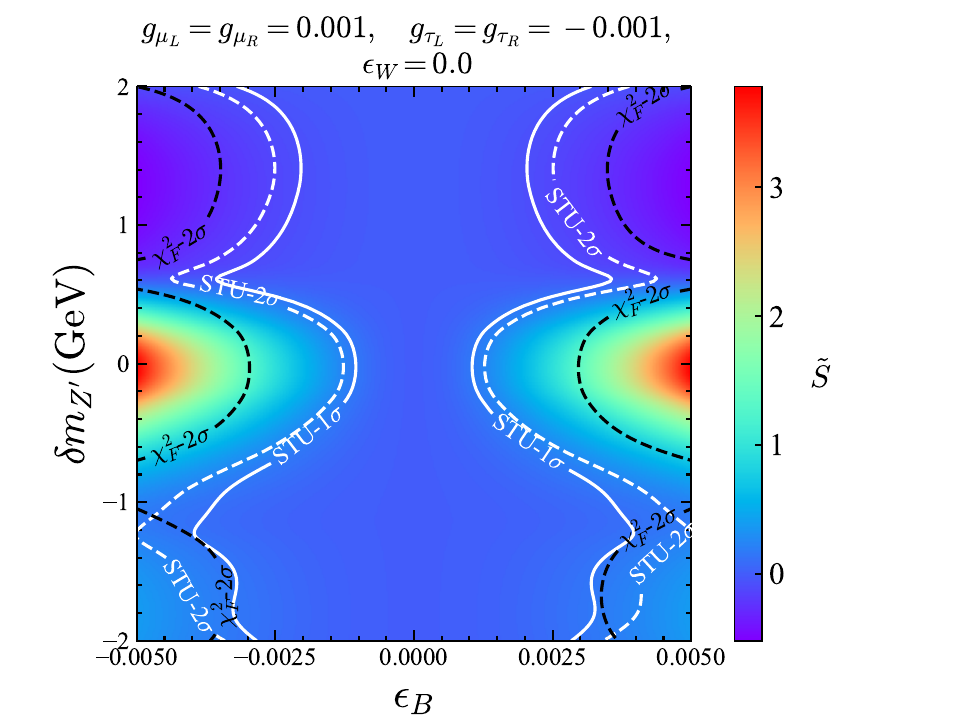}&
    \includegraphics[width=0.32\textwidth]{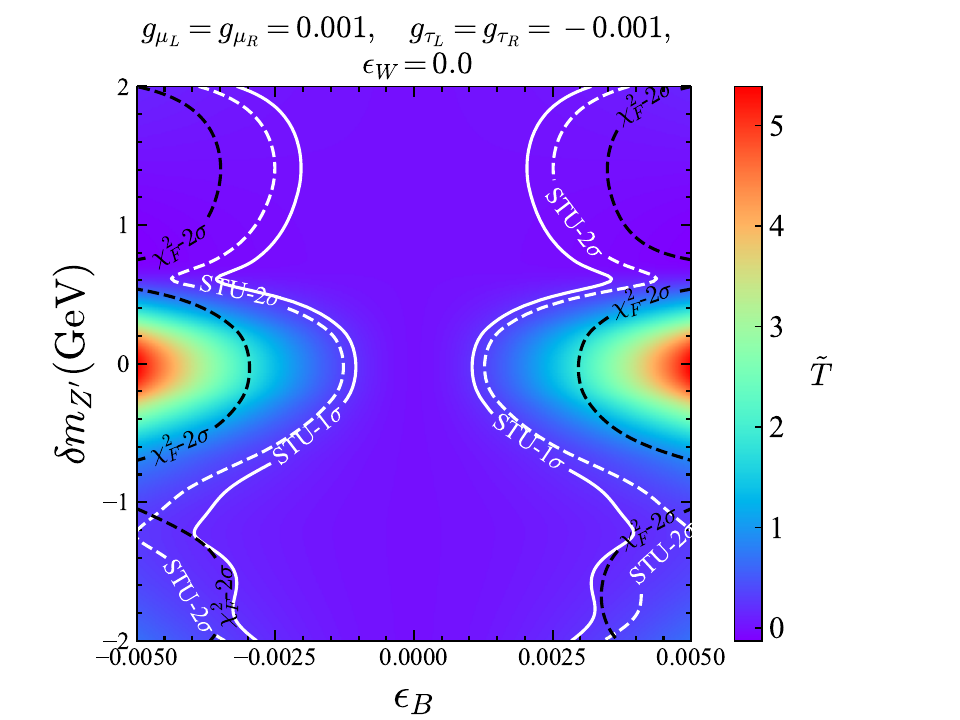}&
    \includegraphics[width=0.32\textwidth]{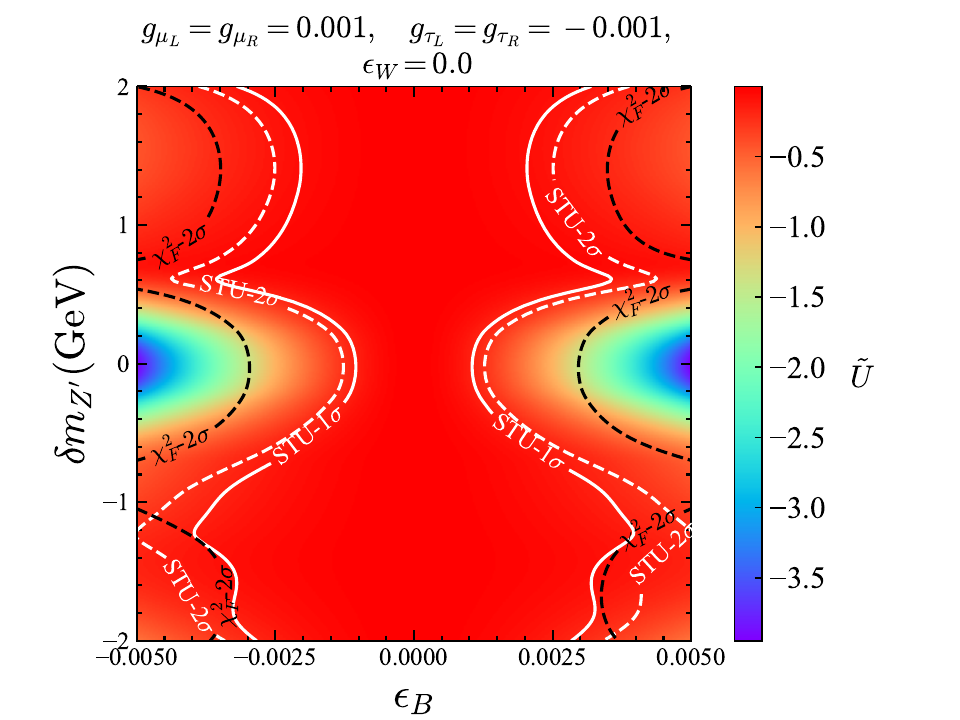}\\
    \includegraphics[width=0.32\textwidth]{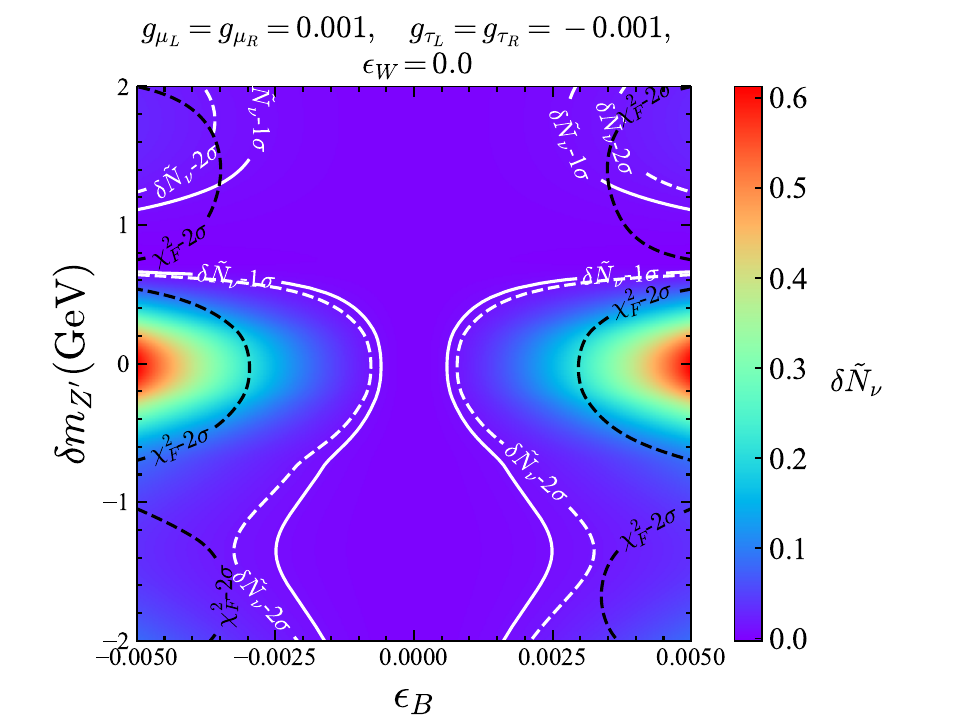}&
    \includegraphics[width=0.32\textwidth]{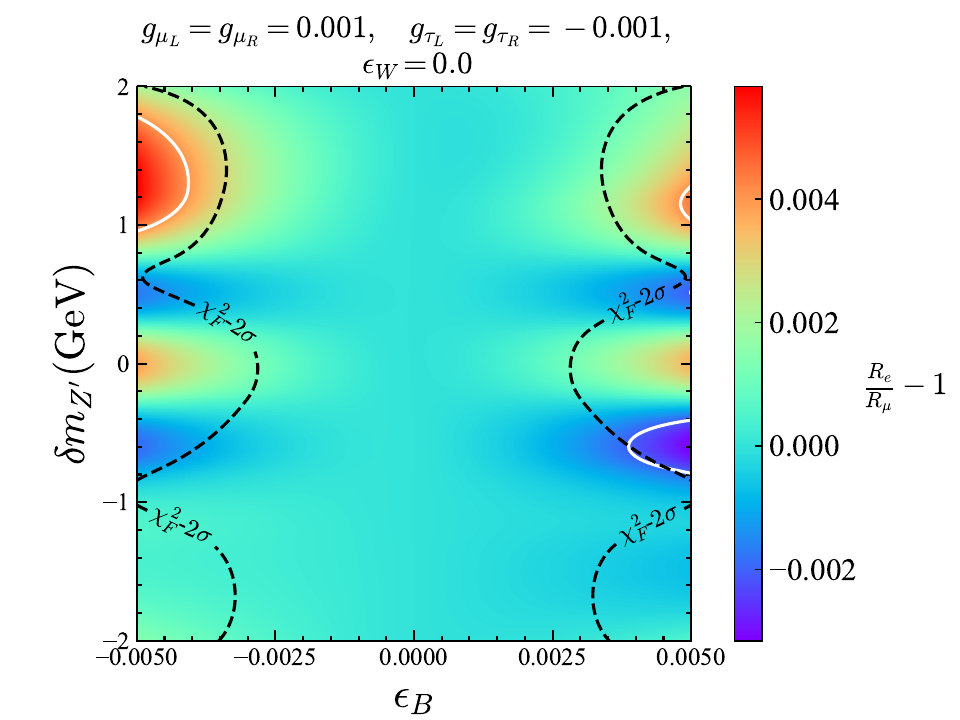}&
   \ \  \includegraphics[width=0.24\textwidth]{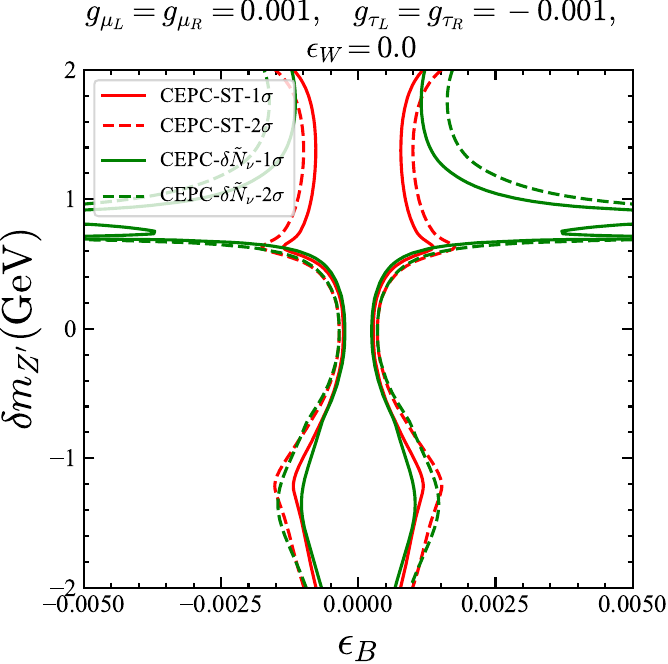}
    \end{tabular}
    \caption{Plots for $g_{\mu_L} = g_{\mu_R} = 0.001$, $g_{\tau_L} = g_{\tau_R} = -0.001$ and $\epsilon_W = \delta m^2  = 0$ on the $\epsilon_B$-$\delta m_{Z^{\prime}}$ plain. Compared with the Fig.~\ref{dmp-0.1_alpha0.9-delZp}, $\frac{R_{e}}{R_{\mu}}-1$ is additionally plotted.} \label{decaymuta0.1-delZp}
\end{figure}

\begin{figure}[!t]
    \begin{tabular}{ccc}
    \includegraphics[width=0.32\textwidth]{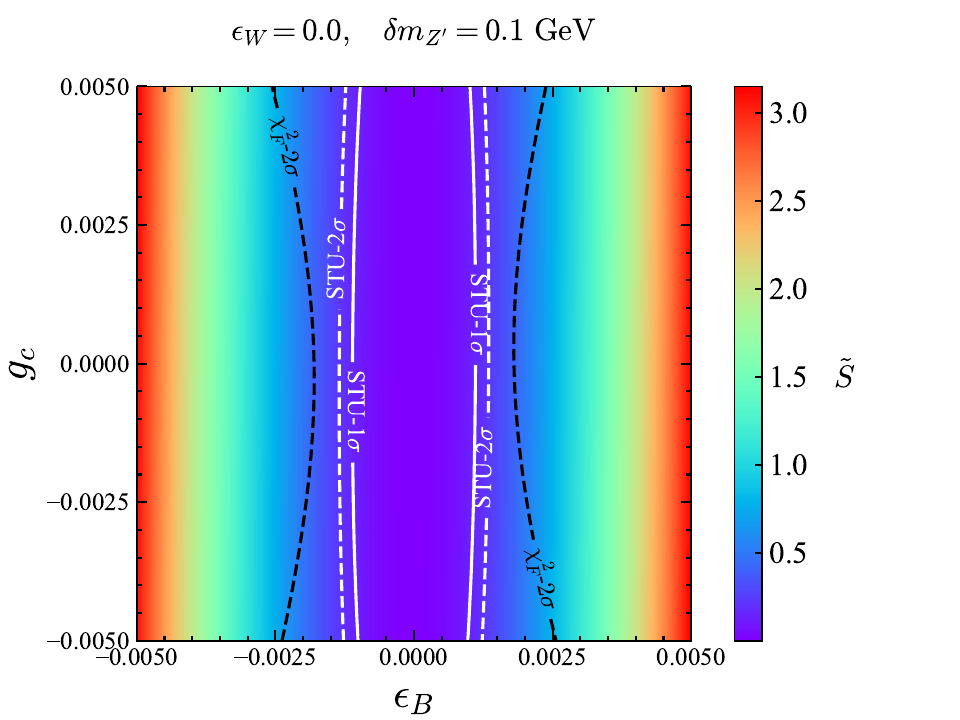} 
    \includegraphics[width=0.32\textwidth]{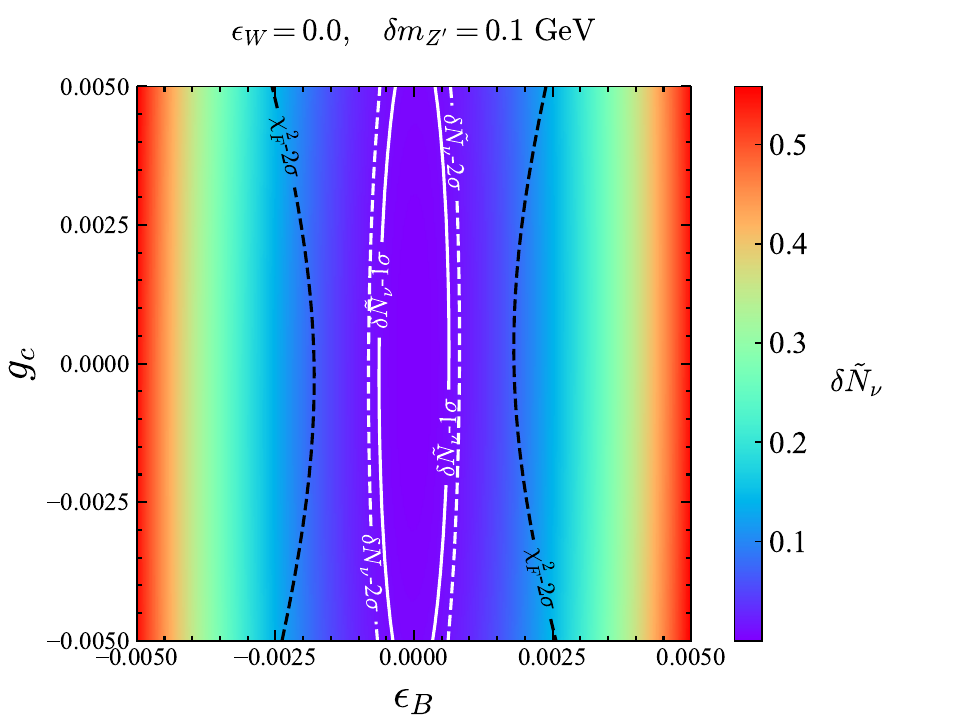}
    \includegraphics[width=0.32\textwidth]{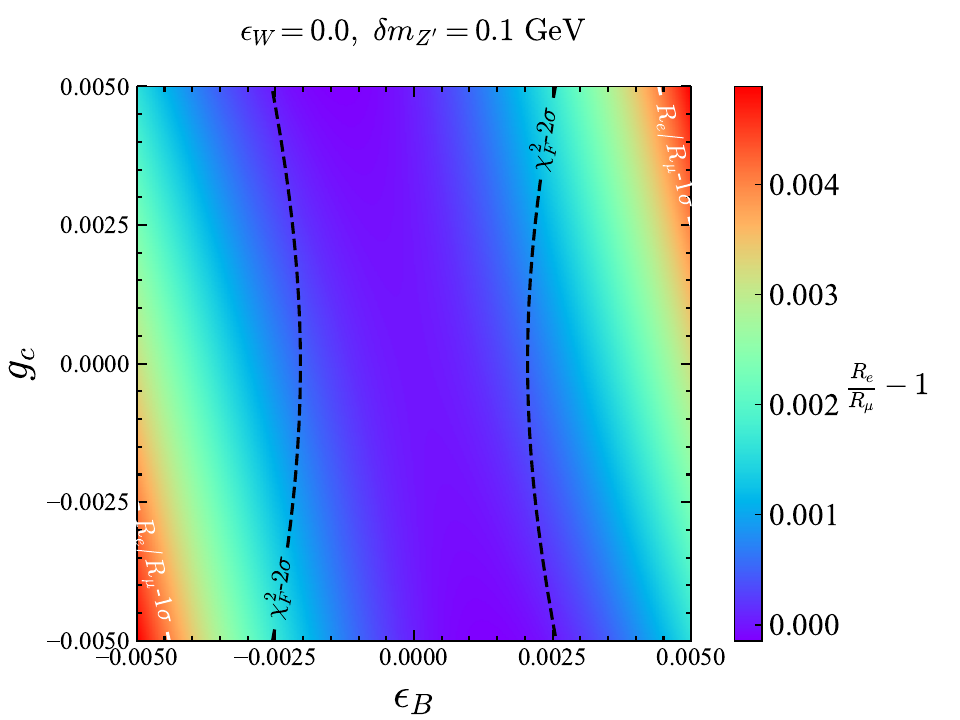}\\
    \includegraphics[width=0.24\textwidth]{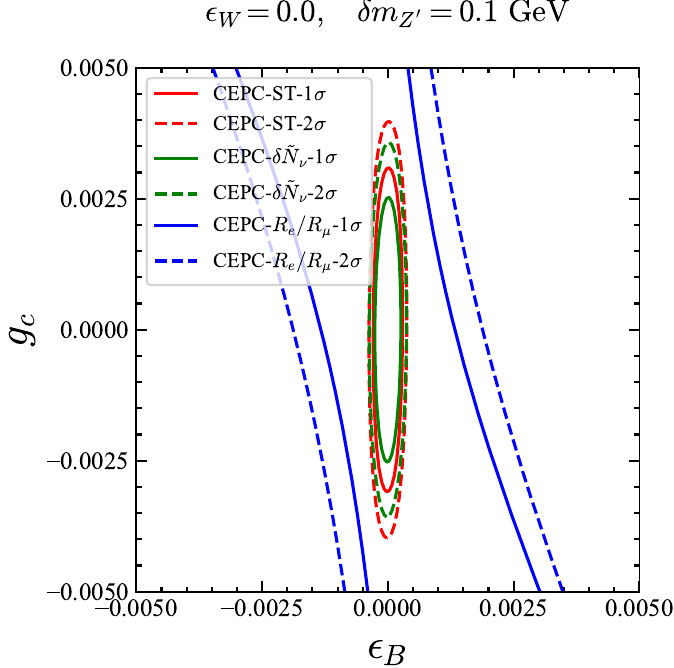}
    \end{tabular}
    \caption{Plots for $\delta m_{Z^{\prime}} = 0.1$~GeV, $g_{\mu_L} = g_{\mu_R} = -g_{\tau_L} = -g_{\tau_R} = g_c$ and $\epsilon_{W} = \delta m^2  = 0$ on the $\epsilon_B$-$g_c$ plain.  The $\tilde{T}$ and $\tilde{U}$ can be obtained by multiplying $\tilde{S}$ by the coefficients $1.5$ and $-1.2$, respectively. The other symbols are the same as in Fig.~\ref{decaymu0.1-gc}} \label{decaymuta0.1-gc}
\end{figure}

Inspired by the models that $Z^{\prime}$ couples with a particular generation of leptons, we show example plains that the $\hat{Z}^{\prime}$ only couples with the second family of the leptons in Fig.~\ref{decaymu0.1}, \ref{decaymu0.1-delZp}, \ref{decaymu0.1-gc}. Inspired by the $U(1)_{L_{\mu}-L_{\tau}}$ models\cite{He:1990pn, He:1991qd, Ma:2001md, Foot:1994vd, Baek:2001kca}, we show example plains in Fig.~\ref{decaymuta0.1}, \ref{decaymuta0.1-delZp}, \ref{decaymuta0.1-gc}. Notice that in some $\frac{R_{e}}{R_{\mu}}-1$ plots among them, e.g., in the $\frac{R_{e}}{R_{\mu}}-1$ plot in Fig.~\ref{decaymu0.1}, the current $\frac{R_{e}}{R_{\mu}}-1$ bounds are far less sensitive than the $\tilde{S}$, $\tilde{T}$, $\tilde{U}$, $\delta \tilde{N}_{\nu}$ parameters so that it disappears in the plot. We also have to note that our current algorithm cannot reach the reliable CEPC-$\frac{R_{e}}{R_{\mu}}-1$ constraints at Fig.~\ref{decaymu0.1-delZp} and Fig.~\ref{decaymuta0.1-delZp}. Since usually $\frac{R_{e}}{R_{\mu}}-1$ constraints are more relaxed than other parameters in the parameter space that we observe, we neglect the CEPC-$\frac{R_{e}}{R_{\mu}}-1$ constraints in these two figures.


\section{Summary} \label{summary}

In a nearly-degenerate $Z^{\prime}$-$Z$ system, not only the widths, or equivalently, the imaginary part of the self-energy diagram of each fields, but also the indispensable ``cross terms'', that is to say, the imaginary part of the self-energy diagram connecting two different fields play important roles in calculating the line-shape observables. After diagonalizing the ``mass matrix'' with the imaginary contributions included, sometimes ``SM-like $Z$'' and ``$Z^{\prime}$'' cannot be well discriminated and they overlap coherently to form a single resonance-like object which might be recognized as a single particle.



Relying on the effective field theory model in which many of the $Z^{\prime}$ models can be accommodated, we simulate the shape of this resonance-like object, and follow the usual literature to utilize ``oblique parameters'' $\tilde{S}$, $\tilde{T}$, $\tilde{U}$, and ``neutrino species deviation'' $\delta \tilde{N}_{\nu}$ to describe the shape of it. Comparing the results with the current data mainly contributed by the LEP, one can estimate the sensitivity of the LEP $\epsilon_{B}$ and $\epsilon_{W}$, $\delta m_{Z^{\prime}}$, and coupling constant parameter space if the LEP data can be reanalyzed. As a paradigm of the future high-luminosity lepton colliders, the predicted sensitivity of this model at the CEPC is also evaluated. Besides, we also estimate the $R_e/R_{\mu}-1$ of the non-universality models, and compare them with the LEP and CEPC sensitivities.

\begin{acknowledgements}
    We thank to Chengfeng Cai,  Zhao-Huan Yu, Hong-Hao Zhang, Weiming Song and Shao-Feng Ge for helpful discussions. This work is supported in part by the National Natural Science Foundation of China under Grants No. 12005312, the Guangzhou Science and Technology Program under Grant No.202201011556, and the Sun Yat-Sen University Science Foundation. Part of the calculation was performed on TianHe-2, and we thank for the support of National Supercomputing Center in Guangzhou (NSCC-GZ).
\end{acknowledgements}

\newpage

\appendix

\section{Resummation of the imaginary parts of the self-energy diagrams} \label{AppResum}

Let us start from a group of real scalar particles for simplicity, e.g., $\phi_i$, $i=1,2,\dots,n$, with the mass matrix to be $\mathcal{M}^2_s = \lbrace m^2_{i j} \rbrace$ where $m^2_{i j} = m^{2}_{j i}$ so $\mathcal{M}_s^{2 T} = \mathcal{M}_s^2$, and the mass terms being $\mathcal{L}_m = -\frac{1}{2} \phi_i m^2_{i j} \phi_j$. The complete propagator of these scalar particles can be written in the form of a matrix
\begin{eqnarray}
    \frac{i}{p^2 I_{n \times n} - \mathcal{M}^2_s} \equiv i (p^2 I_{n \times n} - \mathcal{M}^2_s)^{-1}.
\end{eqnarray}
The usual diagonalization processes to find the mass eigenstates of $\phi_i$'s are equivalent to finding an orthogonal matrix $V$ to diagonalize the propagator
\begin{eqnarray}
    i V  (p^2 I_{n \times n} - \mathcal{M}^2_s)^{-1} V^{-1} = \text{diag}\left[ \frac{i}{p^2-m_1^2}, \frac{i}{p^2-m_2^2}, \dots, \frac{i}{p^2-m_n^2} \right], \label{DiagonalizePropagator}
\end{eqnarray}
where $V$ is a real matrix to satisfy $V V^T = V^T V=I$. Expand the left-hand side of (\ref{DiagonalizePropagator}) according to $\mathcal{M}^2_s$, we acquire
\begin{eqnarray}
    & & i V (p^2 I_{n \times n} - \mathcal{M}^2_s)^{-1} V^{-1} = i V \frac{1}{p^2} \left( I + \sum_{t=1}^{n} (\frac{\mathcal{M}^2_s}{p^2})^n \right) V^{-1} \nonumber \\
    &=& i \frac{1}{p^2} \left[ I + \sum_{t=1}^{n} \left(\frac{V \mathcal{M}^2_s V^{-1}}{p^2}\right)^n \right] = i \left( p^2 I_{n \times n} - V \mathcal{M}_s^2 V^{-1} \right)^{-1}. \label{DiagonalizePropagatorProcess}
\end{eqnarray}
It is then clear to see that $\mathcal{M}_s^2$ and the propagator \eqref{DiagonalizePropagator} must be diagonalized at the same time.

Including all the imaginary parts of the $\phi_i$-$\phi_j$ self-energy diagrams introduces a resummation
\begin{eqnarray}
    \sum_{t=0}^{\infty} \frac{i}{p^2 I_{n \times n} - \mathcal{M}^2_s} \left[ (-i) \text{Im}( \Sigma(p^2)) \frac{i}{p^2 I_{n \times n} - \mathcal{M}^2_s} \right]^t = \frac{i}{p^2 I_{n \times n} - \mathcal{M}^2_s - i\text{Im}( \Pi(p^2) ) }, \label{ScalarResum}
\end{eqnarray}
where $-i \Pi(p^2)$ is a $n \times n$ symmetric matrix with each of its element $-i \Pi_{j k} (p^2)$ the (cross) self-energy diagram connecting $\phi_j$ and $\phi_k$. Therefore, a complete diagonalization process described in \eqref{DiagonalizePropagatorProcess} should be replaced by
\begin{eqnarray}
     i V (p^2 I_{n \times n} - \mathcal{M}^2_s - i\text{Im}( \Pi(p^2) ) )^{-1} V^{-1} = i \left\lbrace p^2 I_{n \times n} - V [\mathcal{M}_s^2 + i\text{Im}( \Pi(p^2) )] V^{-1} \right\rbrace^{-1}. \label{DiagonalizePropagatorProcess_WithWidth}
\end{eqnarray}
It is now clear that for the complete diagonalization processes including all the width informations, one should in turn diagonalize $\mathcal{M}_s^2 + i\text{Im}( \Pi(p^2) )$ at each momentum $p$ instead of a mere $\mathcal{M}_s^2$. Again, it is easy to verify that $\mathcal{M}_s^2 + \text{Im}( \Pi(p^2) )$ is a complex symmetric diagram, which guarantees the possibility of a successful diagonalization by a complex orthogonal diagram $V=(V^T)^{-1}$.

If all the scalars are nearly-degenerate around the mass $m_d$, a good approximation can be $p^2=m_d^2$ to preserve the accuracy of the near-shell performances of the propagators. This is a generalization of the Breig-Wigner propagator for the single particle into a nearly-degenerate multiple particle group.

Since we are discussing about the real scalars, and usually the complex orthogonal $V$ contains not only real numbers, if we treat $V$ as the matrix to recombine $\phi_{1,2,\dots,n}$ into ``mass eigenstates'' $\phi^{\prime}_j=V_{j k} \phi_k$ as usual, then $\phi^{\prime}_j$ is something ``complex'' but cannot be regarded as a ``complex scalar field''. Therefore, in this paper, we remind the reader that \eqref{DiagonalizePropagatorProcess_WithWidth} cannot be understood as a equivalence to diagonalizing the scalar fields, although we sometimes still apply this less rigorous terminology for brevity.

Similar discussions about the scalar fields also appeared in Ref.~\cite{Boyanovsky:2017esz}, while in this paper, we should change to the vector bosons. For the $W^3$-$B$-$Z^{\prime}$ system, the propagators should be accompanied with a Lorentz term
\begin{eqnarray}
    \frac{i g_{\mu \nu}}{p^2 I_{3 \times 3} - \mathcal{M}^2_V} \equiv i g_{\mu \nu} (p^2 I_{3 \times 3} - \mathcal{M}^2_V)^{-1},
\end{eqnarray}
if Feynmann gauge is adopted.  The complete discussions on a general $R_{\xi}$ gauge can be found in Ref.~\cite{Cacciapaglia:2009ic}, and note that the $\xi$-dependence does not make a significant effect in this paper. In principle Goldstone propagators should also be considered. However, around the $Z$-pole, only light leptons and quarks can be produced on-shell. Even the heaviest $b$ quark takes the $m_B \sim 3$-$5$ GeV mass, so the Goldstone/Higgs contributions are suppressed by a $\left( \frac{m_b}{v} \right)^2 \ll 1$ factor, and can be safely neglected.

Due to the Lorentz covariance, the one-loop self-energy diagram of the vector bosons can be decomposed and parametrized by
\begin{eqnarray}
    i \Pi^{\mu \nu}_{3 \times 3}(p^2) = i \Pi_{3 \times 3}(p^2) g^{\mu \nu} + i \Pi^{\prime}_{3 \times 3}(p^2) p^{\mu} p^{\nu}. \label{PiMuNu}
\end{eqnarray}
Then we can follow \eqref{ScalarResum} to resum the vector boson's propagators.  Similar to Ref.~\cite{Cacciapaglia:2009ic}, in the general $R_{\xi}$ gauge, the result is
\begin{eqnarray}
    &-& i (g^{\mu \nu} - \frac{p^{\mu} p^{\nu}}{p^2}) \left[ p^2 I_{3 \times 3} - \mathcal{M}^2_V - i\text{Im}( \Pi_{3 \times 3}(p^2) ) \right]^{-1} \nonumber \\
    &-& i \xi \frac{p^{\mu} p^{\nu}}{p^2} \left[p^2 I_{3 \times 3} - \xi(\mathcal{M}^2_V - i\text{Im}( \Pi_{3 \times 3}(p^2) - i p^2 \text{Im}( \Pi_{3 \times 3}^{\prime}(p^2) )\right]^{-1}, \label{Propagator_TL}
\end{eqnarray}
where $\xi$ is the $R_{\xi}$ gauge parameter. Here the first and second line of (\ref{Propagator_TL}) indicate the transverse and longitudinal contributions respectively. Then, we can again set $\xi$=1 for the Feynmann gauge.  Notice that all the masses of our initial and final state particles are ignorable since they are extremely relativistic particles, so once a $p^{\mu}$ in \eqref{PiMuNu} appears, it will finally dots into the fermionic propagators of the external legs. The Ward-Takahashi identity version in the broken phase\cite{Chanowitz:1985hj} transmute this into a Goldstone propagator, and its contributions are again suppressed by the smallness of the Yukawa couplings. Therefore, we are able to neglect the $p^{\mu} p^{\nu}$ terms in \eqref{PiMuNu}, and finally write down the resummed propagators when only imaginary parts of the \eqref{PiMuNu} are considered
\begin{eqnarray}
    -i g^{\mu \nu} \left[ p^2 I_{3 \times 3} - \mathcal{M}^2_V - i\text{Im}( \Pi_{3 \times 3}(p^2) ) \right]^{-1}. \label{Propagator_WithWidth}
\end{eqnarray}

In principle, $\text{Im}( \Pi_{3 \times 3}(p^2) )$ should be computed for each $p$. However, practical event generators are not designed for such kind of propagators.  Fixing $p^2 \approx m_Z^{\text{SM} 2}$ in \eqref{Propagator_WithWidth} moves the pole of the photon propagator to an unphysical non-zero point. Since we are talking about a $Z^{\prime}$-$Z$ system, and the photon lies too far away from them at the spectrum, it is convenient for us to rotate into the $\gamma$, $\hat{Z}$, $\hat{Z}^{\prime}$ basis if all the mixing terms are shut down. Then, we have to simply diagonalize
\begin{eqnarray}
    \mathcal{M}_V^{\text{SM}2}&=& \begin{pmatrix}
m_{\hat{Z}^{\prime}}^2 &  0 & 0 \\
0 &  \frac{\hat{g}^{\prime 2}}{4} \hat{v}^2&  -\frac{\hat{g}^{\prime} \hat{g}}{4} \hat{v}^2 \\
 0&  -\frac{\hat{g}^{\prime} \hat{g}}{4} \hat{v}^2  & \frac{\hat{g}^2}{4} \hat{v}^2  \end{pmatrix}. \label{MVSM}
\end{eqnarray}
Usually shutting down the mixing terms does not affect the self-energy calculations up to the one-loop order, since all the widths are acquired only from the couplings.  Diagonalizing \eqref{MVSM} requires $V_{\rm SM}$ defined in \eqref{VSM}, $V_{\rm SM}^T \mathcal{M}_V^{\text{SM}2} V_{\rm SM}$ gives
\begin{eqnarray}
    \text{diag}[m_{\hat{Z}^{\prime}}^2, m_{\hat{Z}}^2, 0], \label{SMDiagonalized}
\end{eqnarray}
corresponding to the masses of the ``$\hat{Z}^{\prime}$'', ``SM-like $\hat{Z}$'' with the ``hat'' symbols to distinguish them from the true SM mass-eigenstates, and the photon ``eigenstate'' masses respectively. Here $m_{\hat{Z}}^2 = \frac{\hat{g}^2+\hat{g}^{\prime 2}}{4} \hat{v}^2$.  By principle, we have to insert the imaginary part of all $\Pi_{\hat{Z} \hat{Z}}(p^2)$, $\Pi_{\hat{Z}^{\prime} \hat{Z}^{\prime}}(p^2)$, $\Pi_{\gamma \gamma}(p^2)$, $\Pi_{\gamma \hat{Z}^{\prime}}(p^2)$, $\Pi_{\gamma \hat{Z}}(p^2)$ and $\Pi_{\hat{Z}^{\prime} \hat{Z}}(p^2)$ into (\ref{MVSM}). However, due to the large difference between the photon and the $\hat{Z}^{\prime}$-$\hat{Z}$ mass spectrum, the contributions from the $\Pi_{\gamma \hat{Z}^{\prime}}(p^2)$ and $\Pi_{\gamma \hat{Z}}(p^2)$ are suppressed by a factor of $\sim \left[ \frac{{\rm Im}\left(\Pi_{\gamma \hat{Z}^{(\prime)}}\right)}{m_{\hat{Z}}^2} \right]^2$ during the diagonalization processes. Neglecting them also eliminate the $\Pi_{\gamma \gamma}(p^2)$ effects. Therefore, with the definition $c_{XY} \equiv \mathrm{Im}\left\lbrace  \Pi_{X \leftrightarrow Y} (p^2 \approx m_{X,Y}^2)\right\rbrace $, we only need to consider the $X,Y=\hat{Z},\hat{Z}^{\prime}$ terms, so that \eqref{SMDiagonalized} becomes
\begin{eqnarray}
    V_{\rm SM}^T \mathcal{M}_V^{\text{SM}2} V_{\rm SM} + i C_{\hat{Z}/\hat{Z}^{\prime}} = \begin{pmatrix}
      m_{\hat{Z}^{\prime}}^2 + i c_{\hat{Z}^{\prime} \hat{Z}^{\prime}} & i c_{\hat{Z}^{\prime} \hat{Z}} & 0 \\
      i c_{\hat{Z}^{\prime} \hat{Z}} & m_{\hat{Z}}^2 + i c_{\hat{Z} \hat{Z}} & 0\\
      0 & 0 & 0 \end{pmatrix}, \label{SMDiagonalized_WithWidth}
\end{eqnarray}
where $i C_{\hat{Z}/\hat{Z}^{\prime}}$ indicates the corrections from the imaginary parts of the self-energy diagrams. Then we utilize $V_{\rm SM}^{-1}=V_{\rm SM}^T$ to restore \eqref{SMDiagonalized_WithWidth} into the form under the ``interacting eigenstate basis'',
\begin{eqnarray}
    & & V_{\rm SM}(V_{\rm SM}^T \mathcal{M}_V^{\text{SM}2} V_{\rm SM} + i C_{\hat{Z}/\hat{Z}^{\prime}})V_{\rm SM}^T \nonumber \\
    &=& \begin{pmatrix}
      m_{\hat{Z}^{\prime}}^2 + i c_{\hat{Z}^{\prime} \hat{Z}^{\prime}} & -i \frac{\hat{g}^{\prime}}{\sqrt{\hat{g}^{\prime 2} + \hat{g}^2}} c_{\hat{Z}^{\prime} \hat{Z}} & i \frac{\hat{g}}{\sqrt{\hat{g}^{\prime 2} + \hat{g}^2}} c_{\hat{Z}^{\prime} \hat{Z}} \\
      -i \frac{\hat{g}^{\prime}}{\sqrt{\hat{g}^{\prime 2} + \hat{g}^2}} c_{\hat{Z}^{\prime} \hat{Z}} & \frac{\hat{g}^{\prime 2}}{4} \left( \hat{v}^2 + \frac{4 i c_{\hat{Z} \hat{Z}}}{\hat{g}^{\prime 2} + \hat{g}^2} \right) & -\frac{\hat{g}^{\prime} \hat{g}}{4} \left( \hat{v}^2 + \frac{4 i c_{\hat{Z} \hat{Z}}}{\hat{g}^{\prime 2} + \hat{g}^2} \right)\\
      i \frac{\hat{g}}{\sqrt{\hat{g}^{\prime 2} + \hat{g}^2}} c_{\hat{Z}^{\prime} \hat{Z}} & -\frac{\hat{g}^{\prime} \hat{g}}{4} \left( \hat{v}^2 + \frac{4 i c_{\hat{Z} \hat{Z}}}{\hat{g}^{\prime 2} + \hat{g}^2} \right) & \frac{\hat{g}^2}{4} \left( \hat{v}^2 + \frac{4 i c_{\hat{Z} \hat{Z}}}{\hat{g}^{\prime 2} + \hat{g}^2}\right) \end{pmatrix}. \label{SMRestored}
\end{eqnarray}
Supplement (\ref{SMRestored}) with the mixing terms $\delta m^2$ appeared in \eqref{MassOrigin}, \eqref{mass-squares} is then acquired.

\begin{figure}
    \centering
    \includegraphics[width=0.3\textwidth]{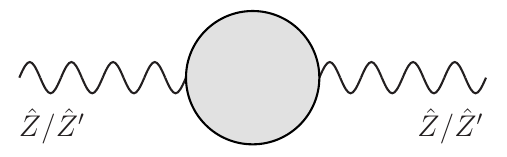}
    \includegraphics[width=0.3\textwidth]{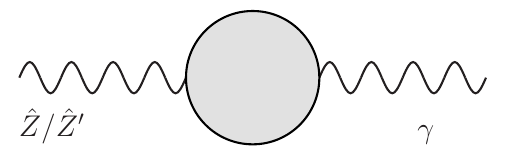}
    \includegraphics[width=0.3\textwidth]{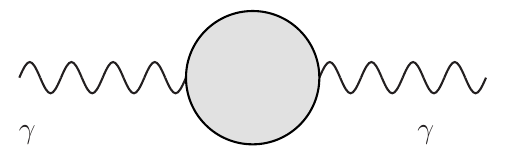}
    \caption{(Cross) self-energy terms connecting among $\hat{Z}^{\prime}$, $\hat{Z}$ and photon fields.}
    \label{CrossDiagrams}
\end{figure}

In the practical event simulation processes that we have performed, we at first shut down all the mixing terms to calculate the width of the SM-Z boson by the event generator to extract the $c_{\hat{Z} \hat{Z}}$ from it. Then, we compute $c_{\hat{Z}^{\prime} \hat{Z}}$ and $c_{\hat{Z}^{\prime} \hat{Z}^{\prime}}$ by hand relying on different model setups. After diagonalizing \eqref{mass-squares} by our own programs, we acquire the ``masses'' and ``widths'' of the ``mass eigenstates'', as well as the rotated ``coupling constants'' to be input into the event generator for further simulations, which is equivalent to diagonalizing the propagator matrix (\ref{Propagator_WithWidth}) to calculate the amplitudes.

\bibliographystyle{utphys}
\bibliography{ref}
\end{document}